\documentclass{article}
\usepackage{iclr2027_conference}
\usepackage{tabularx}

\usepackage{wrapfig,times}  
\usepackage[utf8]{inputenc}
\usepackage[T1]{fontenc}
\usepackage{booktabs}
\usepackage{amsfonts}

\usepackage[ruled,vlined]{algorithm2e}
\usepackage{algcompatible}

\makeatletter

\algnewcommand\algorithmicinput{\textbf{Input:}}
\algnewcommand\algorithmicoutput{\textbf{Output:}}
\algnewcommand\INPUT{\item[\algorithmicinput]}
\algnewcommand\OUTPUT{\item[\algorithmicoutput]}

\algnewcommand\algorithmicfunction{\textbf{Function}}
\algblockdefx[FUNCTION]{FUNCTION}{ENDFUNCTION}%
  [2]{\algorithmicfunction\ #1}{}
\algtext*{ENDFUNCTION} 

\usepackage{nicefrac}

\usepackage{amsmath, amssymb, amsthm}
\usepackage{graphicx}
\usepackage{svg}
\usepackage{subcaption}
\usepackage{float}
\usepackage{float}

\usepackage{multirow}
\usepackage{tikz}
\usepackage{wrapfig}
\usepackage{booktabs}
\usepackage{xcolor}
\usepackage{hyperref}
\usepackage{mysymbol}
\usepackage{setspace}
\usepackage{enumitem}
\usepackage{placeins}
\usepackage{wrapfig}

\usepackage{placeins}

\usepackage{caption}     

\usetikzlibrary{arrows.meta, positioning, shapes, calc, fit}
\usepackage{tcolorbox}
\definecolor{my-gray}{cmyk}{0.05, 0.0, 0.0, 0.05, 1.00}
\definecolor{my-blue}{cmyk}{0.15, 0.0, 0.0, 0.9, 1.00}
\usepackage{xcolor}

\providecommand{\td}[1]{}
\renewcommand{\td}[1]{\tilde{#1}}

\usepackage[patch=none]{microtype}
\makeatletter
\newcommand{\storelabel}[1]{%
  \label{#1}%
  \expandafter\gdef\csname labelname@#1\endcsname{#1}%
}
\newcommand{\labref}[1]{\csname labelname@#1\endcsname}
\makeatother

\newif\ifdraft
\draftfalse  

\ifdraft
  \newcommand{\menda}[1]{\textcolor{cyan}{[tony: #1]}}
\else
  \newcommand{\igna}[1]{}
  \newcommand{\menda}[1]{}
\fi

\newtheorem{proposition}{Proposition}
\newtheorem{theorem}{Theorem}
\newtheorem{lemma}{Lemma}
\newtheorem{corollary}{Corollary}
\newtheoremstyle{upright}
  {3pt}{3pt}
  {\normalfont}
  {}
  {\bfseries}
  {.}
  { }
  {}
\theoremstyle{upright}
\newtheorem{example}{Example}

\newtheorem{remark}{Remark}

\title{Infinity Search: Approximate Vector Search with Projections on q-Metric Spaces}

\author{
\begin{tabular}{cc}
\parbox{0.45\textwidth}{\normalfont\raggedright
\textbf{Antonio Pariente} \\
Department of Electrical and Systems Engineering \\
University of Pennsylvania \\
Philadelphia, PA, USA \\
\texttt{pariente@seas.upenn.edu}
}
&
\parbox{0.45\textwidth}{\normalfont\raggedright
\textbf{Ignacio Hounie} \\
Department of Electrical and Systems Engineering \\
University of Pennsylvania \\
Philadelphia, PA, USA \\
\texttt{ihounie@seas.upenn.edu}
}
\\[5em]
\parbox{0.45\textwidth}{\normalfont\raggedright
\textbf{Santiago Segarra} \\
Department of Electrical and Computer Engineering \\
Rice University \\
Houston, TX, USA \\
\texttt{ssegarra@rice.upenn.edu}
}
&
\parbox{0.45\textwidth}{\normalfont\raggedright
\textbf{Alejandro Ribeiro} \\
Department of Electrical and Systems Engineering \\
University of Pennsylvania \\
Philadelphia, PA, USA \\
\texttt{aribeiro@seas.upenn.edu}
}
\end{tabular}
}

\svgpath{{Images/}}
\graphicspath{{Images/}}

\iclrfinalcopy
\begin{document}

\maketitle
\lhead{Preprint.}
\begin{abstract}
An ultrametric space or infinity-metric space is defined by a dissimilarity function that satisfies a strong triangle inequality in which every side of a triangle is not larger than the larger of the other two. We show that search in ultrametric spaces with a vantage point tree has worst-case complexity equal to the depth of the tree. Since datasets of interest are not ultrametric in general, we employ a projection operator that transforms an arbitrary dissimilarity function into an ultrametric space while preserving nearest neighbors. We further learn an approximation of this projection operator to efficiently compute ultrametric distances between query points and points in the dataset. We proceed to solve a more general problem in which we consider projections in $q$-metric spaces -- in which triangle sides raised to the power of $q$ are smaller than the sum of the $q$-powers of the other two. Notice that the use of learned approximations of projected $q$-metric distances renders the search pipeline approximate. We show in experiments that increasing values of $q$ result in faster search but lower recall. Overall, search in q-metric and infinity metric spaces is competitive with existing search methods.
\end{abstract}



\section{Introduction}

Given a dataset of vector embeddings, a dissimilarity function and a query, nearest neighbor search refers to the problem of finding the point in the dataset that is most similar to the query, or, in its stead, a vector that is among the most similar \citep{IndykMotwani98}. Exact search requires comparing against all points in the dataset in the worst case \citep{murtagh2012futuresearchdiscoverybig} but it is well known that this search complexity can be reduced in the particular case of metric dissimilarities. In this case we can organize data in a vantage point (VP) tree that we can search with a smaller average number of comparisons \citep{vptree}. This happens because we can use triangle inequalities to prune \emph{some} branches of the VP tree \citep{vptree}.

In this paper we observe that metric spaces are a particular case of what we term here as $q$-metric spaces. These spaces satisfy more restrictive triangle inequalities in which the $q$-th power of each side of a triangle is not larger than the sum of the $q$th powers of the other two. The advantage of $q$-metric spaces is that the inequalities that make pruning of metric trees in standard metric spaces possible (Section \ref{sec_VP_trees_and_metric_structure}) become more lenient (Section \ref{sec_VP_trees_and_q_metric_structure}). This fact substantiates the first contribution of our paper:

\begin{itemize}

\item [{(C1)}] We demonstrate empirically that search in a $q$-metric space requires a smaller number of comparisons than search in a standard metric space (Section \ref{section:experiments}).

\end{itemize}

The limit of growing $q$ yields $\infty$-metric spaces in which each side of a triangle is not larger than the maximum of the other two (Section \ref{section:NNS}). The value of $\infty$-metric spaces is that the conditions that allow pruning of VP trees are always met (Section \ref{sec_VP_trees_and_infty_metric_structure}). This fact leads to our second contribution:

\begin{itemize}

\item [{(C2)}] The number of comparisons needed to find a nearest neighbor of a query in an $\infty$-metric space is, at most, the depth of the VP tree (Theorem \ref{theo_log_complexity}). We further demonstrate empirically that, as it would be expected from this observation, search in $\infty$-metric spaces requires a number of comparisons close to the base-2 logarithm of the dataset's cardinality (Section \ref{sec_VP_trees_and_infty_metric_structure}).

\end{itemize}

We point out that $\infty$-metric spaces are often called ultrametric spaces and that the $\infty$-triangle inequality is often called the strong triangle inequality \citep{dovgoshey2025totally}. Ultrametric spaces are equivalent to the dendrograms used in hierarchical clustering \citep{draganov2025want}, a fact that may help to understand why search in ultrametric spaces exhibits complexity close to logarithmic.

Because of (C1) we prefer to solve search problems in $q$-metric spaces with larger $q$. Because of (C2) we most prefer to solve search in $\infty$-metric spaces, thus the title of this paper. However, the data is what the data is and most problems in vector search involve dissimilarity functions that are not even metric \citep{zezula2006similarity}. We thus seek projection operators to map datasets into general $q$-metric or $\infty$-metric spaces. We propose to adopt the \emph{canonical projections} developed first for hierarchical clustering \citep{hierarchical-quasi-clustering, smith2016hierarchical, carlsson2017admissible} and generalized later to metric representations of network data \citep{segarra2019metricoriginal, 9054190}.

These canonical projections generate $q$-metric spaces by computing shortest $q$-norm paths linking any pair of points (Section \ref{section:metric_representation}). In particular, they generate $\infty$-metric spaces by computing shortest $\infty$-norm paths \footnote{We emphasize that the $q$-norm of a path is unrelated to the $q$-norm of the vector embeddings. As all norms do, the $q$-norm of the vector embeddings satisfies the \emph{standard} triangle inequality.}. These projections are deemed canonical because they are the only ones that abide to the (very justifiable) Axioms of Projection and Transformation \citep{segarra2019metricoriginal}. The Axiom of Projection states that the projection of a $q$-metric space should be the same $q$-metric space. The Axiom of Transformation states that the projection of a $q$-metric space must respect the partial ordering of original spaces. I.e., if the dissimilarities of a dataset dominate \emph{all} the dissimilarities of another, the $q$-metric distances of the projected datasets must satisfy the same relationship. To strengthen the case for using canonical projections, we provide a third contribution:

\begin{itemize}

\item [{(C3)}] The canonical projections respect the nearest neighbors of any query (Section \ref{sec_nn_is_preserved}): A point that is a nearest neighbor of a query with respect to the original dissimilarity function is also a nearest neighbor with respect to its canonical $q$-metric or $\infty$-metric projections. 

\end{itemize}

To the extent that the Axioms of Projection and Transformation are reasonable, the canonical projections are the only reasonable methods that we can use to generate a $q$-metric space \cite{segarra2019metricoriginal}. Regardless, (C3) states that the canonical projections respect nearest neighbors, which is all we need for search.

Contributions (C1)-(C3) dictate that we can project datasets into $q$-metric and $\infty$-metric spaces with a canonical projection to search with low complexity. The hitch is that canonical projections require computation of $q$-shortest paths or $\infty$-shortest paths. This can be precomputed for the data but doing so for a query requires that we first compute distances between the query and \emph{all} the points in the dataset. This defeats the purpose of using $q$-metric or $\infty$-metric distances to reduce search complexity. We work around this problem with our fourth contribution:

\begin{itemize}

\item [{(C4)}] We learn a map from original embeddings to auxiliary embeddings whose Euclidean distances estimate $q$-metric or $\infty$-metric distances. We train this map on the dataset and generalize it to queries (Section \ref{section:approximation}).

\end{itemize}

Combining (C1)-(C4) we propose to use canonical $q$-metric and $\infty$-metric projections of a dataset to generate spaces where we expect nearest neighbor search to be more efficient. We incur this cost once and reutilize it for all subsequent searches. We then use the learned embedding in (C4) to estimate $q$-metric distances between queries and points in the dataset. The use of approximate computation of $q$-metric and $\infty$-metric projections renders the resulting search algorithm approximate. Numerical experiments indicate that this approach is competitive with state-of-the-art approximate search algorithms in terms of search complexity and recall (Section \ref{section:experiments}). We refer the reader to Appendix \ref{sec_related_work} for a discussion of related work.

\begin{figure*}[t]

\centering

\includegraphics[width = \linewidth]{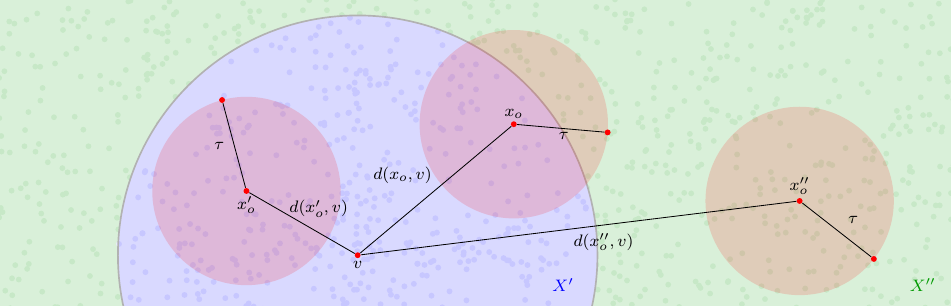}

\caption{Search in metric spaces requires fewer comparisons than search in arbitrary spaces because for \emph{some} queries -- such as $x_o'$ and $x_o''$ -- the triangle inequality allows us to restrict comparisons to subsets of the dataset $X$. Queries, such as $x_o$, for which triangle inequality bounds are inconclusive, also exist \eqref{eqn_vp_fail_condition}. This latter eventuality is impossible when the strong triangle inequality \eqref{eqn_strong_triangle_inequality} holds, leading to a more efficient search in ultrametric spaces (Theorem \ref{theo_log_complexity}).}

\label{fig_vp_tree_inequalities}

\end{figure*}


\def \xnn {\hhatx_o}


\section{Nearest Neighbor Search in \texorpdfstring{$q$}{q}-Metric Spaces}
\label{section:NNS}
We are given a set $X$ containing $m$ data vectors $x\in\reals^n$ along with a nonnegative dissimilarity function $d:\reals^n\times\reals^n \to\reals_+$ so that smaller values of $d(x,y)$ represent more similarity between data vectors $x$ and $y$. The set $X$ along with the set $D$ containing all dissimilarities $d(x,y)$ for all points $x,y\in X$ defines a fully connected weighted graph $G=(X,D)$. We assume here that $d(x,y) = d(y,x)$ for all $x,y\in \reals^n$, which, in particular, implies that the graph $G$ is symmetric. Examples of dissimilarity functions used in vector search are the Euclidean distance, Manhattan (1-norm) distance, cosine dissimilarity, and dissimilarities based on the Jaccard index \citep{zezula2006similarity}; see \ref{app:approximation}. 

We are further given a vector $x_o$ which is not necessarily an element of $X$ but for which it is possible to evaluate the dissimilarity function $d(x_o,x)$ for all $x\in X$. The nearest neighbors of $x_o$ in the set $X$ are the elements that are most similar to $x_o$,
\begin{align}\label{eqn_nearest_neighbor}
    \xnn \equiv \argmin_{x\in X} d(x, x_o) .
\end{align}
The problem of finding \emph{a} vector $\xnn$ is termed nearest neighbor search, the vector $x_o$ is called a query and the vector $\xnn$ is the query's answer \citep{wang2014hashing}. It may be that there are several nearest neighbors. In such case we overload $\xnn$ to denote an arbitrary nearest neighbor and also the set of all nearest neighbors. We make the distinction clear when needed. 

We evaluate the performance of a search algorithm by the number $c(x_o)$ of comparisons against elements of $X$ that are needed to find $\xnn$. Without further assumptions on the dissimilarity function $d$, we have $c(x_o) = m$ because we need to compare the query $x_o$ to all points in $X$. This number of comparisons can be reduced if we assume that $d$ has some metric structure \citep{murtagh2012futuresearchdiscoverybig}. For instance, it may be that $d$ is a proper metric or pseudometric that satisfies the triangle inequality so that for any three points $x,y,z \in \reals^n$ we have that
\begin{align}\label{eqn_triangle_inequality}
    d(x, y) \leq d(x, z) + d (y, z) .
\end{align}
Alternatively, we may consider ultrametric dissimilarity functions $d$ \citep{simovici2004ultrametricity}. In this case, triplets of points $x,y,z \in \reals^n$ satisfy the strong triangle inequality,
\begin{align}\label{eqn_strong_triangle_inequality}
    d(x, y) \leq \max\Big[\, d(x, z) , \, d (y, z) \, \Big] .
\end{align}
In this paper, we also have an interest in $q$-metric spaces \citep{greenhoe2016power}. These spaces arise when the dissimilarity function is such that any three points $x,y,z \in \reals^n$ satisfy the $q$-triangle inequality\footnote{Notice that the $q$-norm $||x||_q$ for $x\in\reals^n$ satisfies the regular triangle inequality, \emph{not} the $q$-triangle inequality.},
\begin{align}\label{eqn_q_triangle_inequality}
    d^q(x, y) \leq d^q(x, z) + d^q (y, z) ,
\end{align}
for some given $q \geq 1$. For $q=1$, \eqref{eqn_q_triangle_inequality} reduces to \eqref{eqn_triangle_inequality} and as $q\to\infty$, \eqref{eqn_q_triangle_inequality} approaches \eqref{eqn_strong_triangle_inequality}. Thus, we can think of $q$-metric spaces as interpolations between regular metric spaces that sastify the standard triangle inequality and ultrametric spaces that satisfy the strong triangle inequality. Henceforth, we may refer to \eqref{eqn_triangle_inequality} as the 1-triangle inequality and to \eqref{eqn_strong_triangle_inequality} as the $\infty$-triangle inequality.


\subsection{Vantage Point Trees and Metric Structure}\label{sec_VP_trees_and_metric_structure}

A classical structure that exploits metric structure to simplify search is a vantage point (VP) tree \cite{vptree}. As we show in Figure \ref{fig_vp_tree_inequalities}, a VP tree is constructed from the partition of the set $X$ into a vantage point $v$ and points other than $v$ that are closer or farther apart from $v$ than a threshold $\mu$
\begin{alignat}{2}\label{eqn_vp_children_main}
    X'  &~:=~ \Big\{\, x \in X \,:\, d(v,x) < \mu \,,\, x \neq v \, \Big\}, \quad
    X'' &~:=~ \Big\{\, x \in X \,:\, \mu \leq d(v,x) \, \Big\}.
\end{alignat}
A VP tree is built by recursive partition of the inside set $X'$ and the outside set $X''$ using corresponding vantage points and radiuses. I.e., we select a vantage point $v'\in X'$ and a radius $\mu'$ to divide $X'$ into inside and outside sets and do the same for $X''$. We then partition the new group of inside and outside sets and repeat until we reach indivisible sets with only one point; see Appendices \ref{app:Vp-Trees} and \ref{sec_vp_trees_algorithms} for details. 

When coupled with metric structure, VP trees may reduce the number of comparisons $c(x_o)$ needed to find a nearest neighbor. This is because comparing the query $x_o$ to a vantage point $v$ allows us to guarantee that to find a nearest neighbor it suffices to explore either the inside or outside set. Indeed, suppose that an upper bound $\tau$ on the distance between query and nearest neighbor is available and consider for the query $x_o'$ in Figure \ref{fig_vp_tree_inequalities}. It is important for forthcoming discussions that $\tau' \leq d(v, x_o')$ because we can otherwise set $\tau' = d(v, x_o')$. If it happens that $x_o'$, $\tau$ and $\mu$ are such that 
\begin{align}\label{eqn_vp_CI_condition}
    d(v, x_o') + \tau < \mu, \tag{CI}
\end{align}
there is no point in the outside set $X''$ with a distance to $x_o'$ smaller than $\tau$. Thus, we can discard the outside set $X''$ and search only the inside set $X'$ -- see Proposition \ref{prop:q-vp-pruning} for a formal proof. Likewise, the query $x_o''$ and the nearest neighbor distance upper bound $\tau$ in Figure \ref{fig_vp_tree_inequalities} are such that
\begin{align}\label{eqn_vp_CO_condition}
    \mu + \tau \leq d(v, x_o''). \tag{CO}
\end{align}
If this condition holds we know that there is no point in the inside set $X'$ with a distance to $x_o'$ smaller than $\tau$. Thus, we can discard the inside set $X'$ and search only the outside set $X''$ -- see Proposition \ref{prop:q-vp-pruning} for a formal proof. While this argument seems to indicate logarithmic complexity of nearest neighbor search in metric spaces, this is not quite so. The reason is that we may have queries such as $x_o$ in Figure \ref{fig_vp_tree_inequalities} for which neither condition is true. That is, there may be points $x_o$ such that,
\begin{align}\label{eqn_vp_fail_condition}
    \mu - \tau ~\leq~ d(v, x_o) ~<~ \mu + \tau,
\end{align}
If $d(x_o, v)$ is such that \eqref{eqn_vp_fail_condition} holds, the nearest neighbor $\xnn$ may be in $X'$ or $X''$ and we therefore do not have a reduction in the number of comparisons needed to find $\xnn$.


\subsection{Vantage Point Trees and \texorpdfstring{$q$}{q}-Metric Structure}
\label{sec_VP_trees_and_q_metric_structure}

The argument we build in Figure \ref{fig_vp_tree_inequalities} can be generalized to $q$-triangle inequality comparisons in $q$-metric spaces. I.e., we consider inequalities analogous to those in \eqref{eqn_vp_CI_condition} and \eqref{eqn_vp_CO_condition} in which respective quantities are raised to the power of $q$,
\makeatletter
\begin{align*}
d^q(v,x_o')+\tau^q &< \mu^q
\quad
\textup{($q$-CI)}
\begingroup
  \def\@currentlabel{$q$-CI}%
  \ltx@label{eqn_vp_q_CI_condition}%
\endgroup
&&
\mu^q+\tau^q \le d^q(v,x_o'')
\quad
\textup{($q$-CO)}
\begingroup
  \def\@currentlabel{$q$-CO}%
  \ltx@label{eqn_vp_q_CO_condition}%
\endgroup
\end{align*}
\makeatother
We show in Proposition \ref{prop:q-vp-pruning} in Appendix \ref{sec_vp_trees_q_metric_spaces} that if we are searching in a $q$-metric space we can prune branches of VP trees in a similar way to how we prune branches in metric spaces. If Condition \eqref{eqn_vp_q_CI_condition} holds, we can discard the outside set $X''$ and focus the search in the inside set $X'$. If Condition \eqref{eqn_vp_q_CO_condition} holds, we can discard the inside set $X'$ and focus the search in the outside set $X''$.

As is the case of metric spaces, it is not always possible to prune VP trees using conditions \eqref{eqn_vp_q_CI_condition} and \eqref{eqn_vp_q_CO_condition}. There may be queries $x_o$ for which 
\begin{align}\label{eqn_vp_q_fail_condition}
    \mu^q - \tau^q ~\leq~ d^q(v, x_o) ~<~ \mu^q + \tau^q,
\end{align}
that require exploration of the inside and outside sets. Observe that for given $\mu$ and $\tau$ the range of dissimilarity values $d(v, x_o)$ that satisfy \eqref{eqn_vp_q_fail_condition} for $q>1$ is larger than the range of dissimilarity values $d(v, x_o)$ that satisfy \eqref{eqn_vp_fail_condition}. Furthermore, the range of dissimilarity values $d(v, x_o)$ that satisfy \eqref{eqn_vp_q_fail_condition} shrinks monotonically with increasing $q$. One then would expect that pruning of VP trees in $q$-metric spaces is more efficient for larger $q$. This expectation is confirmed by experiments (Section \ref{section:Experiments}).


\subsection{Vantage Point Trees and \texorpdfstring{$\infty$}{infinity}-Metric Structure}
\label{sec_VP_trees_and_infty_metric_structure}

As $q$ grows, the $q$-metric dissimilarity function approaches an ultrametric dissimilarity function and Conditions \eqref{eqn_vp_q_CI_condition} and \eqref{eqn_vp_q_CO_condition} reduce to conditions that involve maxima instead of sums of $q$-powers ,
\makeatletter
\begin{align*}
\max\big(\,d(v,x_o),\,\tau\,\big) &< \mu
\quad
\textup{($\infty$-CI)}
\begingroup
  \def\@currentlabel{$\infty$-CI}%
  \ltx@label{eqn_vp_infty_CI_condition}%
\endgroup
&&
\max\big(\,\mu,\,\tau\,\big) \leq d(v,x_o'')
\quad
\textup{($\infty$-CO)}
\begingroup
  \def\@currentlabel{$\infty$-CO}%
  \ltx@label{eqn_vp_infty_CO_condition}%
\endgroup
\end{align*}
\makeatother
These conditions can be further simplified if we recall that $\tau \leq d(v, x_o')$ in \eqref{eqn_vp_infty_CI_condition} and that $\tau \leq d(v, x_o'')$ in \eqref{eqn_vp_infty_CO_condition}. These observations imply that the presence of $\tau$ in these inequalities is moot as conditions \eqref{eqn_vp_infty_CI_condition} and \eqref{eqn_vp_infty_CO_condition} can be reduced to
\begin{align}\label{eqn_infty_vp_prune}
    d(v, x_o') ~ <    ~ \mu,        \qquad
    \mu       ~ \leq ~ d(v, x_o''). 
\end{align}
We show in Proposition \ref{prop:inf-vp-pruning} in Appendix \ref{sec_vp_trees_infty_metric_spaces} that if we are searching in an $\infty$-metric space we can prune branches of VP trees using these two conditions. For queries $x_o'$ such that $d(v, x_o') < \mu$ [cf. \eqref{eqn_vp_infty_CI_condition}] we can restrict the search for a nearest neighbor to the inside set $X'$. For queries $x_o''$ such that $\mu \leq d(v, x_o'')$ [cf. \eqref{eqn_vp_infty_CO_condition}] we can restrict the search for a nearest neighbor to the outside set $X''$.

Remarkably, the conditions in \eqref{eqn_infty_vp_prune} are mutually exclusive. Every query must satisfy one and exactly one of these two conditions. This fact implies that in each vantage point comparison we discard one branch of the VP tree and that the cost of finding a nearest neighbor is upper bound by the depth of the depth of the VP tree. We summarize this claim in the following theorem

\begin{wrapfigure}{R}{0.45\columnwidth}
    \centering
    \vspace{-2em}
    \includegraphics[width=1\linewidth]{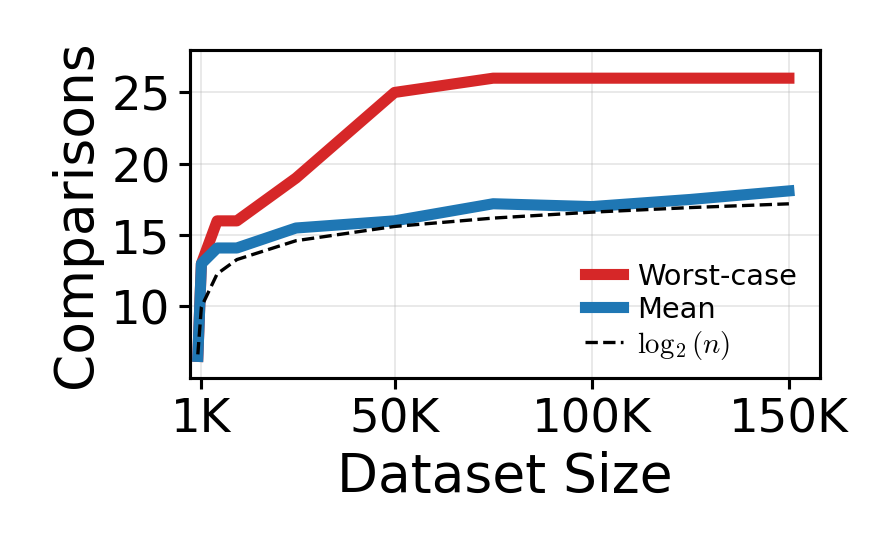}
    \caption{\textbf{VP Tree search complexity on a $\infty$-metric space}. The worst-case bound in comparisons corresponds to the depth of the tree.}
    \label{fig:depth_scaling}
    \vspace{-0.8em}
\end{wrapfigure}

\begin{theorem}\label{theo_log_complexity}

Consider a dataset $X$, a query $x_o$, a dissimilarity function $d$ satisfying the strong triangle inequality \eqref{eqn_strong_triangle_inequality} and a vantage point tree $T(X)$ constructed by the recursive partition of $X$ into vantage points and their corresponding inside and outside sets [cf. \eqref{eqn_vp_children_main}]. The number of comparisons $c(x_o)$ needed to find a nearest neighbor of $x_o$ in the set $X$ is bounded by the depth $h(T(X))$ of the VP tree $T(X)$,
\begin{align}\label{eqn_theo_log_complexity}
    c(x_o) \leq h(T(X)) .
\end{align}

\end{theorem}

\begin{proof} See Appendix \ref{app:Vp-Trees}. \end{proof}


When we split the set $X$ into a vantage point $v$ and corresponding inside and outside sets, we can choose the radius $\mu$ as the median of the distances 
\begin{align}\label{eqn_medians}
    \mu ~= ~ \median_{x \in X,\, x \neq v} d(v,x).
\end{align}
If there is only one point $x \in X$ that attains the median, this choice results in a balanced assignment of points to the inside and outside sets. They each contain the same number of points plus or minus one -- to account for odd numbers of points to be assigned. If this happens throughout the recursive construction of the VP tree, the resulting tree is said to be balanced and has a depth equal to the ceiling of the base-2 logarithm of the number of points in the original dataset. I.e. the depth of a balanced tree is $h(T(X))) = \lceil \log_2 m \rceil$. According to Theorem \ref{theo_log_complexity}, this quantity is also a bound for the worst case number of comparisons for any query in the case of ultrametric dissimilarity functions. 

In practice, we do observe that several points may achieve the median in \eqref{eqn_medians}. As per \eqref{eqn_vp_children_main}, all of these points are assigned to the outside set $X''$ resulting in unbalanced VP trees. While this yields trees with depth larger than $h(T(X))) = \lceil \log_2 m \rceil$, we still observe a logarithmic grow in the depth of the VP tree. A typical empirical evaluation is shown in Figure \ref{fig:depth_scaling}. We also observe that the mean number of comparisons stays close to the ideal depth $h(T(X))) = \lceil \log_2 m \rceil$ suggesting that the path to most leaves of the VP tree is close to this ideal value.


\subsection{Approximate Search with \texorpdfstring{$q$}{q}-Metric Spaces}
\label{sec_approximate_search_preview}

It follows from Theorem \ref{theo_log_complexity} that, given a choice, we would like to solve nearest neighbor search in ultrametric spaces. 
Alas, dissimilarity metrics of interest do not satisfy the strong triangle inequality. Some are not even metric. Due to this mismatch, we propose here an approach to \emph{approximate} nearest neighbor search based on the development of projection and embedding operators to compute and approximate $q$-metric distances:

\begin{description}

\item[Projection Operator.] The projection operator $P_q$ maps dissimilarities $d(x,x') \in D$ that do not necessarily satisfy the $q$-triangle inequality into distances $d_q(x,x') \in D_q$ that do satisfy the $q$-triangle inequality (Section \ref{section:metric_representation}). 

\item [Embedding Operator.] The embedding operator $\Phi_q$ is a learned map from vectors $x \in \reals^n$ into vectors $x_q\in\reals^s$ such that 2-norms $\|x_q - x_q'\|_2$ approximate $q$-distances $d_q(x,x')$. The map is trained on the dataset $X$ and applied to queries $x_o$ (Section \ref{section:approximation}).

\end{description}

We use the projection operator $P_q$ to process the dataset $G=(X,D)$ to produce a graph $G_q = (X, D_q)$ such that any three points $x,y,z \in X$ satisfy the $q$-triangle inequality. This is a one-time preprocessing cost that we leverage for all queries. We use the learned embedding operator $\Phi_q$ to approximate the $q$-metric distances between a query $x_o$ and points $x\in X$ as $d_q(x_o,x) \approx \|x_{oq} - x_q \|_2$ with $x_{oq} = \Phi_q(x_o)$. This is needed because computing the true $q$-metric distance $d_q(x_o,x)$ requires comparing $x_o$ to all points in $X$ (Section \ref{section:metric_representation}). Approximating $q$-metric distances with the embedding operator $E_q$ renders this search methodology approximate. Experiments in Section \ref{section:experiments}, show that we are competitive with existing approximate search methods.


\section{Projecting Dissimilarity Functions on \texorpdfstring{$q$}{q}-Metric Spaces}
 \label{section:metric_representation}

We seek to design a \emph{$q$‑metric projection} $P_q :\;(X,D)\;\to\;(X,D_q)$ where, for arbitrary input dissimilarities $D$, the distances in $D_q$ satisfy the $q$‑triangle inequality given in~\eqref{eqn_q_triangle_inequality}.
Observe that finding a feasible projection is trivial.
For example, consider a projection that assigns all distances in $D_q$ to be $1$, regardless of the input $D$.
In this case, the image space $(X,D_q)$ becomes an ultrametric (hence, a valid $q$-metric for all $q$).
However, performing nearest neighbor search in $(X,D_q)$ would be a poor proxy for searching in $(X,D)$, as all distance information was lost in the projection.
Motivated by this example, we build on the work of~\cite{segarra2019metricoriginal}, which imposes conditions on $P_q$ to ensure that meaningful distance information is preserved through the projection. In their work, these requirements were formalized as the following two axioms:

\textbf{(A1) Axiom of Projection.} The \( q \)-metric graph \(G_q = (X, D_q)\) is a fixed point of the projection, i.e.,
\begin{align}\tag{A1}
    P_q(G_q) = G_q. \label{eq:projection_axiom}
\end{align}

\textbf{(A2) Axiom of Transformation.} Consider any two graphs \( G = (X,D) \) and \( G' = (X',D') \) and a dissimilarity reducing map \( \varphi : X \rightarrow X' \) such that $D(x,y)\geq D'(\varphi(x),\varphi(y))$ for all $x, y \in X$. 
Then, the output \( q \)-metric graphs \( (X, D_q) = P_q(G) \) and
\( (X', D_q') = P_q(G') \) satisfy, for all $x, y \in X$,
\begin{align}
    \quad &D_q(x, y) \geq D_q'(\varphi(x), \varphi(y)).\tag{A2} \label{eq:transformation_axiom}
\end{align}

Axiom (A1) is natural in our setting: if the graph under study already satisfies the $q$-triangle inequality, then the projection should introduce no distortion and simply return the same graph.
Axiom (A2) enforces a notion of monotonicity: if the distances in one graph dominate those in another, this dominance should be preserved under projection.
Though seemingly lax, Axioms (A1) and (A2) impose significant structure on the projection $P_q$. 
In fact, we leverage the fact that \emph{there exists a unique projection $P_q$ that satisfies Axioms (A1) and (A2)}~\cite{segarra2019metricoriginal}. 
Moreover, we show that \emph{this projection preserves nearest neighbors}.
To formally state these results, we introduce the canonical $q$-metric projection.

\textbf{Canonical $q$-metric projection $P^{\star}_q$.}
Consider a graph equipped with a dissimilarity $G = (X, D)$ and let $C_{xy}$ denote the set of all paths from $x$ to $y$, where $x,y \in X$.
For a given path $c=[x=x_0, x_1, \ldots, x_l = y] \in C_{xy}$, we define its $q$-length as $\ell_q(c) = \|[d(x, x_1), \ldots, d(x_{l-1}, y)]  \|_q$.
That is, the $q$-length of a path is the $q$-norm of the vector containing the dissimilarities along the path’s edges.
We define the canonical $q$-metric projection $(X, D_q) = P^{\star}_q(X, D)$ as
\begin{equation}\label{E:canonical_q_metric}
    d_q(x, y) = \min_{c \in C_{xy}} \ell_q(c).
\end{equation}
In words, the canonical $q$-metric projection computes the all-pairs shortest paths using the $q$-norm as the path cost function. A first observation is that $P^{\star}_q$ is a valid $q$-metric projection; i.e., that the output graph is guaranteed to satisfy the $q$-triangle inequality (see Appendix~\ref{app:projection}).
More importantly, as it was already proved in ~\cite{segarra2019metricoriginal}, $P^{\star}_q$ is uniquely characterized by Axioms (A1) and (A2).

\begin{theorem}[Existence and uniqueness] \label{theo_uniqueness}
The canonical projection $P^{\star}_q$ satisfies Axioms (A1) and (A2).
Moreover, if a $q$-metric projection $P_q$ satisfies Axioms (A1) and (A2), it must be that $P_q = P^{\star}_q$. 
\end{theorem}

As we discuss next, Theorem~\ref{theo_uniqueness} has direct practical applications for nearest-neighbor search.

\subsection{Search in Projected \texorpdfstring{$q$}{q}-metric Spaces}\label{sec_nn_is_preserved}

Since the Axiom of Transformation (A2) is satisfied, we expect $P^\star_q$ to preserve up to some extent the distance ordering of the original graph, but this was never proved. In particular, we now show that the nearest neighbors are preserved by this projection. 
To formalize this, consider the graph $H=(Y,E)$ whose node set $Y = X \cup \{x_o\}$ includes the dataset $X$ and the query $x_o$. 
The dissimilarity set \( E \) consists of the original dissimilarities \( D \) as well as the dissimilarities between the query and each data point, i.e., \( E(x_o, x) = d(x_o, x) \) for all \( x \in X \).
We now apply the canonical projection to this extended graph, obtaining \( (Y, E_q) = P^\star_q(H) \), where \( E_q \) includes projected distances of the form \( E_q(x, x_o) \) between each data point \( x \in X \) and the query \( x_o \). We next state that this projection preserves the identity of the nearest neighbor of \( x_o \).

\begin{proposition}\label{prop_neares_neighbor_is_preserved}
Given a graph $G = (X, D)$ and a query $x_o$, define $H = (Y, E)$ with $Y = X \cup \{x_o\}$ and let $(Y, E_q) = P^\star_q(H)$ be the projected graph. 
Then, the set of nearest neighbors satisfies
\begin{align}\label{eqn_nearest_neighbor_is_preserved}
    \xnn 
        ~\equiv~ \argmin_{x\in X} ~ E(x, x_o)
        ~\subseteq~ \argmin_{x\in X} ~ E_q(x, x_o).
\end{align}

Moreover, for the case where $q < \infty$, the result holds with an equality.

\end{proposition}

Proposition \ref{prop_neares_neighbor_is_preserved} shows that the canonical projection $P^\star_q$ in~\eqref{E:canonical_q_metric} preserves the nearest neighbor. 
This is enticing because we have argued that search in $q$-metric spaces is easier (Figure \ref{fig_vp_tree_inequalities}) and proved that search in $\infty$-metric spaces requires a logarithmic number of comparisons (Theorem~\ref{theo_log_complexity}). 
However, Equation~\eqref{eqn_nearest_neighbor_is_preserved} does not immediately help us solve the original nearest neighbor problem in~\eqref{eqn_nearest_neighbor}. The proposition is stated for the projected dissimilarity set \( E_q \), which requires access to the full set \( E \), including all dissimilarities \( E(x_o, x) = d(x_o, x) \) for every \( x \in X \). Thus, computing \( E_q \) requires comparing \( x_o \) with all points in \( X \), which defeats the purpose of reducing the number of comparisons \( c(x_o) \). A broader analysis of this projection complexity is available in Appendix \ref{app:efficient}. In the following section, we address the projection cost challenge by learning an embedding operator that enables us to \emph{approximate} the values \( E_q(x, x_o) \) without computing all pairwise distances explicitly.

\section{Embedding Operator: Learning to Approximate \texorpdfstring{$q$}{q}-Metrics} \label{section:approximation}

We approximate $q$-metrics $E_q(x,x_o)$ with the 2-norm of a parameterized embedding $\Phi(x; \theta)$. Formally, we want to find a function $\Phi(x; \theta):\reals^n \to \reals^{s}$ such that for any $x_o \in \reals^n$ and $x \in X$:
\begin{align}\label{eqn_approximation_requirement}
    E_q(x,x_o)  = \big\| \Phi(x; \theta) - \Phi(x_o; \theta) \big\| 
\end{align}
where $\big\| \cdot  \big\|$ denotes the 2-norm of a vector.
 We fit the parameter $\theta$ to approximate distances $D_q(x,y)$ given by the canonical projection $D_q = P^\star_q(D)$, by minimizing a quadratic loss,
\begin{align}\label{eqn_q_distance_loss}
    \ell_{\text{D}} (x,y) 
        = \Big[ \,  
              D_q(x,y) 
                  - \big\| \Phi(x; \theta) - \Phi(y; \theta) \big\| \, 
                        \Big]^2.
\end{align}
The loss is inspired in type of Stress functional used in metric Multidimensional Scaling (mMDS). In practice, we also enforce neighborhood accuracy by minimizing the contrastive loss
\begin{align}\label{eqn_contrastive_loss}
    \ell_{\text{C}}(x)
        = \tfrac{1}{\tau}\lVert \Phi_\theta(x) - \Phi_\theta(x^{+}) \rVert_2^2
        + \log \sum_{y \in X \setminus \{x\}} e^{-\lVert \Phi_\theta(x) - \Phi_\theta(y) \rVert_2^2 / \tau},
\end{align}
where $x^{+}$ is sampled from the $k$ nearest neighbors of $x$ under $D$ and $\tau > 0$ is a temperature. We minimize the combined objective over the dataset $X$:
\begin{align}\label{eqn_theta_training}
    \theta^\star
        = \argmin_\theta
                 \sum_{x,y \in X} \ell_{\text{D}}(x,y)
                 + \mu \sum_{x \in X} \ell_{\text{C}}(x).
\end{align}
%


 Further notice that although we train $\theta$ over the dataset $X$ we expect it to generalize to estimate the $q$-metric $E_q(x,x_o)$. Our numerical experiments in the next section indicate a successful generalization. For additional details on training see Appendix~\ref{app:experiments}.

\section{Experiments}\label{section:experiments}

In this section, we evaluate our end-to-end approach against exact and approximate vector search methods, including HNSW~\citep{hnsw}, IVF-PQ~\citep{jegou2011product}, and the more recent ScaNN~\citep{scann}. We use 1M points from four datasets spanning a range of dimensionalities and similarity metrics. For cosine similarity, we use DEEP1B, with 96-dimensional representations, and SIFT, with 128-dimensional vectors. For Euclidean distance, we use the 960-dimensional GIST representations. Finally, for Jaccard similarity, we use KOSARAK, which contains sparse web-page interactions in a 41K-dimensional space. KOSARAK presents two additional challenges: it is high-dimensional and relies on the less conventional Jaccard similarity metric. We therefore also consider methods designed to operate effectively in these settings, such as SW-graph \citep{sw-graph}, which is a modification of HNSW, and MinHash-LSH Forest~\citep{lsh}, which is specifically designed for Jaccard similarity.

In this setup, we evaluate methods based on the tradeoff between search accuracy and computational efficiency. Efficiency is measured along three dimensions: index construction time, reported in seconds (s), search throughput, reported as queries per second (QPS) and index size in megabytes (MB). Accuracy is evaluated as Recall@1, defined as the fraction of queries for which the true nearest neighbor is retrieved. Our method learns the embedding offline, prior to index construction, and computes the query projection before search; therefore, we treat both operations as preprocessing steps. Additional results for retrieving $k \in \{5,10\}$ neighbors, as well as a broader comparison with the algorithms included in ANN-Benchmarks~\citep{DBLP:journals/corr/abs-1807-05614}, are provided in Appendix~\ref{app:experiments}.

\begin{figure}[t]
    \centering
    \includegraphics[width=1\linewidth]{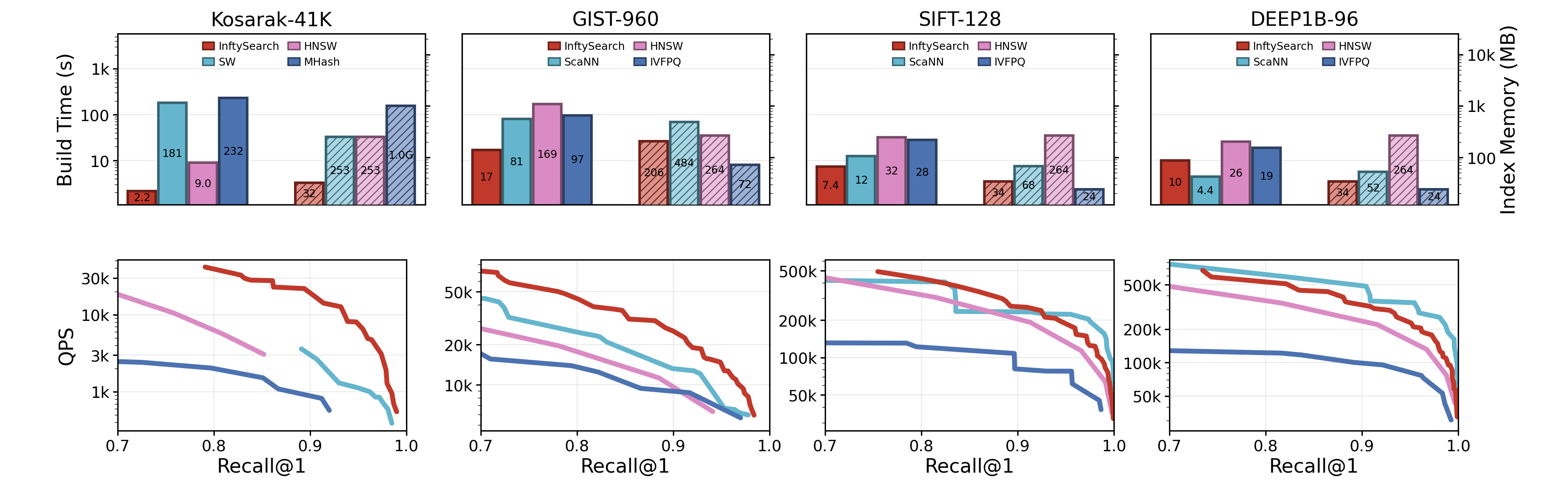}
\caption{\textbf{Infinity Search is competitive in approximate nearest neighbor search.} We learn approximate infinity embeddings on four real-world datasets and perform nearest-neighbor search using VP-Trees. When compared with efficient approximate nearest neighbor methods, Infinity Search achieves competitive query throughput (bottom), with particularly strong performance on high-dimensional datasets such as KOSARAK and GIST. Because trees can be constructed efficiently, Infinity Search leads in index-construction times (top-left) in three of the four datasets and ranks second in memory usage (top-right).
}
    \label{fig:paretos}
\end{figure}

\textbf{Infinity Search is competitive in approximate nearest neighbor search.} Across the evaluated datasets, our method achieves performance comparable to the baseline methods for nearest-neighbor search. Each point on the throughput--accuracy curve in Figure~\ref{fig:paretos} represents a different hyperparameter setting in the benchmark, showing that Infinity Search achieves speed gains with minimal accuracy trade-offs, thereby satisfying (C1). These gains are particularly pronounced on high-dimensional embeddings, such as KOSARAK-41K and GIST960. KOSARAK is high-dimensional (\(41\text{K}\)), sparse, and evaluated using Jaccard similarity, making it unsupported by many ANN implementations. Under this setting, Infinity Search achieves its strongest performance, which highlights the ability of our method to accommodate arbitrary dissimilarities, while scaling to higher-dimensional data.

Infinity Search achieves the fastest index construction on three of the four datasets, ranking second on DEEP1B. We attribute this efficiency to the VP-Tree index our method builds upon, whose simple structure requires only a handful of computations to construct. Infinity Search also ranks second in index memory, just behind IVF-PQ, which requires no additional index structure. The reader will find a more detailed analysis of this construction phase in Appendix \ref{app:Vp-Trees}.

\begin{table}[H]
\centering
\begin{tabular}{l|cccc}
\toprule
$\lceil\log_2(n)\rceil=20$ & SIFT-128 & DEEP1B-96 & GIST-960 & KOSARAK-41K \\
\midrule
Mean / Max & 20.75 / 28 & 20.90 / 29 & 20.92 / 28 & 21.07 / 32 \\
\bottomrule
\end{tabular}
\caption{\textbf{VP-Tree depth remains close to logarithmic.} We build VP-Trees over approximate infinity embeddings on large real-world datasets. Across datasets, the resulting trees exhibit near-logarithmic depth, close to the optimal value of a balanced tree.}

\label{tab:tree_depth}
\end{table}

\textbf{VP-Tree depth remains close to logarithmic.} Theorem~\ref{theo_log_complexity} claims that nearest neighbor search can be solved with a number of comparisons equal to the depth of a VP-Tree. However, this depth need not be close to its optimal logarithmic value, which is attained only by balanced trees. Table~\ref{tab:tree_depth} shows the depths of VP-Trees constructed on real-world datasets, demonstrating that approximate balanced trees can be built with near-logarithmic depth even on millions of points.

\begin{wrapfigure}{r}{0.45\textwidth}
\vspace{-2em}
    \centering
    \includegraphics[width=\linewidth]{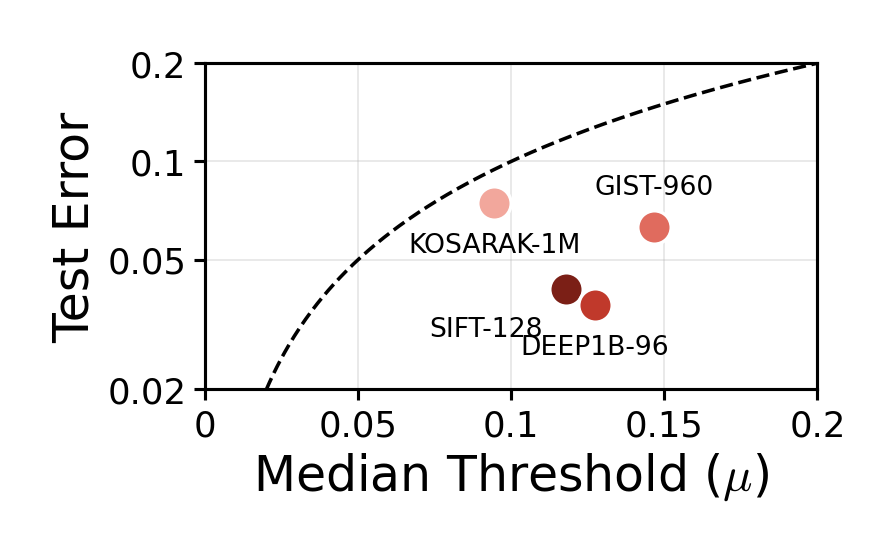}
    \vspace{-2em}
\caption{\textbf{Learning error falls below pruning threshold.} When trained on 1M DEEP1B points, embeddings exhibit errors that fall below the minimum threshold across vantage points and the average tree traversal exhibits no pruning errors.
}
    \label{fig:margin}
        \vspace{-1em}
\end{wrapfigure}
\textbf{Learning error falls below the pruning threshold.} When $q=\infty$, the canonical projection $P_{\infty}^\star$ maps points into an $\infty$-metric space. Computing this projection directly is impractical when fast search is required. In Section~\ref{section:experiments}, we propose a learning approach that obtains approximate $\infty$-metric embeddings by minimizing their pairwise distance to the canonical projections. If the approximation error is large, VP-Tree pruning may discard true nearest neighbors, impairing recall. Conversely, if the approximation error remains below the distance threshold $\mu$ between a vantage point and its children, tree traversal is expected to remain accurate under the approximate embeddings.

In Figure~\ref{fig:margin}, we compare the minimum margin in the tree with the average test error of the learned embeddings. Across datasets, the approximation error remains sufficiently small relative to the minimum threshold of the tree, allowing effective pruning while preserving search accuracy. We note that, since the reported error is an average quantity, perfect traversal cannot be guaranteed for every query. Nevertheless, the results suggest that the learned embeddings provide sufficiently accurate distances for reliable search.

\newpage

\begin{wrapfigure}{r}{0.45\textwidth}
\vspace{-1em}
    \centering
    \includegraphics[width=1\linewidth]{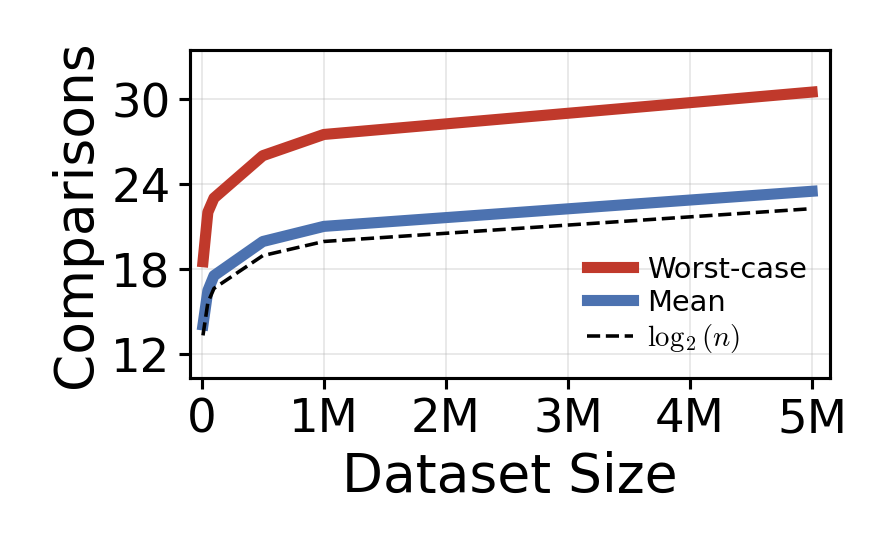}
\caption{\textbf{Infinity Search scales.} We train Infinity Search on a subset of $40$K points from DEEP1B and then use it to search over up to $5$M points, where our method exhibits a favorable scaling behavior.}
    \label{fig:scale_comparisons}
    \vspace{-1em}
\end{wrapfigure}

\textbf{Infinity Search scales.} In Section~\ref{section:NNS}, Figure~\ref{fig:depth_scaling} reported the growth in the number of comparisons under exact $\infty$-metric embeddings, obtained by applying the canonical projection $P_{\infty}^\star$. Since this projection is computationally expensive, the scaling law was originally tested only up to 150K points, showing a near-logarithmic average depth. By approximating the canonical projection, we can leverage the near-$\infty$-metric structure at larger scales. Figure~\ref{fig:scale_comparisons} shows the scaling of the number of comparisons for learned embeddings on up to 5M points. We highlight that the scaling law observed for the approximate embeddings closely resembles  that of exact projections. Furthermore, the gap between the maximum number of comparisons and the logarithmic optimum appears to remain constant across scales.

\section{Conclusion}
\label{sec:conclusion}
Driven by the insight that search in ultrametric spaces remains close to logarithmic, in this paper we leverage a projection that endows a set of vectors with arbitrary dissimilarities with $q$-metric structure, while preserving the nearest neighbor. This projection of a dataset is computed as the shortest path in a graph using the $q$-norm as path length. To address the challenge of computing $q$-metric distances for queries efficiently, we developed a learning approach that embeds vectors into a space where Euclidean distances approximate $q$-metric distances. Our experiments real-world datasets, which included high-dimensional sparse data with less common dissimilarity functions, demonstrated that our method is competitive with strong approximate search algorithms at low scale.



\clearpage

\clearpage

{
\small
\bibliographystyle{apalike}
\bibliography{references}
}

\appendix

\section{Related Work}\label{sec_related_work}

The literature on vector‐based nearest–neighbour search is vast and spans many decades. We briefly review the families that are most germane to our contributions and direct the reader to existing surveys~\citep{survey, ukey2023survey, li2019approximatesurvey} for a broader analysis.

For data sets endowed with a proper metric, pruning rules derived from the standard triangle inequality have long been exploited through pivot- or vantage-point trees, $k$-d trees, ball trees, and their variants~\citep{vptree,Uhlmann91,beygelzimer2006cover}. Although these methods can yield exact results, they typically struggle with dimensionality and rely on existing metric structure of the date have been largely ouperformed by inverted file indexes~\citep{babenko2015inverted}, hashing~\citep{IndykMotwani98, wang2014hashing}, quantisation~\citep{kalantidis2014lopq} and graph based methods~\citep{hnsw} such as HNSW~\citep{hnsw}. Approaches—such as SPANN~\citep{chen2021spannhighlyefficientbillionscaleapproximate}, and ONNG~\citep{arai2018onnng}— have optimized memory layout, link degree, or storage hierarchy to scale to billion-point datasets.

While many of these methods can offer strong performance, recent works increasingly seek to embed learning into the indexing process for data-aware improvements.  ScaNN~\citep{scann} jointly optimizes clustering and quantization, SOAR adds orthogonality-amplified residuals to further boost ScaNN’s efficiency~\citep{sun2023soar}. Learned policies can also steer the search itself: AdaptNN predicts early-termination thresholds online~\citep{li2020adaptnn}, and Tao learns to set them \emph{before} query execution using only static local-intrinsic-dimension features~\citep{yang2021tao}. On the other hand, LoRANN replaces PQ scoring with a supervised low-rank regressor that outperforms traditional IVF on billion-scale data~\citep{jaasaari2024lorann}.

Industrial adoption of large language models has driven the emergence of ``vector DBMSs'', see~\citep{pan2024dbmssurvey} and references therein. Despite many developments in compression, partitioning, graph navigation, and hybrid attribute-vector query planning, most if not all systems index the \emph{original} dissimilarities - or an approximation -- and thus inherit their metric limitations. In contrast, we propose to pre-process data to obtain a $q$-metric space whose structure is provably more favorable for search.


\section{Vantage Point Trees}\label{app:Vp-Trees}
To search in $q$-metric and $\infty$-metric spaces we adopt vantage point (VP) trees \citep{vptree}. To explain VP trees consider a set of points $X$ and a dissimilarity function $d$. Select a vantage point $v \in X$ and an associated radius $\mu$. Proceed then to partition the space $X$ into $v$ itself and sets $X'$ and $X''$ containing points (other than $v$) that are inside and outside the ball of radius $\mu$ centered at $v$, respectively,
\begin{alignat}{2}\label{eqn_vp_children_apx}
    X'  &~:=~ \Big\{\, x \in X  \,:\,  d(v,x) <    \mu    \, , \, x \neq v  && \, \Big\}, \nonumber \\
    X'' &~:=~ \Big\{\, x \in X  \,:\,  \mu    \leq d(v,x)                   && \, \Big\}.
\end{alignat}
A VP tree is built by recursive partition of the sets $X'$ and $X''$ using corresponding vantage points and radiuses. I.e., we select a vantage point $v'\in X'$ and a radius $\mu'$ to divide $X'$ into inside and outside sets and do the same for $X''$. We then partition the new group of inside and outside sets and repeat until we reach indivisible sets with only one point (see Section \ref{sec_vp_trees_algorithms} and Figure \ref{alg:vp-build} for details). We emphasize, as this will be important later, that the border points at which $\mu = d(v,x)$ are assigned to the outside set $X''$.

VP trees were developed for efficient search of metric spaces (Section \ref{sec_vp_trees_metric_spaces}). We will see here that they can also be used for efficient search of $q$-metric (Section \ref{sec_vp_trees_q_metric_spaces}) and $\infty$-metric (Section \ref{sec_vp_trees_infty_metric_spaces}) spaces. We will further see that we expect exploration of VP trees to be more efficient in $q$-metric spaces with larger $q$ (Section \ref{sec_vp_trees_q_metric_spaces}). In particular, when $q=\infty$ we will show that we are guaranteed to find a nearest neighbor $\xnn$ of any query $x_o$ with a number of comparisons that is, at most, equal to the depth of the VP tree (Section \ref{sec_vp_trees_infty_metric_spaces}).


\subsection{Vantage Point Trees in Metric Spaces}\label{sec_vp_trees_metric_spaces}

The inside and outside sets $X'$ and $X''$ reduce the complexity of finding the nearest neighbor of a query $x_o$ when the dissimilarity function $d(x,y)$ corresponds to a (standard) metric space (Figure \ref{fig_vp_tree_inequalities}). This fact is well known, see e.g., \cite{vptree}, but we repeat the relevant facts here to facilitate understanding of the material in Sections \ref{sec_vp_trees_q_metric_spaces} and \ref{sec_vp_trees_infty_metric_spaces}.

Start by assuming that we have available an upper bound $\tau \geq d(x_o, \xnn)$ on the distance between the query $x_o$ and a nearest neighbor $\xnn$. This upper bound is typically a distance to an earlier vantage point and must satisfy $\tau \leq d(v,x_o)$ because we can otherwise set  $\tau = d(v,x_o)$ as an upper bound on $d(x_o, \xnn)$. We can then show that the following proposition holds.


\medskip

\begin{proposition} \label{prop_prune_conditions_metric}

Consider a set of points $X$, a vantage point $v \in X$, and a radius $\mu$ to define inside and outside sets $X'$ and $X''$ as in \eqref{eqn_vp_children_apx}. We are given a query $x_o$ and an upper bound $d(v,x_o) \geq \tau \geq d(x_o, \xnn)$ on the nearest neighbor distance $d(x_o, \xnn)$. If the dissimilarity function $d(x,y)$ satisfies the triangle inequality \eqref{eqn_triangle_inequality}, we have that

\begin{description} [labelwidth=0.9cm,  leftmargin=!]

\item[~\textbf{(CI)}] 
    If $d(v, x_o) + \tau < \mu$, then, either $v$ is a nearest neighbor or there exist a nearest neighbor in the inside set; i.e., 
    $\xnn \cap ( X' \cup v ) \neq \emptyset$.

\item[~\textbf{(CO)}] 
    If $\mu + \tau \leq d(v, x_o)$, then, there exist a nearest neighbor in the outside set; i.e., $\xnn \cap X'' \neq \emptyset$.

\end{description}

\end{proposition}


\begin{proof} 
We show first that if the condition $d(v, x_o) + \tau < \mu$ in (CI) holds, there is no point $x''$ in the outside set $X''$ that can be a bearest neighbor of $x_o$. To see that this is true, recall that for any point $x''\in X''$ we have $\mu \leq d(v,x'')$ and apply the triangle inequality to the triangle $(v, x_o, x'')$ to write
\begin{align}\label{eqn_prooof_prune_conditions_metric_10}
    \mu ~\leq~ d(v,x'') ~\leq~ d(v, x_o) + d(x_o, x'') .
\end{align}
Because we are assuming that the condition in (CI) holds, we also know that $d(v, x_o) + \tau < \mu$. This inequality is compatible with \eqref{eqn_prooof_prune_conditions_metric_10} if and only if,
\begin{align}\label{eqn_prooof_prune_conditions_metric_20}
    \tau < d(x_o, x'') .
\end{align}
Further recall that according to the definition of the upper bound $\tau$ we must have $d(x_o, \xnn) \leq \tau$. Consequently,
\begin{align}\label{eqn_prooof_prune_conditions_metric_25}
    d(x_o, \xnn) \leq \tau < d(x_o, x'') .
\end{align}
This means that all points $x'' \in X''$ have distances to $x_o$ larger than the distance to its nearest neighbor $\xnn$. Thus, there is no nearest neighbor in $X''$ and we must therefore have that either $v$ or some other point in $X'$ is a nearest neighbor of $x_o$.

Analogously, we show that if the condition $\mu + \tau \leq d(v, x_o)$ in (CO) holds, we must have $ d(x_o, \xnn) < d(x_o, x')$ for any point $x'$ in the inside set $X'$. To see that this is true, apply the triangle inequality to the triangle $(v, x', x_o)$ and recall that for any point $x'\in X'$ we must have $d(v, x') < \mu$ to write
\begin{align}\label{eqn_prooof_prune_conditions_metric_30}
    d(v, x_o) ~\leq~ d(v, x') + d(x', x_o) ~<~ \mu + d(x', x_o).
\end{align}
Because we are assuming that the condition in (CO) holds we also know that $\mu + \tau \leq d(v, x_o) $. This inequality is compatible with \eqref{eqn_prooof_prune_conditions_metric_30} if and only if,
\begin{align}\label{eqn_prooof_prune_conditions_metric_40}
    \tau < d(x_o, x').
\end{align}
Further recall, as we did in \eqref{eqn_prooof_prune_conditions_metric_25}, that the definition of the upper bound $\tau$ implies that $d(x_o, \xnn) \leq \tau$. We can thus write
\begin{align}\label{eqn_prooof_prune_conditions_metric_45}
    d(x_o, \xnn) \leq \tau < d(x_o, x') .
\end{align}
This means that there is no nearest neighbor in $X'$. We therefore must have that either $v$ is a nearest neighbor or that we have a nearest neighbor in $X''$. In this case we know that $v$ is not a nearest neighbor of $x_o$ because of condition (CO). Indeed, because $\mu > 0$ we have that $d(x_o, \xnn) \leq \tau < \mu + \tau  \leq d(v, x_o)$.
\end{proof}


Because of (CI) and (CO) we can sometimes eliminate entire branches of VP trees during the search for a nearest neighbor of a query $x_o$. If the condition $d(v, x_o) + \tau < \mu$ in (CI) holds, we know that either $v$ is a nearest neighbor or that we can find a nearest neighbor in $X'$. We can therefore discard all points in $X''$ and focus the search on the inside set $X'$. Likewise, if the condition $\mu + \tau \leq d(v, x_o)$ in (CO) holds, we know that a nearest neighbor $\xnn$ can be found in $X''$. We can therefore discard all points in $X'$ and focus the search on the outside set $X''$. If we do this recursively, each comparison with a vantage point has the potential of halving the number of points that remain in contention as nearest neighbor candidates (see Section \ref{sec_vp_trees_algorithms} and Algorithm \ref{alg:vp-search} for details). 

This potential to discard points from consideration as nearest neighbors is not always realized because it can be that neither of the conditions in (CI) or (CO) hold true. Indeed, for queries $x_o$ such that the distance to the vantage point satisfies
\begin{align}\label{eqn_no_pruning_possible_metric}
    \mu - \tau ~\leq~ d(v, x_o) ~<~ \mu + \tau,
\end{align}
neither of the conditions in (CI) or (CO) hold true and we must explore both sets $X'$ and $X''$ further. We will see next that this limitation is mitigated in $q$-metric spaces (Section \ref{sec_vp_trees_q_metric_spaces}) and eliminated in $\infty$-metric spaces (Section \ref{sec_vp_trees_infty_metric_spaces}).


\subsection{Vantage Point Trees in \texorpdfstring{$q$}{q}-Metric Spaces}
\label{sec_vp_trees_q_metric_spaces}

The partition of a set $X$ into the inside and outside sets defined in \eqref{eqn_vp_children_apx} for a vantage point $v$ and radius $\mu$ is independent of the structure of the dissimilarity function $d(x,y)$. The partition is valid even if $d(x,y)$ is not metric. Proposition \ref{prop_prune_conditions_metric} \emph{does} require $d(x,y)$ to be metric and shows that in such a case comparisons of a query $x_o$ with a vantage point $v$ \emph{may} confine the search for the nearest neighbor $\xnn$ to either the inside set $X'$ or the outside set $X''$.

We show here that an analogous result holds for $q$-metric dissimilarity functions. The difference is that the conditions in (CI) and (CO) compare $q$-powers of the respective quantities. This modification matches the use of $q$-powers of distances in the $q$-triangle inequality \eqref{eqn_q_triangle_inequality}.


\medskip

\begin{proposition} \label{prop:q-vp-pruning}

Consider a set of points $X$, a vantage point $v \in X$, and a radius $\mu$ to define inside and outside sets $X'$ and $X''$ as in \eqref{eqn_vp_children_apx}. We are given a query $x_o$ and an upper bound $d(v,x_o) \geq \tau \geq d(x_o, \xnn)$ on the nearest neighbor distance $d(x_o, \xnn)$. If the dissimilarity function $d(x,y)$ satisfies the $q$-triangle inequality \eqref{eqn_q_triangle_inequality}, we have that

\begin{description} [labelwidth=1.1cm,  leftmargin=!]

\item[~\textbf{($q$-CI)}] 
     If $d^q(v, x_o) + \tau^q < \mu^q$, then, either $v$ is a nearest neighbor or there exist a nearest neighbor in the inside set;
     i.e., $\xnn \cap (X' \cup v) \neq \emptyset$.

\item[~\textbf{($q$-CO)}] 
     If $\mu^q + \tau^q \leq d^q(v, x_o)$, then, there exist a nearest neighbor in the outside set;
     i.e., $\xnn \cap X'' \neq \emptyset$.

\end{description}

\end{proposition}


\begin{proof} 
The proof is a ready extension of the proof of Proposition \ref{prop_prune_conditions_metric}. Thus, we show first that if the condition $d^q(v, x_o) + \tau^q < \mu^q$ in ($q$-CI) holds, we must have $ d(x_o, \xnn) < d(x_o, x'') $ for any point $x''$ in the outside set $X''$. To see that this is true, recall that for any point $x''\in X''$ we must have $\mu \leq d(v,x'')$ and, consequently, we must also have $\mu^q \leq d^q(v,x'')$. Writing this latter inequality followed by the $q$-triangle inequality \eqref{eqn_q_triangle_inequality} to the triangle $(v, x_o, x'')$, we can write,
\begin{align}\label{eqn_prooof_prune_conditions_q_metric_10}
    \mu^q ~\leq~ d^q(v,x'') ~\leq~ d^q(v, x_o) + d^q(x_o, x'') .
\end{align}
Because we are assuming that the condition in ($q$-CI) holds we also know that $d^q(v, x_o) + \tau^q < \mu^q$. This inequality is compatible with \eqref{eqn_prooof_prune_conditions_metric_10} if and only if,
\begin{align}\label{eqn_prooof_prune_conditions_q_metric_20}
    \tau^q < d^q(x_o, x'') .
\end{align}
This inequality implies that we must have $ \tau < d(x_o, x'')$ for any $x'' \in X''$. To complete the claim recall that since $\tau$ is an upper bound on the nearest neighbor distance we know that $d(x_o, \xnn) \leq \tau$. Combining this inequality with \eqref{eqn_prooof_prune_conditions_q_metric_20} yields
\begin{align}\label{eqn_prooof_prune_conditions_q_metric_25}
    d(x_o, \xnn) \leq \tau < d(x_o, x'') .
\end{align}
We conclude that there is no nearest neighbor in $X''$. We therefore must have that either $v$ is a nearest neighbor or there is a nearest neighbor in $X'$. 

Analogously, we show that if the condition $\mu^q + \tau^q \leq d^q(v, x_o)$ in ($q$-CO) holds, we must have $d(x_o, \xnn) < d(x_o, x')$ for any point $x'$ in the inside set $X'$. To see that this is true recall that since $d(v, x') < \mu$ for any point $x'\in X'$ we must also have that $d^q(v, x') < \mu^q$. Write now the $q$-triangle inequality for the triangle $(v, x', x_o)$ and use the fact that $d^q(v, x') < \mu^q$ to conclude that 
\begin{align}\label{eqn_prooof_prune_conditions_q_metric_30}
    d^q(v, x_o) ~\leq~ d^q(v, x') + d^q(x', x_o) ~<~ \mu^q + d^q(x', x_o).
\end{align}
Because we are assuming that the condition in ($q$-CO) holds we also know that $\mu^q + \tau^q \leq d^q(v, x_o)$. This inequality is compatible with \eqref{eqn_prooof_prune_conditions_metric_30} if and only if,
\begin{align}\label{eqn_prooof_prune_conditions_q_metric_40}
    \tau^q < d^q(x_o, x').
\end{align}
We therefore must have that $\tau < d(x_o, x') > \tau$ for any $x' \in X'$. Since we also know that $d(x_o, \xnn) \leq \tau$, we conclude that
\begin{align}\label{eqn_prooof_prune_conditions_q_metric_45}
    d(x_o, \xnn) \leq \tau < d (x_o, x').
\end{align}
Thus, there cannot be a nearest neighbor of $x_o$ in $X'$. It follows that we must either have a nearest neighbor in $X''$ or $v$ itself is a nearest neighbor. We know that the latter is not possible because since $\mu>0$ and condition ($q$-CO) holds, we have that $d^q(x_o, \xnn) \leq \tau^q < \mu^q + \tau^q  \leq d^q(v, x_o)$, which in turn implies $d^q(x_o, \xnn) < d (v, x_o)$.
\end{proof}


Because of ($q$-CI) and ($q$-CO) it may be possible to prune branches of the VP tree when searching for the nearest neighbor of a query $x_o$ when the dissimilarity function $d(x,y)$ satisfies the $q$-triangle inequality. 
If the condition $d^q(v, x_o) + \tau^q < \mu^q$ in ($q$-CI) holds, we prune the outside set $X''$ and focus the search on the inside set $X'$. If the condition $\mu^q + \tau^q \leq d^q(v, x_o)$ in ($q$-CO) holds, we prune the inside set $X'$ and focus the search on the outside set $X''$. This is analogous to the pruning of VP trees when the dissimilarity function $d(x,y)$ satisfies the regular triangle inequality -- which is also particular case with $q=1$.

As is the case of standard triangle inequalities, pruning is not always possible. If the distance $d(v, x_o)$ between the query and the vantage point is such that
\begin{align}\label{eqn_no_pruning_possible_q_metric}
    \mu^q - \tau^q ~\leq~ d^q(v, x_o) ~<~ \mu^q + \tau^q,
\end{align}
neither of the conditions in ($q$-CI) or ($q$-CO) hold. We must therefore explore both, the inside set $X'$ and the outside set $X''$ as we do in the case of standard triangle inequalities when \eqref{eqn_no_pruning_possible_metric} holds -- this equation being a particular case of \eqref{eqn_no_pruning_possible_q_metric} corresponding to $q=1$. 

A remarkable fact is that \eqref{eqn_no_pruning_possible_q_metric} is stricter than \eqref{eqn_no_pruning_possible_metric} for any $q>1$. I.e., for given $\mu$ and $\tau$ the range of values between $\mu^q - \tau^q$ and $\mu^q + \tau^q$ is smaller than the range of values between $\mu - \tau$ and $\mu + \tau$. This observation suggests that pruning of VP trees is more effective when data and queries are samples of a $q$-metric space than when they are samples of a standard metric space. Our experiments corroborate that this is true (see Section \ref{app:metric_representation}, Figures \ref{app:haven1} and \ref{app:haven2}). 

A second remarkable fact is that the range of values between $\mu^q - \tau^q$ and $\mu^q + \tau^q$ for given $\mu$ and $\tau$ shrinks with larger $q$ and becomes empty as $q$ approaches infinity. This observation indicates that pruning of a VP tree is likely to be always possible when the strong triangle inequality holds. We explain in the next section that this is indeed true.


\subsection{Vantage Point Trees in \texorpdfstring{$\infty$}{infinity}-Metric Spaces}
\label{sec_vp_trees_infty_metric_spaces}
\label{app:NNS}

When the dissimilarity function $d(x,y)$ satisfies the strong triangle inequality \eqref{eqn_strong_triangle_inequality}, the analogous of the conditions in (CI) and ($q$-CI) is 
\begin{align}\label{eqn_inside_condition_infty_before simplification}
    \max \big( \, d(v, x_o), \, \tau \, \big) ~ < ~ \mu .
\end{align}
Observe however that the upper bound $\tau$ on the distance to the nearest neighbor of the query $x_o$ is not larger than $d(v, x_o)$ by construction. I.e., $d(v,x_o) \geq \tau \geq d(x_o, \xnn)$. Thus, the maximum in \eqref{eqn_inside_condition_infty_before simplification} is attained by $d(v, x_o)$, from where it follows that \eqref{eqn_inside_condition_infty_before simplification} simplifies to,
\begin{align}\label{eqn_inside_condition_infty}
    d(v, x_o) ~ < ~ \mu .
\end{align}
Likewise, the analogous of the conditions in (CO) and ($q$-CO) is the inequality,
\begin{align}\label{eqn_outside_condition_infty_before_simplification}
    \max \big( \, \mu, \, \tau \, \big) ~\leq~ d(v, x_o) .
\end{align}
Again, we know that by construction we have $d(v,x_o) \geq \tau$. Thus \eqref{eqn_outside_condition_infty_before_simplification} holds if and only if,
\begin{align}\label{eqn_outside_condition_infty}
    \mu ~\leq~ d(v, x_o) .
\end{align}
The conditions in \eqref{eqn_inside_condition_infty} and \eqref{eqn_outside_condition_infty} are exact complements. One and only one of them holds for any value of $d(v, x_o)$. It remains for us to show that they are valid conditions for pruning a VP tree when the dissimilarity function $d(x,y)$ satisfies the strong triangle inequality. The following proposition claims that this is true.


\medskip

\begin{proposition} \label{prop_prune_conditions_q_metric}
\label{prop:inf-vp-pruning}
Consider a set of points $X$, a vantage point $v \in X$, and a radius $\mu$ to define inside and outside sets $X'$ and $X''$ as in \eqref{eqn_vp_children_apx}. We are given a query $x_o$. If the dissimilarity function $d(x,y)$ satisfies the $\infty$-triangle inequality \eqref{eqn_q_triangle_inequality}, we have that

\begin{description} [labelwidth=1.2cm,  leftmargin=!]

\item[~\textbf{($\infty$-CI)}] 
     If $d(v, x_o) < \mu$, then, either $v$ is a nearest neighbor or
     there exist a nearest neighbor in the inside set;
     i.e., $\xnn \cap (X' \cup v) \neq \emptyset$.

\item[~\textbf{($\infty$-CO)}] 
     If $\mu \leq d(v, x_o)$, then, either $v$ is a nearest neighbor or
     there exist a nearest neighbor in the outside set;
     i.e., $\xnn \cap (X''\cup v) \neq \emptyset$.

\end{description}

\end{proposition}


\begin{proof} 
Consider first the case in which the condition $d(v, x_o) < \mu$ in ($\infty$-CI) holds and let $x''$ be an arbitrary element of the outside set $X''$. As per the definition of the outside set, any point $x''\in X''$ is such that $\mu \leq d(v,x'')$. Writing this latter inequality followed by the $\infty$-triangle inequality \eqref{eqn_strong_triangle_inequality} applied to the triangle $(v, x_o, x'')$, we can write,
\begin{align}\label{eqn_prooof_prune_conditions_infty_metric_10}
    \mu ~\leq~ d(v,x'') ~\leq~ \max \big(\, d(v, x_o) , \, d(x_o, x'') \, \big).
\end{align}
Because we are assuming that the condition $d(v, x_o) < \mu$ in ($\infty$-CI) holds, we can also conclude that the maximum in the right hand side of \eqref{eqn_prooof_prune_conditions_infty_metric_10} must be achieved by $d(x_o, x'')$. This is the only way in which the maximum can equal or exceed $\mu$ when $d(v, x_o)$ is strictly smaller than $\mu$. We have therefore proven that for any point $x''\in X''$,
\begin{align}\label{eqn_prooof_prune_conditions_infty_metric_20}
    \mu \leq d(x_o, x'').
\end{align}
We also know that the distance $d(x_o, \xnn)$ between the query $x_o$ and a nearest neighbor $\xnn$ cannot exceed the distance $d(v, x_o)$ between the query $x_o$ and the vantage point $v$. Given that condition ($\infty$-CI) is assumed to hold we also have that $d(v, x_o) < \mu$ and we can therefore write,
\begin{align}\label{eqn_prooof_prune_conditions_infty_metric_25}
    d(x_o, \xnn) \leq d(v, x_o) < \mu.
\end{align}
Contrasting \eqref{eqn_prooof_prune_conditions_infty_metric_20} with \eqref{eqn_prooof_prune_conditions_infty_metric_25} we conclude that it is impossible for a nearest neighbor to belong to the outside set $X''$. We must therefore have that either $v$ is a nearest neighbor or there is a nearest neighbor in $X'$. Notice that this proof is analogous to the corresponding proofs of claims (CI) and ($q$-CI) of Propositions \ref{prop_prune_conditions_metric} and \ref{prop:q-vp-pruning}.

Consider now the case in which the condition $\mu \leq d(v, x_o) $ in ($\infty$-CI) holds and let $x'$ be an arbitrary element of the inside set $X'$. Further consider the distance $d(v, x_o)$ and apply the $\infty$-triangle inequality \eqref{eqn_strong_triangle_inequality} to the triangle $(v, x', x_o)$ to write,
\begin{align}\label{eqn_prooof_prune_conditions_infty_metric_30}
    d(v,x_o) 
        ~\leq~ \max \big(\, d(v, x') , \, d(x',x_o) \, \big).
\end{align}
Suppose that the maximum in \eqref{eqn_prooof_prune_conditions_infty_metric_30} is attained by $d(v, x')$. Under this assumption we have that $d(v,x_o) ~\leq~ d(v, x')$. Furthermore, because we assume that $x' \in X'$, we have according to \eqref{eqn_vp_children_apx}, that $d(v, x') < \mu$. These facts together yield,
\begin{align}\label{eqn_prooof_prune_conditions_infty_metric_35}
    d(v,x_o) ~\leq~ d(v, x') ~<~ \mu
\end{align}
This is contradiction with the condition $\mu \leq d(v, x_o) $ in ($\infty$-CI). Thus, it is impossible for the maximum in \eqref{eqn_prooof_prune_conditions_infty_metric_30} to be attained by $d(v, x')$. The maximum must be therefore attained by $d(x',x_o)$ leading to the conclusion that,
\begin{align}\label{eqn_prooof_prune_conditions_infty_metric_40}
    d(v,x_o) ~\leq~ d(x', x_o) .
\end{align}
Since the nearest neighbor distance is $d(x_o, \xnn) \leq d(v, x_o)$, there is no point in $X'$ whose distance to the query $x_o$ is smaller than the distance between $x_o$ and the vantage point $v$. There are now two possibilities: (i) The vantage point $v$ is a nearest neighbor. (ii) The vantage point $v$ is \emph{not} a nearest neighbor. If $v$ is a nearest neighbor the claim in ($\infty$-CO) holds true as the vantage point $v$ is contained in the nearest neighbor set. If $v$ is \emph{not} a nearest neighbor, \eqref{eqn_prooof_prune_conditions_infty_metric_40} implies that there is no nearest neighbor in $X'$. The nearest neighbors must therefore be in $X''$. 

Observe that this proof is \emph{not} analogous to the proof of claims (CO) and ($q$-CO) of Propositions \ref{prop_prune_conditions_metric} and \ref{prop:q-vp-pruning}. Indeed, the claims itself are different. The reason for having a different argument is that in the case of $\infty$-metric spaces the inequality in \eqref{eqn_prooof_prune_conditions_infty_metric_40} does not prevent the existence of nearest neighbors in $X'$. It implies the (weaker) claim that this is possible only if $v$ itself is a nearest neighbor. This difference does not affect our ability to prune the set $X'$ if our goal is to find \emph{a} nearest neioghbor of $x_o$.
\end{proof}

Because of ($\infty$-CI) and ($\infty$-CO) each comparison with a vantage point in an $\infty$-metric space focuses the search on a branch of the VP tree. If the condition in ($\infty$-CI) holds, we continue exploration of the inside branch $X'$ and discard the outside set $X''$. We know that no nearest neighbors are in this set. If the condition in ($\infty$-CO) holds, we continue exploration of the outside branch $X''$ and discard the inside set $X''$. In this case we know that there may be some nearest neighbors in $X'$ if more than one exists, but none of them are closer to $x_o$ than the vantage point $v$.

The remarkable property of $\infty$-metric spaces is that the conditions in ($\infty$-CI) and ($\infty$-CO) are complementary. One and exactly one of them hold. It follows from this observation that to find the nearest neighbor of a query $x_o$ for dissimilarity functions that satisfy the strong triangle inequality is bounded by the depth of the VP tree.

\begin{figure}[t]

\centering

\begin{minipage}{0.98\linewidth}

\begin{algorithm}[H]
\caption{Vantage Point (VP) Tree Construction}
\label{alg:vp-build}

\smallskip
\begin{algorithmic}[1]
\setstretch{1.4}

\INPUT Data set $X$ and dissimilarity function $d$.
\OUTPUT Node $[p,\mu,\text{inside},\text{outside}]$.

\FUNCTION{BuildVPTree}{$X,d$}
    \IF{$X \neq \emptyset$}
        \STATE Choose random vantage point $v \in X$
        \STATE $\text{node}.p \gets v$
        \STATE $\text{node}.\mu \gets \textsc{Median}_{x \in X}(d(x,v))$
        \STATE $X' \gets \{x \in X : d(x,v) < \mu,\ x \neq v\}$
        \STATE $X'' \gets \{x \in X : d(x,v) \ge \mu\}$
        \STATE $\text{node}.\text{inside} \gets \text{BuildVPTree}(X',d)$
        \STATE $\text{node}.\text{outside} \gets \text{BuildVPTree}(X'',d)$
        \STATE \textbf{return} node
    \ELSE
        \STATE \textbf{return} \textit{null}
    \ENDIF
\ENDFUNCTION

\end{algorithmic}
\end{algorithm}

\caption{Construction of a Vantage Point (VP) tree. Nodes are of the form $[p,\mu,\text{inside},\text{outside}]$. Here, $p$ is a vantage point, $\mu$ its associated radius, ``inside'' is a node representing the inside set $X'$ and ``outside'' is a node representing the outside set $X''$. All nodes $[v',\mu',\text{inside}',\text{outside}']$ in the inside subtree satisfy $d(v,v') < \mu$, and all nodes $[v'',\mu'',\text{inside}'',\text{outside}'']$ in the outside subtree satisfy $ \mu \leq d(v,v'')$.}

\label{fig:vp-build}

\end{minipage}

\end{figure}

\setcounter{theorem}{0}

\begin{theorem}\label{app:theo_log_coplexity}

Consider a dataset $X$, a query $x_o$, a dissimilarity function $d$ satisfying the strong triangle inequality \eqref{eqn_strong_triangle_inequality} and a vantage point tree $T(X)$ constructed by the recursive partition of $X$ into vantage points and their corresponding inside and outside sets [cf. \eqref{eqn_vp_children_main}]. The number of comparisons $c(x_o)$ needed to find a nearest neighbor of $x_o$ in the set $X$ is bounded by the depth $h(T(X))$ of the VP tree $T(X)$,
\begin{align}\label{eqn_theo_log_complexity_apx}
    c(x_o) \leq h(T(X)) .
\end{align}

\end{theorem}

\begin{proof}

Consider a VP-tree, constructed on \(X\) using Algorithm~\ref{alg:vp-build}. Proposition \ref{prop:inf-vp-pruning} ensures that making one single comparison -\ref{eqn_prooof_prune_conditions_infty_metric_25} or \ref{eqn_prooof_prune_conditions_infty_metric_20}- for each vantage point will discard a set of points -either $X'$ or $X''$- not containing the nearest neighbor of $x_0$. Notice that, by construction in \ref{alg:vp-build}, the maximum number of tims this can be done is limited by the recursion depth, which is indeed, the tree depth.

\end{proof}






\section{Algorithms to Construct and Search Vantage Point Trees}\label{sec_vp_trees_algorithms}


Given a set of points $X$, Figure\ref{fig_vp_tree_inequalities} depicts how data can be organized to enable efficient search. As a concrete example, Algorithm \ref{alg:vp-build} illustrates how a VPTree can be constructed by recursively partitioning a set of points $X$. This construction begins by randomly selecting a vantage point $v$ (Step 3). Fixing $v$ as one argument of the dissimilarity function $d$, allows to compute the median $\mu$ from the induced set distances $\{d(x,v)\}_{x\in X}$ (Step 5). This median serves as a radius that partitions $X$ into an \textit{inside} subset $X'$ (Step 6) and \textit{outside} subset $X''$ (Step 7). The function then recurses on subsets $X'$ and $X''$, until no further split is possible, at which point the recursion halts and a leaf node is created (Step 12). This process results in a nested list of nodes $[p, \mu, inside, outside]$ storing a vantage point, the corresponding radius and the subsequent inside and outside nodes.

\begin{figure}[H]
\centering
\begin{minipage}{0.98\linewidth}
\begin{algorithm}[H]
\caption{Searching Phase of the $q$-VPTree}
\label{alg:vp-search}

\begin{algorithmic}[1]
\setstretch{1.4}

\INPUT Query point $x_o$, parameter $q$ and a node (initialized to the root). Nearest neighbor $\hat{x}_o$ is set to \textit{null} and its distance $\tau$ to the query is initialized as $\infty$.
\OUTPUT Nearest neighbor $\hat{x}_o$ and its distance $\tau$, after searching the subtree rooted at a node.

\FUNCTION{SearchVPTree$_q(\text{node}, \tau, \hat{x}_o, x_o)$}

    \IF{$\text{node}\neq\textit{null}$}
    \STATE $d \gets d(x_o,\text{node}.p)$
    \IF{$d<\tau$}
        \STATE $\tau \gets d$ 
        \STATE $\hat{x}_o \gets \text{node}.p$
        \ENDIF

    \IF{$d^q < \text{node}.\mu^q - \tau^q$}
        \STATE $\hat{x}_o, \tau \gets \text{SearchVPTree}(\text{node}.inside,\tau, \hat{x}_o, x_o)$ 
    \ELSIF{$d^q \ge \text{node}.\mu^q + \tau^q$}
        \STATE $\hat{x}_o, \tau \gets \text{SearchVPTree}(\text{node}.outside,\tau, \hat{x}_o, x_o)$
    \ELSE
        \STATE $\hat{x}_o, \tau \gets \text{SearchVPTree}(\text{node}.inside,\tau, \hat{x}_o, x_o)$ 
        \STATE $\hat{x}_o, \tau \gets \text{SearchVPTree}(\text{node}.outside\tau, \hat{x}_o, x_o)$\ENDIF
    \ENDIF
    \STATE \textbf{return} $\hat{x}_o, \tau$
\ENDFUNCTION
\end{algorithmic}
\end{algorithm}
\end{minipage}

\vspace{1mm}

\caption{Nearest neighbour search in a $q$-VP tree. Each node is of the form $[p,\mu,\text{inside},\text{outside}]$, where $p$ is a vantage point and $\mu$ its associated radius. Given a query $x_o$, the algorithm maintains the current best candidate $\hat{x}_o$ and its distance $\tau=d(x_o,\hat{x}_o)$. At each visited node, it updates $(\hat{x}_o,\tau)$ and may prune inside or outside subtrees using the $q$-triangle inequality.}
\label{fig:vp-search}
\end{figure}

Notice, that the partition of $X$ into inside and outside sets as defined in \eqref{eqn_vp_children_apx} is independent of the structure of the dissimilarity $d$. This means, we can build the sets $X'$ and $X''$ for any dissimilarity dunction $d$. In particular, we can build these sets when $d$ represents distances of points in a $q$-metric or $\infty$-metric space.

The specific properties of $q$-metric and $\infty$-metric spaces are leveraged when searching for nearest neighbors. This search phase is illustrated in Algorithm~\ref{alg:vp-search} for the $q$-metric case and Algorithm~\ref{alg:vp-search-inf} for $\infty$-metric spaces. Given a node $[v,\mu,\texttt{inside},\texttt{outside}]$, we first compute the distance from the query $x_o$ to the node's vantage point $p$ (Step 3). If this improves the current best distance $\tau$ (Step 4), then the candidate nearest neighbor $\hat{x}_o$ is updated (Steps 5--6). Next, the pruning rules of Propositions~\ref{prop:q-vp-pruning} and~\ref{prop:inf-vp-pruning} are used to decide whether to visit the \texttt{inside} child (Step 8) or the \texttt{outside} child (Step 10), without discarding the true nearest neighbor. Although both algorithms share the same initial steps, their pruning differs. In the \(q\)-metric case, Algorithm~\ref{alg:vp-search} may need to visit both children (Step~12); this additional step is not present in Algorithm~\ref{alg:vp-search-inf}. The procedure recurses until a leaf is reached; it then returns the current best neighbor $\hat{x}_o$ and its distance to $x_o$, namely $\tau$ (Step 14).

\begin{figure}[H]
\centering
\begin{minipage}{0.98\linewidth}
\begin{algorithm}[H]
\caption{Searching Phase of the $\infty$-VPTree}
\label{alg:vp-search-inf}
\begin{algorithmic}[1]
\setstretch{1.4}
\INPUT Query point $x_o$, parameter $q$, and a node (initialized to the root). Nearest neighbor $\hat{x}_o$ is set to \textit{null} and its distance $\tau$ to the query is initialized as $\infty$.
\OUTPUT Nearest neighbor $\hat{x}_o$ and its distance $\tau$, after searching the subtree rooted at a node.

\FUNCTION{SearchVPTree$_q(\text{node},\tau,x_o,\hat{x}_o)$}
    \IF{$v\neq\textit{null}$}
    \STATE $d \gets d(x_o,\text{node}.p)$
    \IF{$d<\tau$}
        \STATE $\tau \gets d$ 
        \STATE $\hat{x}_o \gets \text{node}.p$ 
    \ENDIF

    \IF{$d < \text{node}.\mu$}
        \STATE $\hat{x}_o,\tau\gets \text{SearchVPTree}(\textit{inside},\tau,\hat{x}_o, x_o)$ 
    \ELSE
        \STATE $\hat{x}_o,\tau\gets \text{SearchVPTree}(\textit{outside},\tau,\hat{x}_o,x_o)$ 
    \ENDIF
    \STATE \textbf{return} $\hat{x}_o,\tau$
\ENDFUNCTION
\end{algorithmic}
\end{algorithm}
\end{minipage}

\vspace{1mm}

\caption{Nearest-neighbour search in an $\infty$-VP tree. Each node is of the form $[p,\mu,\text{inside},\text{outside}]$, where $p$ is a vantage point and $\mu$ its associated radius. Given a query $x_o$, the algorithm maintains the current best candidate $(\hat{x}_o,\tau)$. At each visited node, it updates $(\hat{x}_o,\tau)$ and prunes to a single branch using the strong triangle inequality.}
\label{fig:vp-search-inf}
\end{figure}

\begin{wrapfigure}{R}{0.52\columnwidth}
    \centering
    \vspace{-1.0em}
    \includegraphics[width=\linewidth]{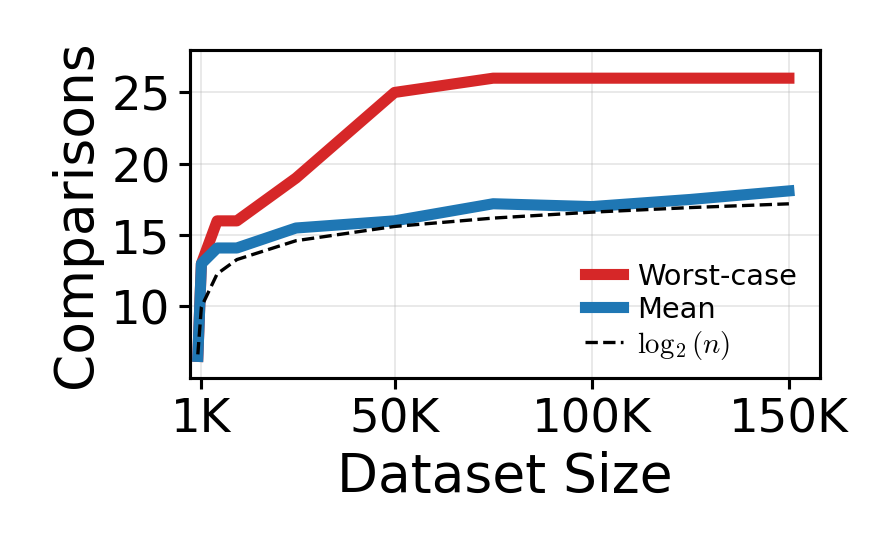}
    \caption{VP Tree search complexity on a $\infty$-metric space with $n\in\{100,500,1\text{K},5\text{K},\,10\text{K},50\text{K},100\text{K},150\text{K}\}$ points. The worst-case bound in comparisons corresponds to the depth of the tree.}
    \label{app:fig:worst_case}
    \vspace{-1.0em}
\end{wrapfigure}

\begin{remark}[Balanced VP trees] \label{rmk_balanced_tree}

A balanced VP node is one in which the number of points in the inside and outside sets differ by at most one, $| |X'|-|X''| | \leq 1$. A balanced VP tree is one in which all nodes are balanced. If the VP tree is balanced its depth satisfies $h \leq \log_2 (|X|) + 1$ and is the lowest possible among all VP trees of the set $X$. This is accomplished in Step 5 of the algorithm in Algorithm \ref{alg:vp-build} where we set the radius to $\mu = \median_{ X }  d(v,x)$. This is a good choice for a balanced tree because precisely half of the points in $X$ are on each side of the median. The tree is not perfectly balanced, however, because we may have several points $x \in X$ with $d(v,x) = \mu$, which, according to \eqref{eqn_vp_children_apx}, must be assigned to the outside set $X''$. Having several points with $d(v,x) = \mu$ almost never happens in metric and $q$-metric spaces but it does sometimes happen in $\infty$-metric spaces. Because of this, the depth of VP trees of samples of an $\infty$-metric space is (slightly) larger than the depth  $h \leq \log_2 (|X|) + 1$ of a balanced tree; see Figure \ref{app:fig:worst_case}. This figure shows that the worst-case complexity -namely, the tree depth- lies within a constant slack of logarithmic scaling, the latter matching the expected number of comparisons.

\end{remark}

In this settings where logarithmic depth is reached, Theorem \ref{app:theo_log_coplexity} guarantees logarithmic search complexity. Remarkably, this is not merely an upper bound: it is an exact characterization. That is, for every query, the search requires exactly a logarithmic number of comparisons -up to a $+1$ slack-, so that best- and worst-case coincide.

\begin{remark} \label{rmk_vp_practicalities}

Note that in Step 2 we select $v$ arbitrarily. Proper choice of $v$ may have a significant effect in practice and the use of heuristic to select good vantage points is common \cite{vptree}. Commonly, the vantage point $v$ is set to maximize the spread of the induced distances. In the original VP-tree work \cite{vptree}, this was done by considering higher moments of the distance distribution, such as the variance.

\end{remark}

\section{Projection}\label{app:projection}

In this Appendix, we expand on the theoretical properties exposed in Section \ref{section:metric_representation}.
The proofs rely on the axioms (\ref{eq:projection_axiom}) and (\ref{eq:transformation_axiom}), that are imposed on the Canonical Projection. These two requirements are enough to imbue the projection with beneficial properties on Nearest Neighbor search.

\subsection{\texorpdfstring{$P_q^\star$}{Pq*} exists, is unique, and satisfies the \texorpdfstring{$q$}{q}-triangle inequality}
\label{app:subproofs}

The results in Section~\ref{app:subproofs} were presented in \citep{segarra2019metricoriginal} and are included here for completeness. 

For a given $q$, the Canonical Projection $P_q^\star$ as defined in \eqref{E:canonical_q_metric}, is guaranteed to satisfy the $q$-triangle inequality. Moreover, it meets axioms of Projection \ref{eq:projection_axiom} and Transformation \ref{eq:transformation_axiom}. 

\begin{lemma}[Satisfaction of $q$-triangle inequality]
    The canonical Projection as defined in \ref{E:canonical_q_metric}, satisfies the $q$-triangle inequality.
    \label{prop:satisfies-q}
\end{lemma}

\begin{proof}
    To verify the \( q \)-triangle inequality, let \( c_{xx'} \) and \( c_{x'x''} \) be paths that achieve the minimum in \eqref{E:canonical_q_metric} for \( d_q(x, x') \) and \( d_q(x', x'') \), respectively. Then, from the definition in \eqref{E:canonical_q_metric}, it follows that:
    \begin{align}
    d_q(x, x'')^q &= \min_{c_{xx''}} \|c\|_q^q, 
    \leq \|c_{xx'} \oplus c_{x'x''}\|_q^q, \notag \\ 
    &= \|c_{xx'}\|_q^q + \|c_{x'x''}\|_q^q \notag \\
    &= d_q(x, x')^q + d_q(x', x'')^q,
    \end{align}
    where the inequality holds because the concatenated path \( c_{xx'} \oplus c_{x'x''} \) is a valid (though not necessarily optimal) path between \( x \) and \( x'' \), while \( d_q(x, x'') \) minimizes the \( q \)-norm across all such paths.
\end{proof}

\newtheorem*{theoremone}{Theorem 1}  
\begin{theoremone}[Existence and uniqueness] 

The canonical projection $P^{\star}_q$ satisfies Axioms \ref{eq:projection_axiom} and \ref{eq:transformation_axiom}.
Moreover, if a $q$-metric projection $P_q$ satisfies Axioms \ref{eq:projection_axiom} and \ref{eq:transformation_axiom}, it must be that $P_q = P^{\star}_q$. 
\end{theoremone}

\begin{proof}
We first prove that \( d_q \) is indeed a \( q \)-metric on the node space \( X \). That \( d_q(x, x') = d_q(x', x) \) follows from the fact that the original graph \( G \) is symmetric, and that the norms \( \|\cdot\|_q \) are symmetric in their arguments for all \( q \). Moreover, \( d_q(x, x') = 0 \) if and only if \( x = x' \), due to the positive definiteness of the \( q \)-norms.

To see that the Axiom of Projection \ref{eq:projection_axiom} is satisfied, let \( M = (X, d) \in \mathcal{M}_q \) be an arbitrary \( q \)-metric space, and denote \( (X, d_q) = P_q^\star(M) \), the output of applying the canonical \( q \)-metric projection. For any pair \( x, x' \in X \), we have:
\begin{align}
d_q(x, x') = \min_{c\in C_{xx'}} \|c\|_q \leq \|[x, x']\|_q = d(x, x'),
\end{align}
for all \( q \), where the inequality comes from choosing the trivial path \( [x, x'] \) consisting of a single edge from \( x \) to \( x' \).

Let \( c_{xx'}^\star = [x = x_0, x_1, \ldots, x_\ell = x'] \) be the path that achieves the minimum in the above expression. Using Lemma \ref{prop:satisfies-q} we know $d_q$ satisfies the \( q \)-triangle inequality:
\begin{align}
d(x, x') \leq \left( \sum_{i=0}^{\ell-1} d(x_i, x_{i+1})^q \right)^{1/q} = \|c_{xx'}^\star\|_q = d_q(x, x').
\end{align}
Substituting this into the previous inequality, we find that:
\begin{align}
d(x, x') = d_q(x, x').
\end{align}
Since \( x \) and \( x' \) were arbitrary, we conclude \( d \equiv d_q \), hence \( P_q^\star(M) = M \), as required.

To show that the Axiom of Transformation \ref{eq:transformation_axiom} holds, consider two graphs \( G =(X,D) \) and \( G' = (X',D') \), and a dissimilarity-reducing map \( \varphi: X \to X' \). Let \( (X, d_q) = P_q^\star(G) \) and \( (X', d_q') = P_q^\star(G') \) be the projected spaces.

For a pair \( x, x' \in X \), let \( c_{xx'}^\star = [x = x_0, x_1, \ldots, x_\ell = x'] \) be a path achieving the minimum for \( d_q(x, x') = \|c_{xx'}^\star\|_q \). Consider the image path in \( X' \):
\begin{align}
P_{\varphi(x)\varphi(x')}^\star = [\varphi(x_0), \varphi(x_1), \ldots, \varphi(x_\ell)].
\end{align}
Since \( \varphi \) is dissimilarity-reducing, we have:
\begin{align}
D'(\varphi(x_i), \varphi(x_{i+1})) \leq D(x_i, x_{i+1}) \quad \text{for all } i.
\end{align}
Hence,
\begin{align}
\|P_{\varphi(x)\varphi(x')}^\star\|_q \leq \|c_{xx'}^\star\|_q = d_q(x, x').
\end{align}
Now, since \( d_q'(\varphi(x), \varphi(x')) \) is defined as the minimum over all such paths in \( X' \), we get:
\begin{align}
d_q'(\varphi(x), \varphi(x')) \leq \|P_{\varphi(x)\varphi(x')}^\star\|_q \leq d_q(x, x'),
\end{align}
which completes the proof of Axiom \ref{eq:transformation_axiom}.
\end{proof}

\subsection{Canonical Projection preserves the Nearest Neighbor}

We  canonical projection maps any graph to a $q$-metric space, as showed in Lemma~\ref{prop:satisfies-q}. Through  Section~\ref{app:Vp-Trees}, we exposed the advantages of searching nearest neighbors in such metric spaces. It remains to be proved, that this search is not only efficient but also accurate, i.e., that the projection preserves the nearest neighbor.

\begin{proposition}\label{app:nn_preserved}
Given a graph $G = (X, D)$ and a query $x_o$, define $H = (Y, E)$ with $Y = X \cup \{x_o\}$ and let $(Y, E_q) = P^\star_q(H)$ be the projected graph. 
Then, the set of nearest neighbors satisfies
\begin{align} \label{eqn_nn_preserved_appendix}
    \hhatx_o 
        ~\equiv~ \argmin_{x\in X} ~ E(x_o,x)
        ~\subseteq~ \argmin_{x\in X} ~ E_q( x_o,x).
\end{align}
\end{proposition}

\begin{figure}[t]
    \centering
    \vspace{-1.0em}
    \includegraphics[width=0.6\columnwidth]{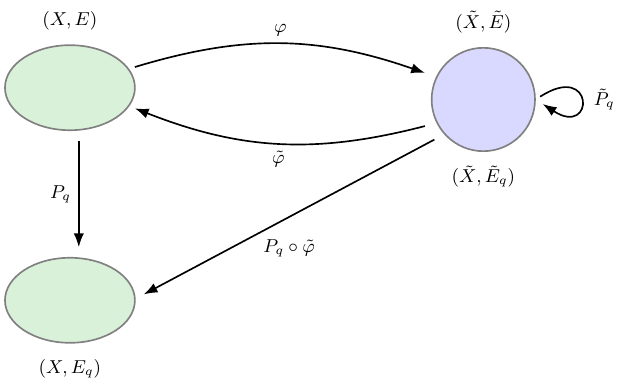}
    \caption{Canonical projection preserves the nearest neighbor. Starting from the original graph $(X,E)$, we construct the graph $(\tilde{X},\tilde{E})$ together with two dissimilarity-reducing maps $\varphi$ and $\tilde{\varphi}$ between them. Applying any \textit{admissible} map $P_q$ and $\tilde{P}_q$ yields the respective $q$-metric spaces $(X,E_q)$ and $(\tilde{X},\tilde{E}_q)$. Note that, since $(\tilde{X},\tilde{E})$ is already a trivial $q$-metric space, the projection leaves it unchanged \eqref{eq:projection_axiom}.}
    \label{app:fig:proof}
    \vspace{-1.0em}
\end{figure}

\begin{proof} We prove the theorem by showing that any \textit{admissible} projection i.e, any map satisfying axioms \eqref{eq:projection_axiom} and \eqref{eq:transformation_axiom}, \emph{cannot} result in larger dissimilarities for any pair of points. We further show that nearest neighbor dissimilarities are preserved by the canonical projection. The combination of these two facts yields \eqref{eqn_nn_preserved_appendix}. A diagram to better follow the proof is provided in Figure~\ref{app:fig:proof}.

To prove that dissimilarities are not increased, construct an arbitrary graph $\tdG = (\tdX,\tdE)$ made up of two nodes $\tdX = \{\tdx, \tdy \}$, along with a unique edge connecting them $\tdE(\tdx,\tdy)$ . We also enforce this graph to have its unique edge weight, equal to the distance between the nearest neighbor $\hat{x}_o$ and the query point $x_o$, i.e $\tdE(\tdx,\tdy)=E(x_o,\hat{x}_o)$. In addition, we trivially set any reflexive edge as null, namely $\tdE(\tdx,\tdx)=\tdE(\tdy,\tdy)=0$.

Notice that $(\tdX,\tdE)$ has only two vertices. As such, it is a trivial $q$-metric space: no triplet can be formed, and the $q$-triangle inequality holds by default. As a direct consequence of axiom~\eqref{eq:projection_axiom}, any admissible projection acting on $(\tdX,\tdE)$ leaves the space unchanged. Therefore, for the remainder of the proof, $(\tdX,\tdE)$ will be used indistinguishably as $(\tdX,\tdE_q)$.
    
Let us now map the original graph $G$ to the new one as in Figure \ref{app:fig:proof},
\begin{align}\label{proof:phi}
    \varphi(x) =
    \begin{cases}
        \tdx & \text{if } x = x_o, \\
        \tdy & \text{otherwise.}
    \end{cases}
\end{align}
Since the only existing edge in $\tdG$ attains the minimum possible value, namely $E(\hat{x}_o,x_o)$, every edge of the original graph $G$ that is incident to $x_o$ will be mapped to a smaller dissimilarity by~\eqref{proof:phi}. On the other hand, edges that are not incident to $x_o$, are mapped to the self-loop $\tilde{E}(\tilde{y},\tilde{y})=0$. Because we can assume $E$ to be a \textit{positive}\footnote{\label{fn:positive} Notice that by definition, dissimilarity measures are lower bounded \cite{dissimilarity}, therefore, it is always possible to obtain a positive version of them.} dissimilarity, this also yields a dissimilarity reduction. Therefore, we can consider $\varphi$ a dissimilarity-reducing map. Consequently, $\forall \ x \in X$,
\begin{align}E(x_o,x)\geq \tdE(\varphi(x_o),\varphi(\hat{x}_o))=E(x_o,\hat{x}_o)\end{align}
Now we can invoke axiom \eqref{eq:transformation_axiom}, which ensures the satisfaction of the previous inequality also in the projected spaces,
\begin{align}
E_q(x_o,x) \geq \tdE_q(\varphi(x_o),\varphi(\hat{x}_o))=E(x_o,\hat{x}_o).
\end{align}
While this may appear to conclude the proof, it does not. Up to this point, we’ve shown that any admissible map cannot decrease edge weights below the original nearest-neighbor distance. It remains to prove that this nearest-neighbor distance is in fact preserved by any admissible map. With this purpose, we define the following map from $\tdG$ to the original graph $G$, illustrated in Figure \ref{app:fig:proof},
\begin{align}
    \td{\varphi}(\tdx) = x_o,  \notag\\
    \td{\varphi}(\tdy) = \hat{x}_o.
\end{align}
This map is a bijection sending $\tdE(\tdx,\tdy)$ to $E(x_o,\hat{x}_o)$. Consequently, we can conclude that $\tilde{\varphi}$ is a trivial dissimilatity-reducing map,
\begin{align}E(x_o,\hat{x}_o)=\tdE(\tdx,\tdy)\geq E(\td\varphi(\tdx),\td\varphi(\tdy))=E(x_o,\hat{x}_o).\end{align} 
We then proceed as with the previous case, invoking Axiom \eqref{eq:transformation_axiom}, which preserves the inequality in the projected space,
\begin{align}
    E(x_o,\hat{x}_o)=\tdE_q(\tdx,\tdy) \geq E_q(\td\varphi(\tdx),\td\varphi(\tdy))=E_q(x_o,\hat{x}_o).
\end{align}
Therefore, the projected distance between $x_o$ and its nearest neighbor $\hat{x}_o$ coincides with the original distance $E(x_o,\hat{x}_o)$, and thus attains the lower bound,
\begin{align}
    E_q(x,x_o)\geq E(x_o,\hat{x}_o) = E_q(x_o,\hat{x}_o)
\end{align}
The proof was made with an arbitrary \textit{admissible} projection, taking into account that the canonical projection is the only admissible map (Theorem \ref{theo_uniqueness}), we conclude the proof.
\end{proof}

Having a preservation of the nearest neighbors makes the projection a valid candidate for search. The previous result can be further extended as follows,

\begin{corollary}\label{app:nn_preserved_equality}
    Under the setting of Proposition \ref{app:nn_preserved}, if $q< \infty $, the nearest neighbor is preserved with the following equality:
    \begin{align}
        \hhatx_o  \equiv \argmin_{x\in X} ~ E(x, x_o) = \argmin_{x\in X} ~ E_q(x, x_o).
    \end{align}
\end{corollary}

\begin{proof}
The proof will follow by contradiction.

Assume there exists \( z \in X \) a nearest neighbor in the projected space $E_q$, such but not in the original space $E$, i.e,
\begin{align}
z \notin \argmin_{x \in X} E(x, x_o) \quad \text{and} \quad z \in \argmin_{x \in X} E_q(x, x_o).
\end{align}
Theorem \ref{theo_uniqueness},  demands \( E_q \) to be obtained with the canonical projection operator $P_q^*$, which is the $q$-shortest path between points. Thus, there exist a shortest chain $c^* = [ x_o, \dots, z]$ linking $x_o$ to $z$ such that
\begin{align}
    E_q(x_o,z) = ||c^*||_q = \min_{c \in \mathcal{C}_{x_oz}} \ell_q(c).
\end{align}
However, since \( z \notin \argmin_{x \in X} E(x, x_o) \), the one-hop chain \( [x_o,z] \), is not a valid minimum path,
\begin{align}
 E(x_o, z) =\sqrt[q]{d(x_o,z)^q}=\ell_q([x_o, z]) \neq \min_{x \in X} d(x_o, x).
\end{align}
Without loss of generality, assume there exists at least one intermediate vertex in the optimal path, i.e  \( \exists \ x_1 \neq z \) such that \( c^*=[x_o, x_1, \dots, z]\). Since $E$ is a positive dissimilarity (see footnote \ref{fn:positive}), \( \ell_q \) is strictly increasing, 
\begin{align}
    \ell_q([x_o, x_1, \dots, z]) > \ell_q([x_o, x_1]) \geq \min_{c \in \mathcal{C}_{x_oz}} \ell_q(c).
\end{align}
This means that an intermediate node must exist; yet in that case it would induce a projected distance smaller than that of $z$, contradicting the assumption that $z$ is a nearest neighbor. Therefore,
\begin{align}
    z \notin \argmin_{x \in X} E_q(x, x_0) .
\end{align}
\end{proof}

Notice that Proposition~\ref{app:nn_preserved} formulates nearest neighbor preservation as an inclusion. In contrast, Proposition~\ref{app:nn_preserved_equality} characterizes preservation via a strict equality between nearest-neighbor sets. This distinction matters: when projecting with $q=\infty$, points that were not nearest neighbors in the original space may become nearest neighbors in the projected space, as illustrated in Example \ref{ex:spurious}. We refer to such points as \textit{spurious neighbors}. From the perspective of vector search, this behavior is undesirable, it enlarges the candidate set, detracting from the advantages discussed in Section \ref{app:Vp-Trees}.

\begin{wrapfigure}{R}{0.52\columnwidth}
  \centering

  \begin{minipage}{\linewidth}
    \centering
    \begin{tikzpicture}[font=\small,scale=1]
      \node[draw,circle, thin, inner sep=1pt] (x3) at (0,0) {$x_3$};
      \node[draw,circle, thin, inner sep=1pt] (x2) at (1,0) {$x_2$};
      \node[draw,circle, thin, inner sep=1pt] (x1) at (0.875,1.75) {$x_1$};
      \draw (x1) -- node[midway,right] {3} (x2)
            (x2) -- node[midway,below] {2} (x3)
            (x3) -- node[midway,left] {5} (x1);
      \node[right=30pt] at ($(x2)!0.5!(x3)$) {$E$};
    \end{tikzpicture}
  \end{minipage}

  \vspace{0.8em}

  \begin{minipage}{\linewidth}
    \centering
    \begin{tikzpicture}[font=\small,scale=1]
      \node[draw,circle, thin, inner sep=1pt] (x3p) at (0,0) {$x_3$};
      \node[draw,circle, thin, inner sep=1pt] (x2p) at (1,0) {$x_2$};
      \node[draw,circle, thin, inner sep=1pt] (x1p) at (0.5,1.75) {$x_1$};
      \draw (x1p) -- node[midway,right] {3} (x2p)
            (x2p) -- node[midway,below] {2} (x3p)
            (x3p) -- node[midway,left] {3} (x1p);
      \node[right=30pt] at ($(x2p)!0.5!(x3p)$) {$E_{\infty}$};
    \end{tikzpicture}
  \end{minipage}

  \caption{The Canonical Projection with $q=\infty$, added nearest neighbors to the original set.}
  \label{app:fig:spurious}
  \vspace{-3.0em}
\end{wrapfigure}

\begin{example}\label{ex:spurious}
    Let \(\{x_1,x_2,x_3\}=X\) be three points equipped with the original distances
    \begin{align}
      E(x_1,x_2)&=3,\quad E(x_2,x_3)=2, \notag \\
      &E(x_1,x_3)=5.
    \end{align}
    Clearly \(E(x_1,x_2)<E(x_1,x_2)\), so the unique nearest neighbor of \(x_1\) in \(G=(\{x_1,x_2,x_3\},E)\) is \(x_2\).  We now enforce the strong triangle inequality by projecting:
    \begin{align}
      E_{\infty}(x_2,x_3)\;=\;\max\Big[\{E_{\infty}(x_1,x_2),\,E_{\infty}(x_2,x_3)\Big]\;=\;3.
    \end{align}
    while leaving \(E_{\infty}(x_1,x_2)=3\) and \(E_{\infty}(x_2,x_3)=2\).  Under the new distance \(E_{\infty}\), one finds
    \begin{align}
      E_{\infty}(x_1,x_2)=3 \quad\text{and}\quad E_{\infty}(x_1,x_3)=3.
    \end{align}
    so \(x_1\) is equidistant from \(x_2\) and \(x_3\).  As showed in Figure \ref{app:fig:spurious}, the original nearest neighbor \(x_2\) is no longer uniquely closest, i.e, any nearest‐neighbor search on \(G'=(\{x_1,x_2,x_3\},E_{\infty})\) may return \(x_3\) instead of \(x_2\), demonstrating that ultra‐projecting a metric can extend original nearest neighbor set.
\end{example}

Proposition \ref{app:nn_preserved_equality} guarantees the nearest neighbor to be preserved under the Canonical Projection. Therefore, the Nearest Neighbor Search problem in the transformed space is a relaxation of the original problem. Although this relaxation provides speedup, it can also make the problem more difficult if many solutions are added to the optimal set. This problem is addressed in Section \ref{app:two_stage}, by enriching the VPTree with a two-stage search.

\subsection{Efficient Approximation of \texorpdfstring{\( P_q^* \)}{Pq} }\label{app:efficient}

Calculating the canonical projection $P_q^\star$ over the training points can incur prohibitive computing times. Algorithms~\ref{alg:fermat-gpu-q} and~\ref{alg:fermat-gpu-inf} describe the computation of the canonical projection, inspired by a matrix-form variant of the classical Floyd-Warshall algorithm \cite{floyd,warshal}. These procedures are highly paralellizable and can be efficiently executed on GPUs, which significantly reduces the projection time. Nevertheless, computing the canonical projection for $n$ points still requires $O(n^3)$ operations in the worst case, which remains prohibitive for large datasets.

Instead of computing the shortest $q$-norm path in the complete graph \( D \in \mathbb{R}^{n \times n} \), we restrict the search to only consider the $k$-nearest neighbor graph. Doing this reduces the complexity of computing all pairwise distances from $O(n^3)$ to a factor $O(n^2 k^2)$. Moreover, \cite{facundo} showed that when the original space is euclidean and $k\approx log(n)$, the projected distances remain unchanged with high probability, with worst-case complexity $O(n^2 log^2(n))$.

\begin{figure}[H]
  \centering
  \begin{minipage}[t]{0.48\textwidth}
    \begin{algorithm}[H]
    \caption{Canonical Projection $P_q^*$}
    \label{alg:fermat-gpu-q}
    \begin{algorithmic}[1]
    \setstretch{1.4}
    \INPUT Distance matrix $D\in\mathbb{R}_+^{n\times n}$ and parameter $q$.
    \OUTPUT Projected matrix $P_q^*(D)$.

    \FUNCTION{$P_q^*(D,q)$}
      \STATE $M \gets D^q$
      \STATE $\mathbf{1} \gets (1,\dots,1) \in \mathbb{R}^n$
      \FOR{$i = 1$ : $n$}
        \STATE $C_i \gets M[:,i]\mathbf{1} + \mathbf{1}^\top M[i,:]$
        \STATE $M \gets \min(M, C_i)$
      \ENDFOR
      \STATE \textbf{return} $M^{1/q}$
    \ENDFUNCTION
    \end{algorithmic}
    \end{algorithm}
  \end{minipage}\hfill
  \begin{minipage}[t]{0.48\textwidth}
    \begin{algorithm}[H]
    \caption{Canonical Projection $P_\infty^*$}
    \label{alg:fermat-gpu-inf}
    \begin{algorithmic}[1]
    \setstretch{1.4}
    \INPUT Distance matrix $D\in\mathbb{R}_+^{n\times n}$.
    \OUTPUT Projected matrix $P_\infty^*(D)$.

    \FUNCTION{$P_\infty^*(D)$}
      \STATE $M \gets D$
      \STATE $\mathbf{1} \gets (1,\dots,1) \in \mathbb{R}^n$
      \FOR{$i = 1$ : $n$}
        \STATE $C_i \gets \max\!\big(M[:,i]\mathbf{1},\;\mathbf{1}^\top M[i,:]\big)$
        \STATE $M \gets \min(M, C_i)$
      \ENDFOR
      \STATE \textbf{return} $M$
    \ENDFUNCTION
    \end{algorithmic}
    \end{algorithm}
  \end{minipage}
  \caption{Algorithms for Canonical Projection computation. Complexity: $O(n^3)$. The $\max$ and $\min$ operators are applied element-wise to each element of the matrix. }
\end{figure}

These modifications are incorporated in Algorithms~\ref{alg:approx-gpu-q} and~\ref{alg:approx-gpu-inf}. Algorithm~\ref{alg:approx-gpu-q} begins by raising each entry of the distance matrix $D$ to the power $q$ (Step~2), thereby producing edge weights consistent with the $q$-path metric. However, this operation does is not performed in Algorithm \ref{alg:approx-gpu-inf}, since in ultrametric spaces, the Canonical Projection only accounts for the maximum value along a path. 

Next, for each node $x_i$, we restrict attention to its $k$ nearest neighbors in the original space via the adjacency matrix $A$. Concretely, we introduce a masking matrix $S_A$ (Step 3), which is later added to the working distance matrix $M$ (Step 4), in order to render non-neighbor edges effectively unreachable, i.e., assigned a prohibitively large cost.

At each iteration, for every node \( x_i \) and each of its neighbors \( x_j \), we update the path cost \( D(x_i, x') \) to a third node \( x' \), to remain the minimum possible cost. In the $q$-metric case, Algorithm~\ref{alg:approx-gpu-q} combines previously $q$-powered distances via addition (Step~5), as in
\begin{align}
M(x_i, x') \leftarrow \min\left\{ M(x_i, x'),~ M(x_i, x_j)^q + M(x_j, x')^q \right\}.
\end{align}

In contrast, Algorithm~\ref{alg:approx-gpu-inf} applies an entrywise $\max$ update (Step~5), which corresponds to the canonical projection in ultrametric spaces.

Notice that, without loss of generality (see footnote~\eqref{fn:positive}), we may assume a \textit{positive} dissimilarity matrix $D\in\mathbb{R}^{n\times n}$. Consequently, raising $D$ to any power $q$ yields a positive matrix as well. This observation is relevant when applying $\min$ or $\max$ operators, since it preserves the interpretation of a minimum path.

Finally, the result is root-transformed to return to the original distance scale (Step~12):
\begin{align}
d_q^*(x_i, x_j) \approx \left( M(x_i, x_j) \right)^{1/q}.
\end{align}

\begin{figure*}[t]
  \centering

  \begin{minipage}[t]{0.48\textwidth}
    \begin{algorithm}[H]
      \caption{Sparse Canonical Projection $P_q^*$}
      \label{alg:approx-gpu-q}
      \begin{algorithmic}[1]
      \setstretch{1.4}
      \INPUT Distance matrix $D$, adjacency matrix $A$, number of pivots $l$, parameter $q$.
      \OUTPUT Projected approximate matrix $P_q^*(D;A,l)$.

      \FUNCTION{$P_q^*(D,A,l,q)$}
        \STATE $M \gets D^q$
        \STATE $(S_A)_{ij} \gets
          \begin{cases}
            0,      & \text{if } A_{ij}=1,\\
            \infty, & \text{otherwise}
          \end{cases}$
        \STATE $M \gets M + S_A$
        \STATE $\mathbf{1} \gets (1,\dots,1)^\top \in \mathbb{R}^n$
        \FOR{$t = 1$ : $l$}
          \FOR{$i = 1$ : $n$}
            \STATE $C_i \gets M[:,i]\mathbf{1}^\top + \mathbf{1}M[i,:]$
            \STATE $M \gets \min\!\big(M,\; C_i + S_A\big)$
          \ENDFOR
        \ENDFOR
        \STATE \textbf{return} $M^{1/q}$
      \ENDFUNCTION
      \end{algorithmic}
    \end{algorithm}
  \end{minipage}
  \hfill
  \begin{minipage}[t]{0.48\textwidth}
    \begin{algorithm}[H]
      \caption{Sparse Canonical Projection $P_\infty^*$}
      \label{alg:approx-gpu-inf}
      \begin{algorithmic}[1]
      \setstretch{1.4}
      \INPUT Distance matrix $D$, adjacency matrix $A$, number of pivots $l$.
      \OUTPUT Projected approximate matrix $P_\infty^*(D;A,l)$.

      \FUNCTION{$P_\infty^*(D,A,l)$}
        \STATE $M \gets D$
        \STATE $(S_A)_{ij} \gets
          \begin{cases}
            0,      & \text{if } A_{ij}=1,\\
            \infty, & \text{otherwise}
          \end{cases}$
        \STATE $M \gets M + S_A$
        \STATE $\mathbf{1} \gets (1,\dots,1)^\top \in \mathbb{R}^n$
        \FOR{$t = 1$ : $l$}
          \FOR{$i = 1$ : $n$}
            \STATE $C_i \gets \max\!\big(M[:,i]\mathbf{1}^\top,\;\mathbf{1}M[i,:]\big)$
            \STATE $M \gets \min\!\big(M,\; C_i + S_A\big)$
          \ENDFOR
        \ENDFOR
        \STATE \textbf{return} $M$
      \ENDFUNCTION
      \end{algorithmic}
    \end{algorithm}
  \end{minipage}

  \caption{Sparse Canonical Projection using adjacency $A$ (restricting updates to $k$ local neighbors) and early stopping after $l$ pivots (paths of length $\leq l$). Complexity: $O(lnk^2)$.}
  \label{fig:sparse-canonical-proj}
\end{figure*}
In addition, paths may be truncated to length $l$, so that at most $l$ intermediate components can participate in any path (Step 6). This additional masking step reduces the runtime to $O(ln\log^2(n)\,)$. At present, however, we do not have theoretical guarantees on the approximation error it induces on the projected distances. Still, since the $q=\infty$ case depends only on the maximum entry along a path, it is plausible that limiting the path length yields a small approximation error in practice.

After the aforementioned complexity reduction, the approximated algorithm remains highly paralellizable in GPU and now benefits from sparse matrix multiplications due to the sparsity of $A$.

\newpage

\section{Experiments \& Results}\label{section:Experiments}\label{app:experiments}
\subsection{Experimental settings}
\textbf{Vector Datasets}

The experiments were conducted on two commonly used datasets and one with distinct characteristics. We split the data randomly, using 80\% for indexing and the remaining 20\% is used as queries. We provide a summary of the datasets below, additional details can be found in the references provided.

\begin{itemize}
    \item \textbf{\citep{fashion}{Fashion-MNIST-784} Euclidean}: Image samples from the Fashion-MNIST collection, each flattened into a 784-dimensional vector.
    \item \textbf{\citep{pennington2014glove}{GloVe-200 Cosine}}: Text embeddings extracted from the GloVe (Global Vectors for Word Representation) model, each of dimension 200.

      \item \textbf{\citep{oliva2001modeling}{Gist-960 Euclidean}}: A set of real-valued GIST descriptors of natural scene images, each represented as a 960-dimensional vector. A GIST descriptor is a global, low-dimensional representation of an image’s spatial envelope.
      
    \item \textbf{\citep{nytimes256}{NYTimes-256 Cosine}}: Document embeddings from the UCI NYTimes Bag-of-Words corpus. We convert term-count vectors to TF–IDF and apply truncated SVD to 256 dimensions; evaluation uses cosine (angular) distance.

    \item \textbf{\citep{bodon2003kosarak}{Kosarak-41,000} Jaccard}: A real-world sparse binary transaction dataset derived from click-stream data of a Hungarian news portal. Each transaction is represented as a 41,000-dimensional vector.

\item \textbf{\citep{deep1b}{Deep1B-96} Euclidean}: A billion-scale collection of CNN-based image descriptors which are later PCA-reduced to 96 dimensions.

\end{itemize}

\textbf{Dissimilarities}

Apart from searching using the euclidean distance, as customary for Glove and FashionMNIST, to showcase how our approach can accommodate arbitrary dissimilarity functions, we also evaluate our method and results searching with other dissimilarities on the same datasets. The dissimilarities used are specified in Table~\ref{tab:dissimilarity}.

\begin{table}[h]
\renewcommand{\arraystretch}{1.4} 
\centering
\caption{Distance and Dissimilarity Metrics used.}\label{tab:dissimilarity}
\begin{tabular}{@{}ll@{}}
\toprule
\textbf{Metric} & \textbf{Formula} \\
\midrule
Euclidean Distance & \( d(x, y) = \sqrt{\sum_{i=1}^{d}(x_i - y_i)^2} \) \\
Manhattan Distance & \( d(x, y) = \sum_{i=1}^{d} |x_i - y_i| \) \\
Cosine Dissimilarity & \( d(x, y) = 1 - \frac{x \cdot y}{\|x\| \, \|y\|} \) \\
Correlation & \( d(x, y) = 1 - \frac{(x - \bar{x}) \cdot (y - \bar{y})}{\|x - \bar{x}\| \, \|y - \bar{y}\|} \) \\
\bottomrule
\end{tabular}
\end{table}

\textbf{Metrics} 

Approximate search algorithms are evaluated along two key dimensions: retrieval quality and search efficiency.

\emph{Search efficiency} is assessed primarily by the number of comparisons needed to retrieve a result for a given query. Unlike throughput, this metric is agnostic to implementation and hardware, offering a fair basis for comparing algorithmic efficiency.

\begin{itemize}
    \item \textbf{Number of Comparisons}: Every time the $q$-Metric VP-Tree visits a node. Reflects the computational cost related to search speed. 
    \item \textbf{Queries Per Second (QPS)}: Measures the throughput of the algorithm, indicating how many queries can be processed per second under the current configuration.
\end{itemize}

\emph{Retrieval quality} is evaluated using metrics that capture not just whether relevant results are retrieved, but how well their ordering is preserved. In particular, we rely on Recall@k and Rank Order, which -- unlike recall -- penalizes deviations in the relative ranking of retrieved results. This is crucial in downstream tasks such as recommendation, where the order of results matters. The metrics used in our evaluation are summarized below:

\begin{itemize}
    \item \textbf{\citep{bodon2003kosarak}{Kosarak-41,000} Jaccard}: A real-world sparse binary transaction dataset derived from click-stream data of a Hungarian news portal. Each transaction is represented as a 41,000-dimensional vector.

    \item \textbf{\citep{deep1b}{Deep1B-96} Euclidean}: A billion-scale collection of CNN-based image descriptors which are later PCA-reduced to 96 dimensions.
    \item \textbf{RankOrder@k}: Measures how well the approximate method preserves the original ordering of the true nearest neighbors. Let \( \mathcal{N}_k^{\text{true}}(y)  \) be the true \(k\)-nearest neighbors of a query \( y \), and \( \mathcal{N}_k^{\text{approx}}(y)= \{x_1, \ldots, x_k\} \) the corresponding approximate result. Let \( \pi(x_i, \mathcal{N}_k^{\text{true}}(y)) \) denote the position of \( x_i \) in the true result $\mathcal{N}_k^{\text{true}}(y)$ (or \(k+1\) if not found). The metric is defined as:
    \begin{align}
    \textbf{Absolute RankOrder@k} \notag \\ (y) =  \sum_{i=1}^{k} \left| i - \pi(x_i, \mathcal{N}_k^{\text{true}}(y))\right|\cdot 
    \frac{1}{k}\end{align}    
    Lower values indicate better rank preservation, with \(0\) being optimal.
    
    In addition, a variation of the rank order that takes into account the total number of points in the dataset is also used:

    \begin{align}
    \textbf{Relative RankOrder@k} \notag \\ (y) =  \sum_{i=1}^{k} \left| i - \pi(x_i, \mathcal{N}_k^{\text{true}}(y)) \right|
    \cdot\frac{100}{nk} \end{align}

    where $n$ is the size of the indexed points, i.e $|X|$. This measure expresses rank order but now as a percentage of the points available for retrieval.\\

    \item \textbf{Recall@k}: Measures the proportion of true \(k\)-nearest neighbors that are successfully retrieved by the approximate method. Let \( \mathcal{N}_k^{\text{true}}(y) \) be the true \(k\)-nearest neighbors of a query \( y \), and \( \mathcal{N}_k^{\text{approx}}(y) \) the corresponding approximate result. The recall is defined as:

    \begin{align}
    \texttt{Recall@k}(y) = \frac{|\mathcal{N}_k^{\text{true}}(y) \cap \mathcal{N}_k^{\text{approx}}(y)|}{k}\end{align}
    This metric ranges from \(0\) to \(1\), where \(1\) indicates that all true neighbors were retrieved. It reflects the \textit{coverage} of the ground-truth neighbors in the approximate result.

\textbf{Setup}

Experiments were run on a workstation equipped with a 64-core (128-thread) AMD CPU, 256GB of system RAM, and two NVIDIA RTX A5000 GPUs with 24GB of VRAM each. This configuration provides sufficient CPU parallelism and GPU acceleration to support the projection step in Algorithms~\ref{alg:approx-gpu-q} and~\ref{alg:approx-gpu-inf}, as well as the embedding-model training performed in the experiments.

\end{itemize}

\subsection{Canonical Projection \texorpdfstring{$P_q^\star$}{Pq*}}\label{app:metric_representation}

We validate the theoretical properties of searching in $q$-metric and ultrametric spaces presented in Section~\ref{section:NNS}. In particular, we confirm our complexity claims (C2), the preservation of nearest neighbors, and the stability of rank order under projection (C3).

In order to do so, we use the Canonical Projection $P^\star_q$ presented in Section \ref{section:metric_representation} to project distances imposing $q$-metric structure on FashionMNIST and GloVe. For these two datasets we use four different dissimilarities. After projecting dissimilarities, we search using the $q$-metric VP tree as described in Appendix~\ref{app:Vp-Trees}.

Figures ~\ref{app:haven1} and \ref{app:haven2} (first row), show that the number of nodes visited during search decreases monotonically with increasing $q$, reaching the theoretical minimum of $\log_2(n)$ at $q=\infty$. This directly confirms Theorem~\ref{theo_log_complexity} and supports claim (C2). Additionally, a rank order of zero across all queries--for a wide range of \( q \) values (excluding \( q = \infty \)) and across all dissimilarities-- confirms claim~ and~(C3), as established in Lemma~\ref{prop:satisfies-q} and Proposition~\ref{prop_neares_neighbor_is_preserved}. At the same time, this nearest neighbor preservation is also observed at recall, showing perfect matches at $k=1$ for moderate $q$ values.

\begin{figure*}[!htbp]

    \centering
    \includegraphics[width=1\textwidth]{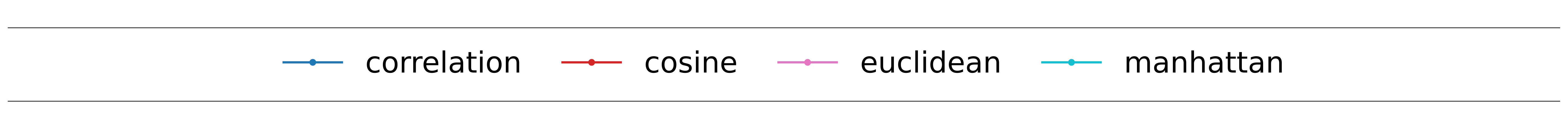}
    \vspace{-0.3cm} 
    \begin{subfigure}[b]{1\textwidth}
        \centering
        \includegraphics[width=\linewidth,height=0.25\textheight,keepaspectratio]{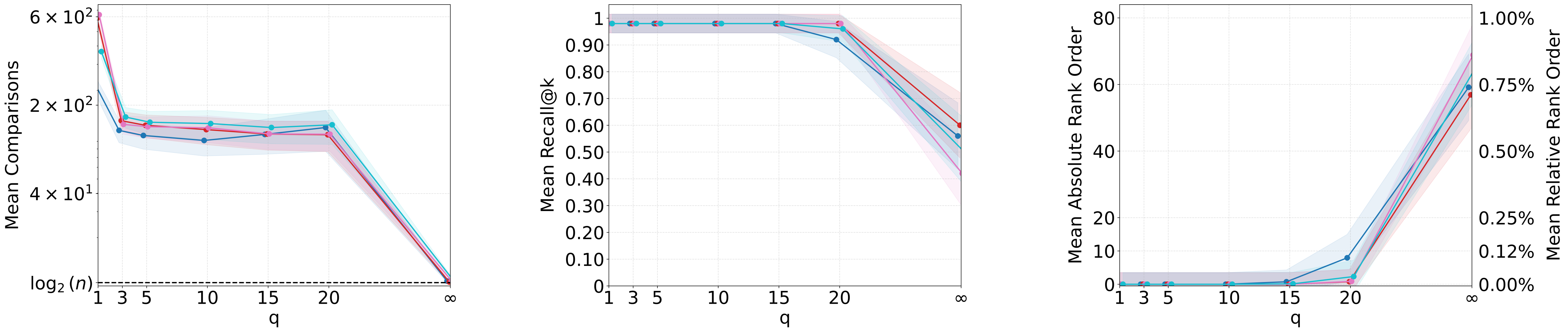}
        \vspace{0.3cm}
        \includegraphics[width=\linewidth,height=0.25\textheight,keepaspectratio]{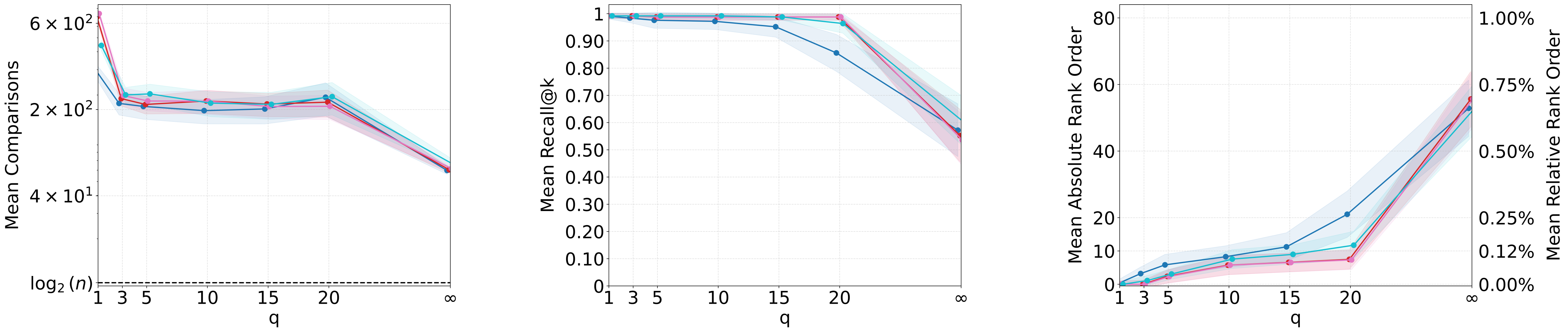}
        \vspace{0.3cm}
        \includegraphics[width=\linewidth,height=0.25\textheight,keepaspectratio]{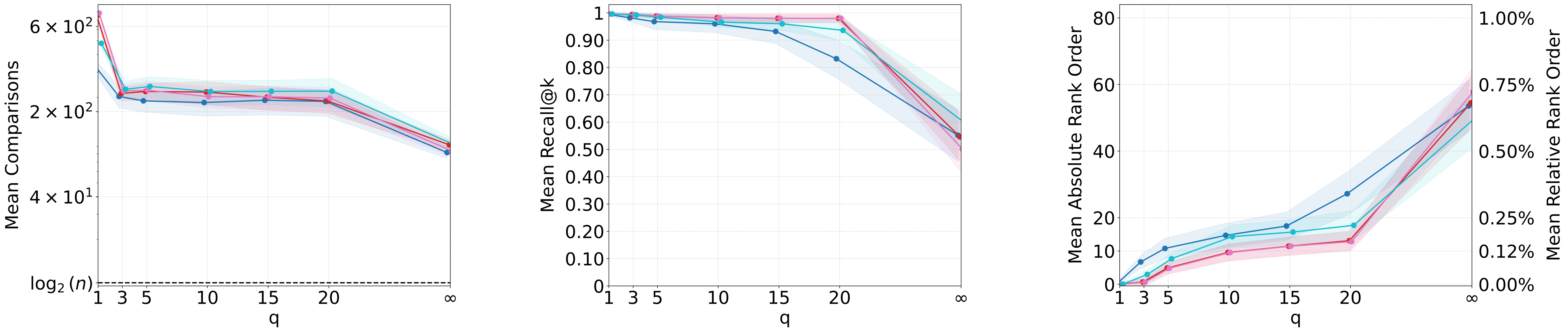}
        \caption{$n=1{,}000$ points of Fashion-MNIST}
        \label{app:haven1}
    \end{subfigure}%
        \caption{Number of comparisons and rank order across different dissimilarities when searching after applying Canonical Projection when a query point is added ($E_q$). The search was performed with a $q$-VPTree. Solid lines denote the mean and shading the standard deviation computed across queries. Each row shows results for $k$-nearest neighbors, with $k=1,5,10$ from top to bottom.}

\end{figure*}
    
\begin{figure*}[!htbp]
    \ContinuedFloat
    \centering
    \includegraphics[width=1\textwidth]{legend_only.png}
    \vspace{-0.3cm}
    \begin{subfigure}[b]{1\textwidth}
        \centering
        \includegraphics[width=\linewidth,height=0.25\textheight,keepaspectratio]{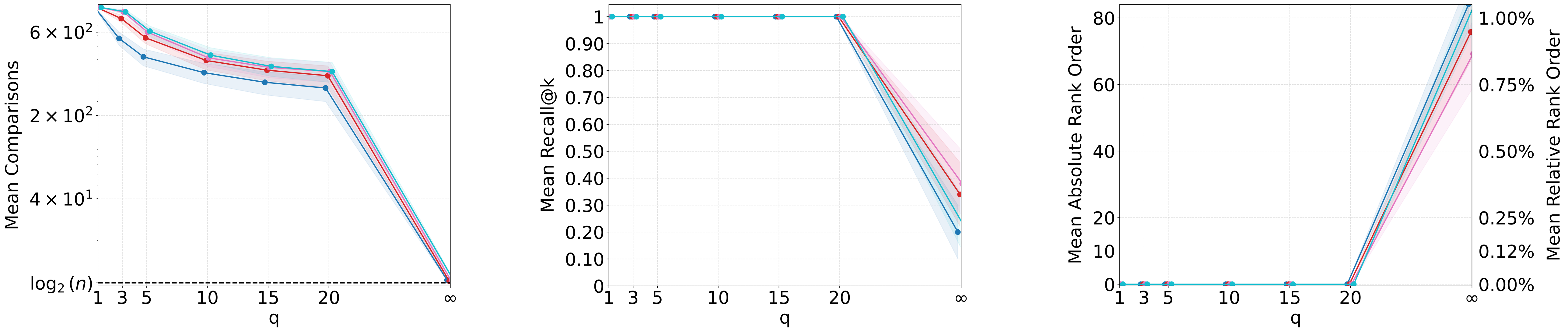}
        \vspace{0.3cm}
        \includegraphics[width=\linewidth,height=0.25\textheight,keepaspectratio]{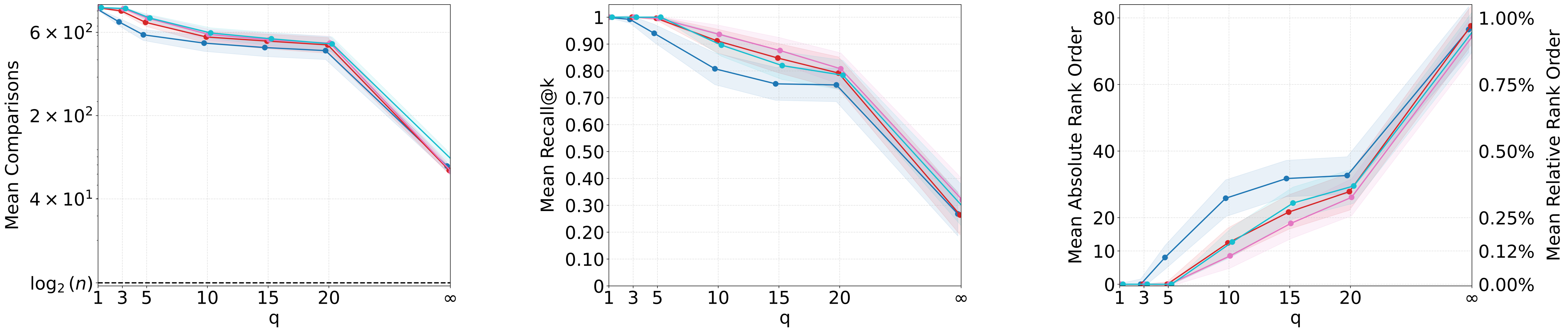}
        \vspace{0.3cm}
        \includegraphics[width=\linewidth,height=0.25\textheight,keepaspectratio]{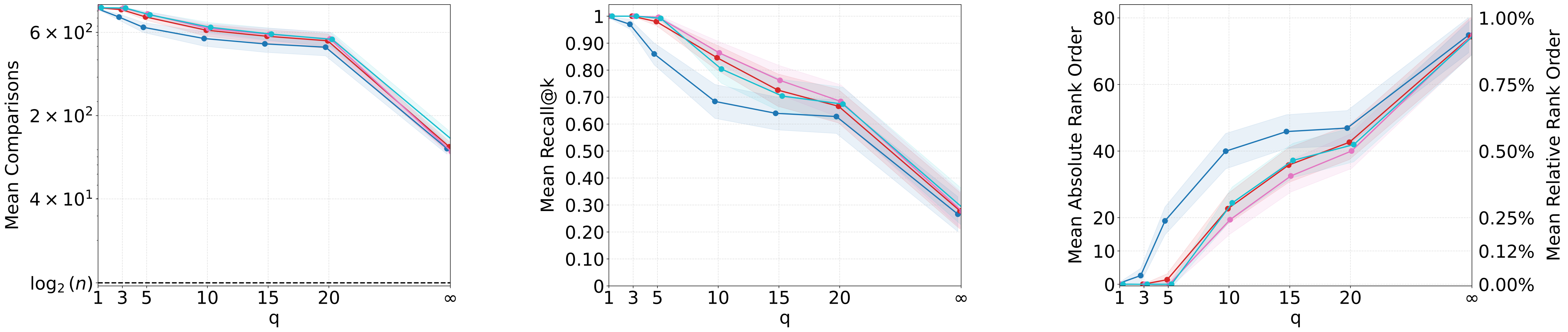}
        \caption{$n=1{,}000$ points of GloVe}
        \label{app:haven2}
    \end{subfigure}%
    \vspace{0.2cm}
\caption{Number of comparisons and rank order across different dissimilarities when searching after applying Canonical Projection when a query point is added ($E_q$). The search was performed with a $q$-VPTree. Solid lines denote the mean and shading the standard deviation computed across queries. The $k$-nearest neighbors are listed from top to bottom for $k=1,5,10$}
\end{figure*}

At $q=\infty$, the projection may introduce spurious optima not included in the original nearest neighbor set. Although the original nearest neighbours are still a solution in the transformed space, in practice we observe that the spurious optima at $q=\infty$ indeed affect accuracy. A similar effect occurs for large $q$, as distances between points can become artificially close, imitating this behavior.

Although our theoretical guarantees focus on the $k=1$ nearest neighbor case, an empirical preservation of locality observed in Figures \ref{app:haven1} and \ref{app:haven2}. We evaluate this by searching for $k=5$ and $k=10$ neighbors using the projected distances, as shown in the second and third rows of Figures \ref{app:haven1} and \ref{app:haven2}. While the number of comparisons does not decrease as rapidly as in the $k=1$ case, the method consistently yields improvements across all values of $q$.

\subsection{Approximating the canonical projection with \texorpdfstring{$\Phi(x;\theta^\star)$}{Phi}}\label{app:learning}

We analyze how well the learned distances $\hat{E}_q\left(x, x^{\prime}\right)=\left\|\Phi\left(x ; \theta^{\star}\right)-\Phi\left(x^{\prime} ; \theta^{\star}\right)\right\|$, described in Section~\ref{section:approximation}, reproduce the properties of the true $ \mathrm{q} $-metric distances $E_q\left(x, x^{\prime}\right)$.

\begin{figure*}[!htbp]
\centering

\makebox[\textwidth][c]{%
  \begin{subfigure}[b]{0.8\textwidth}
    \centering
    \includegraphics[width=\linewidth,keepaspectratio]{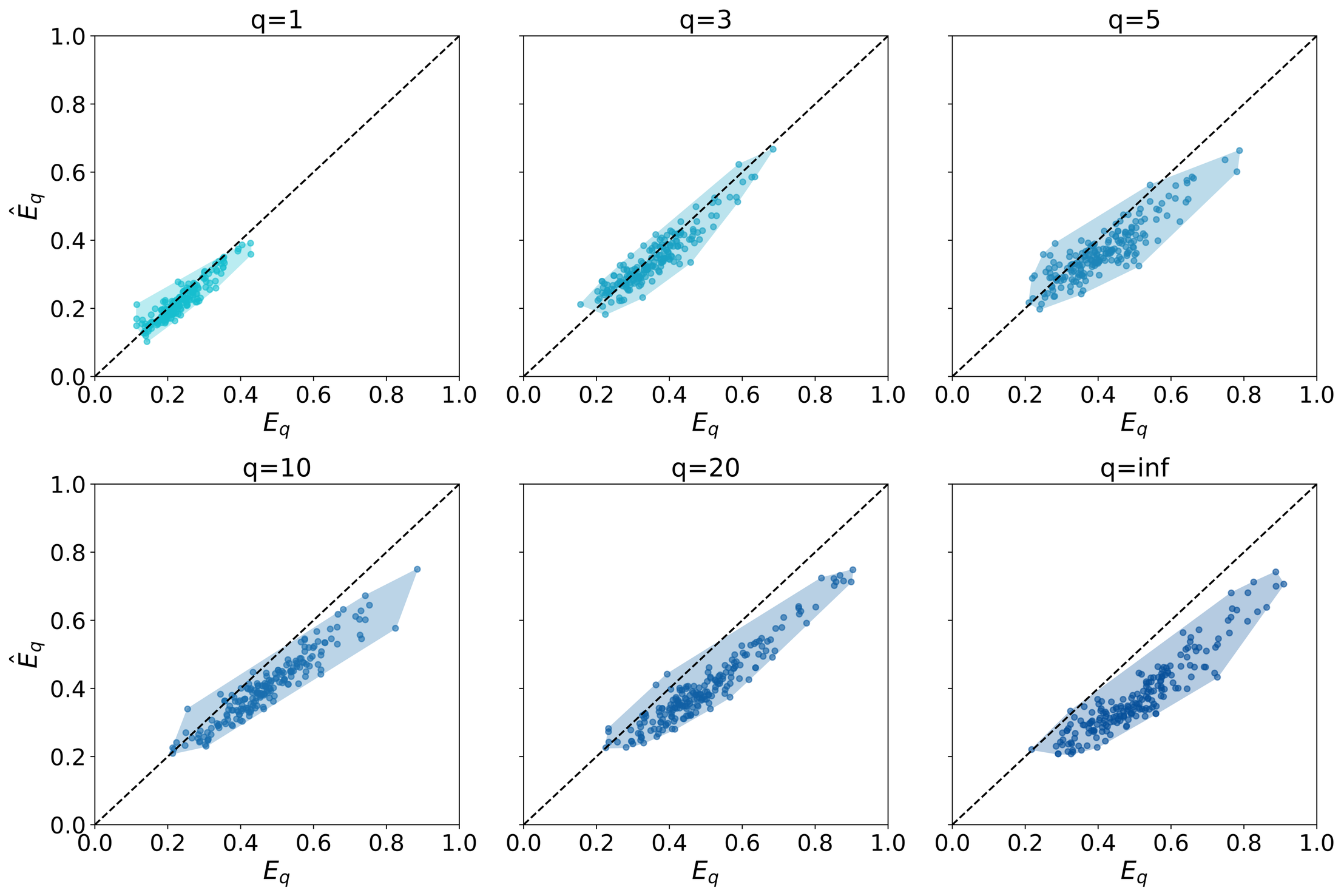}
    \caption{Train}
  \end{subfigure}%
}

\vspace{0.5cm}

\makebox[\textwidth][c]{%
  \begin{subfigure}[b]{0.8\textwidth}
    \centering
    \includegraphics[width=\linewidth,keepaspectratio]{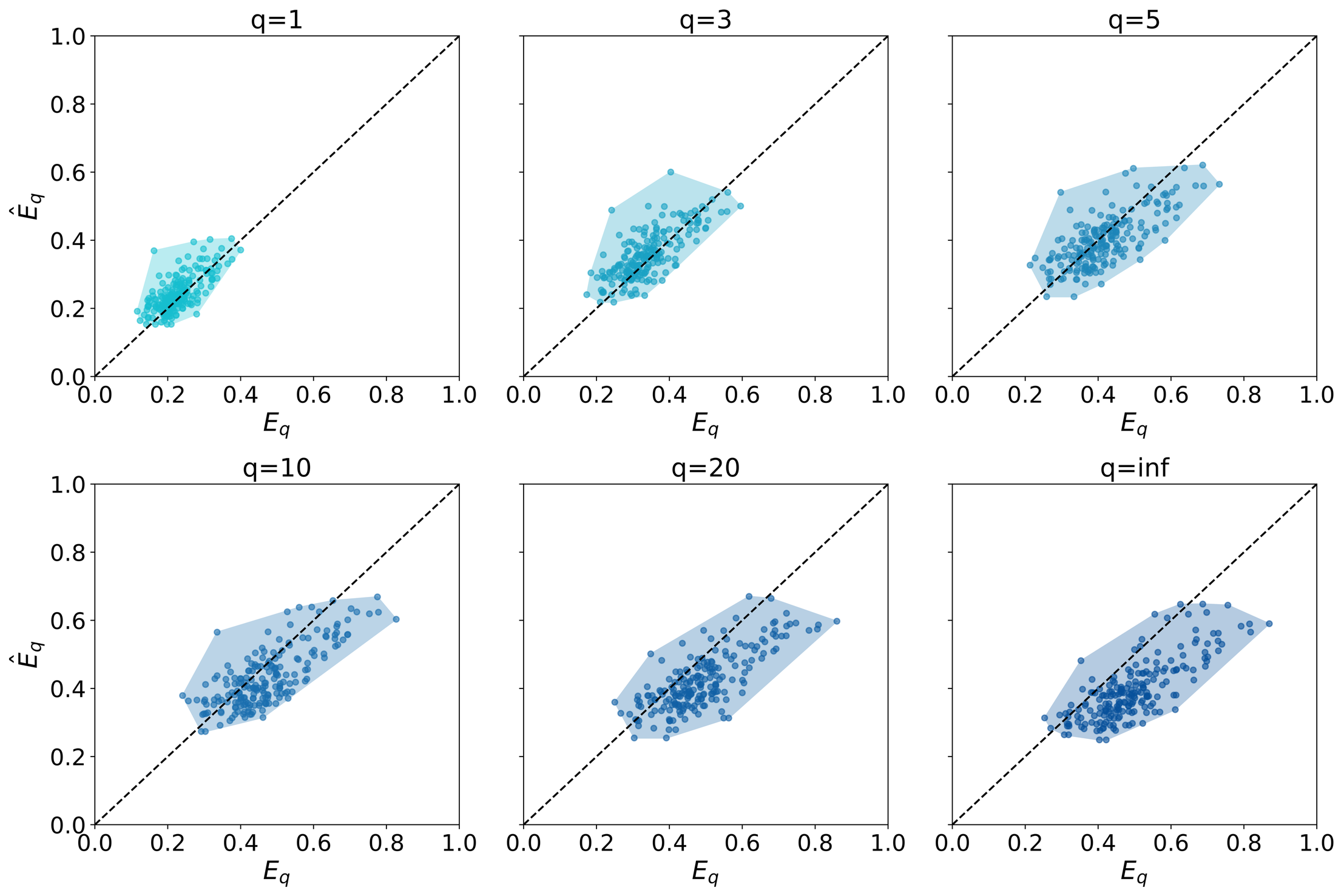}
    \caption{Test}
  \end{subfigure}%
}

\caption{Distribution of the distance to Nearest Neighbor. In the X-axis $E_q$ depicts distance values obtained after projecting when a query point is added. The Y-axis shows the learned approximation of the canonical projection $\hat{E}_q$. The dashed line represents the perfect match between projected and approximated $x=y$.}
\label{app:scatter}
\end{figure*}

Predicted and ground-truth distances in Figure~\ref{app:scatter} exhibit similar distributions for training and testing, albeit with a clear generalization gap. As $q$ increases, the approximation error increases accordingly. The model architecture is the multilayer perceptron shown in Table~\ref{tab:embedding-arch}; an Optuna study \cite{optuna} was used to select the hyperparameters of each experiment, including the number of layers, intermediate widths, and the final embedding dimension.

\begin{table}[t]
\centering
\caption{Embedding network (MLP) architecture. Optuna is used to tune the number of blocks $L$ and their hidden width.}
\label{tab:embedding-arch}
\renewcommand{\arraystretch}{1.25}
\setlength{\tabcolsep}{8pt}
\begin{tabular}{ll}
\toprule
\textbf{Model} & \textbf{EmbNet} \\
\midrule
Block & Linear $\rightarrow$ GELU $\rightarrow$ Dropout \\
Output layer & Linear \\
\bottomrule
\end{tabular}
\end{table} 

\begin{wrapfigure}{R}{0.58\columnwidth}
    \centering
    \vspace{-1.0em}
    \includegraphics[width=\linewidth]{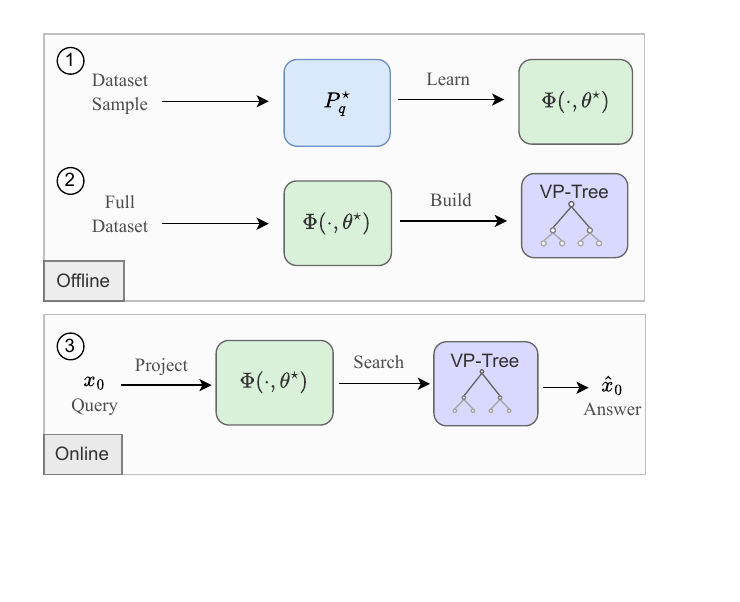}
    \vspace{-0.8cm}
    \caption{Infinity Search pipeline. Offline (up): Dataset samples are projected into a $q$-metric space using the canonical projection, and used to fit an embedding projection (1). Then, the dataset is transformed with the learned projection to build the VP-tree (2). Online (down): A query $x_o$ is transformed with the learned projection and then search is conducted using the VP-tree index (3).}
    \label{fig:infinity_pipeline}
    \vspace{-1.0em}
\end{wrapfigure}
\subsection{Searching with \texorpdfstring{$\phi_q(\cdot, \theta^\star)$}{Phiq}}\label{app:approximation}

In this section we demonstrate that the speedup observed in the exact projection experiments of Section~\ref{app:metric_representation} is also attained when replacing the Canonical projection with the learned map $\phi_q(\cdot, \theta^\star)$, at the cost of small errors in search results. The final $\infty$-Search framework is summarized in Figure~\ref{fig:infinity_pipeline}.

Figures~\ref{app:inductive_haven1} and~\ref{app:inductive_haven2} show a consistent reduction in the number of comparisons as $q$ increases, mirroring the trend observed in the exact projection experiments. This reduction is accompanied by a moderate increase in rank error and a decrease in recall, suggesting that the retrieved neighbors remain close but are not always exact. Nevertheless, cases with recall above 0.9 still yield substantial speedups. Since recall captures only exact matches and the method guarantees order preservation only for the 1-nearest neighbor, rank order can provide a complementary view of performance.

Despite the degradation of the approximation quality for high $q$ values, the retrieved neighbors remain within the top 80 nearest, which corresponds to less than $1\%$ of the indexed dataset. When higher precision is required, setting $q=10$ can yield a two-orders-of-magnitude speedup (Figure \ref{app:inductive_haven1}) while returning neighbors ranked around 10th, corresponding to a relative error close to $0.12\%$. These results demonstrate that the learned embeddings preserve local structure to a satisfactory  extent.

The results also extend to $\mathrm{k} $-nearest neighbor search with $k>1$. While the reduction in comparisons is less pronounced than for $k=1$, the method maintains a consistent speedup across values of $q$. Rank Order remains low, and in some cases improves as $k$ increases, suggesting that the learned map preserves small-scale neighborhood structure reasonably well even on these datasets.

Examples of retrieval results were also generated. For each dataset, queries were selected from varied categories. In the Fashion-MNIST case (Figure~\ref{fig:retrieval_main}), each panel shows the original query image, its true nearest neighbor, and the result returned by Infinity Search. While the exact nearest neighbor was not retrieved in some cases—such as those involving sandals or sneakers—the returned items consistently belonged to the same category as the query. For GloVe-200 text embeddings, Figure~\ref{fig:retrieval_glove} includes several exact matches, as well as examples where the retrieved word preserved the semantic meaning of the query.

\begin{figure*}[t]

    \centering
    \includegraphics[width=1\textwidth]{legend_only.png}
    \vspace{-0.3cm}

    \begin{subfigure}[b]{1\textwidth}
        \centering
        \includegraphics[width=\linewidth]{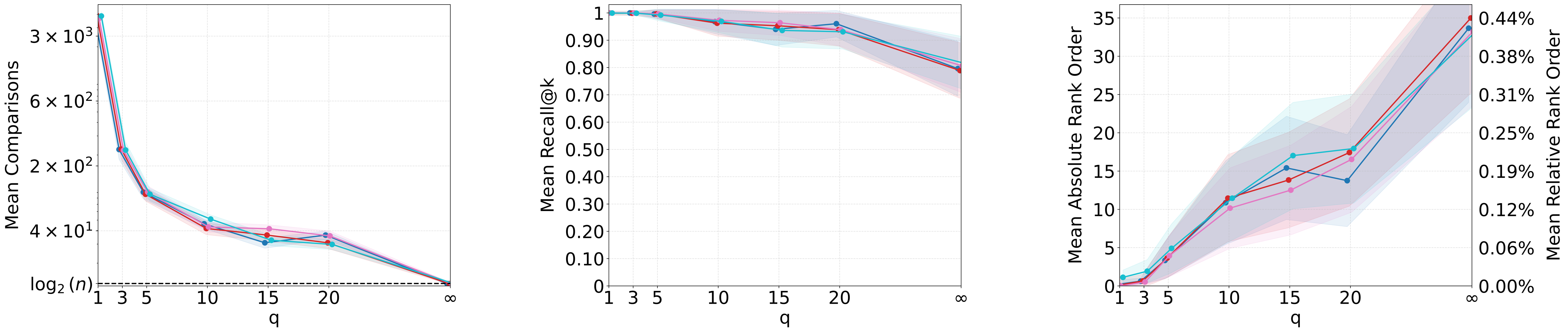}
        \vspace{0.3cm}
        \includegraphics[width=\linewidth]{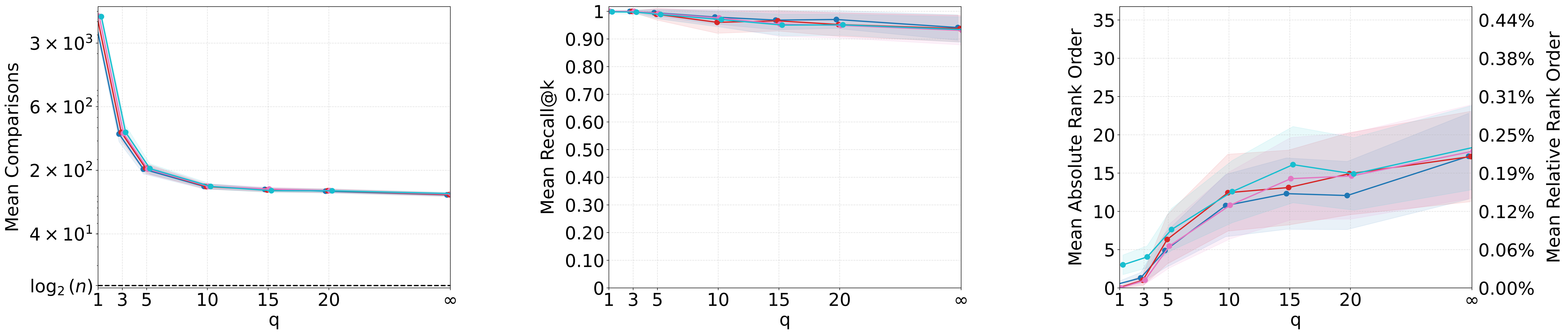}
        \vspace{0.3cm}
        \includegraphics[width=\linewidth]{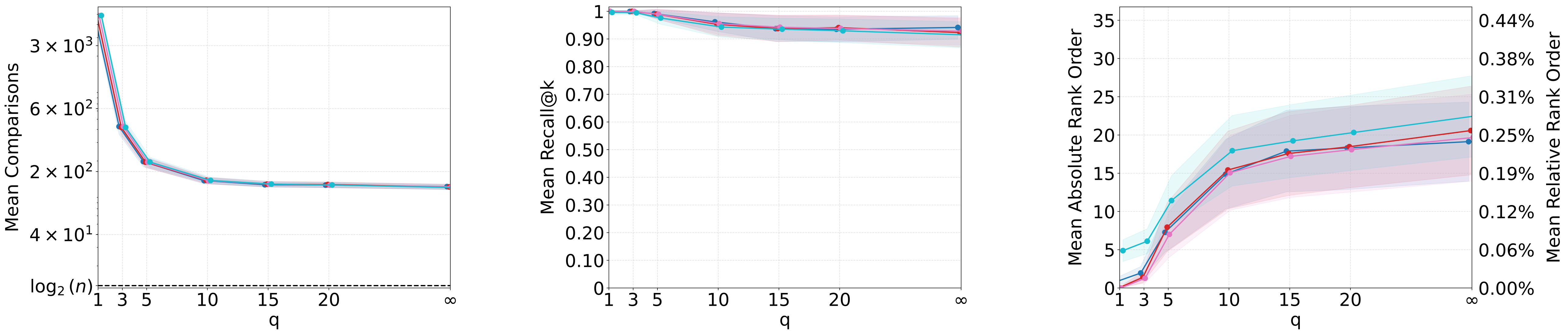}
        \caption{$n=10{,}000$ points of Fashion-MNIST}
        \label{app:inductive_haven1}
    \end{subfigure}%
    \vspace{0.5cm}

    \caption{Number of comparisons and rank order when searching after approximating the Canonical Projection $\hat{E}_q$, with the learned map $\Phi(x;\theta)$. Solid lines denote the mean and shading the standard deviation computed across queries. The $k$-nearest neighbors are listed from top to bottom for $k=1,5,10$}

\end{figure*}

\begin{figure*}[!htbp]
    \ContinuedFloat
    \centering
    \includegraphics[width=1\textwidth]{legend_only.png}
    \vspace{-0.3cm}

    \begin{subfigure}[b]{1\textwidth}
        \centering
        \includegraphics[width=\linewidth,keepaspectratio]{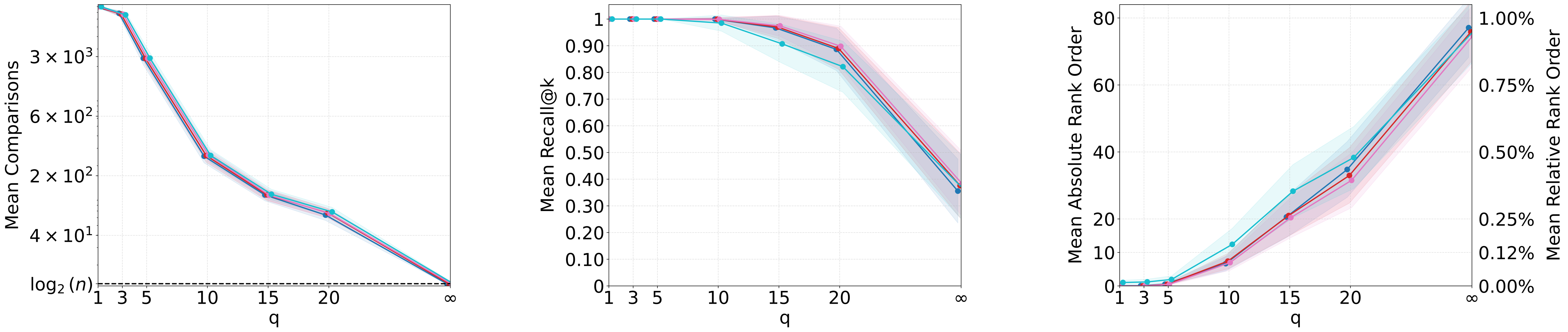}
        \vspace{0.3cm}
        \includegraphics[width=\linewidth,keepaspectratio]{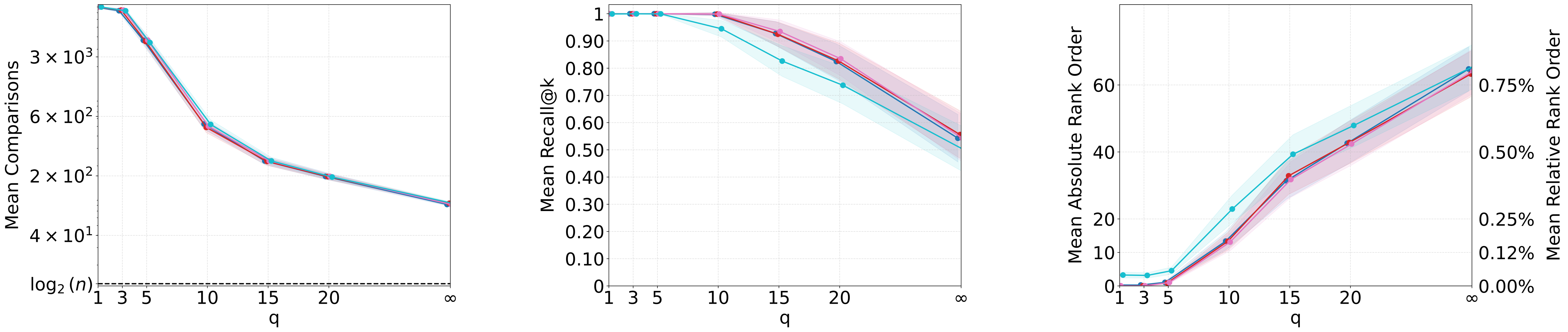}
        \vspace{0.3cm}
        \includegraphics[width=\linewidth,keepaspectratio]{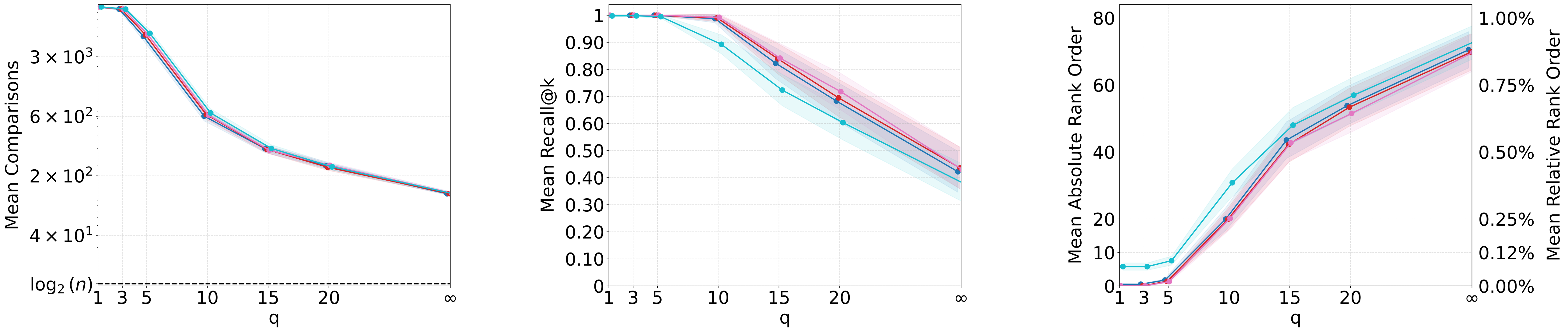}
        \caption{$n=10{,}000$ points of GloVe}
        \label{app:inductive_haven2}
    \end{subfigure}%

    \caption{Number of comparisons and rank order when searching after approximating the Canonical Projection $\hat{E}_q$, with the learned map $\Phi(x;\theta)$. Solid lines denote the mean and shading the standard deviation computed across queries. The $k$-nearest neighbors are listed from top to bottom for $k=1,5,10$}

\end{figure*}

\begin{figure*}[!htbp]
    \centering
    \begin{subfigure}[b]{0.8\textwidth}
        \centering
        \includegraphics[
            width=\linewidth,
            height=1\textheight,
            keepaspectratio
        ]{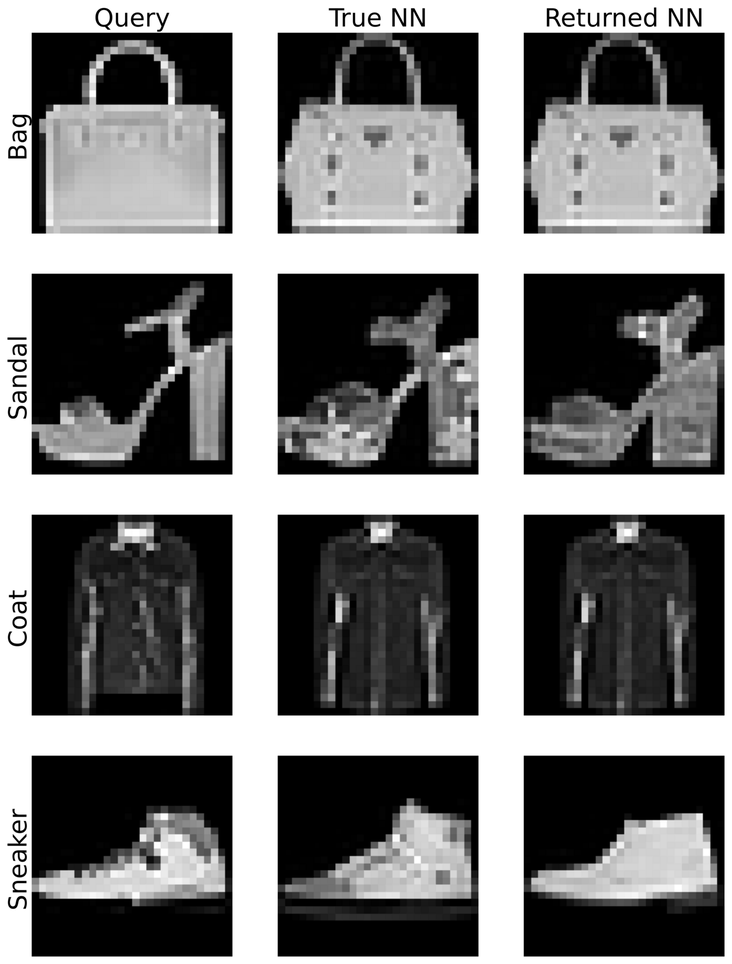}
        \caption{Fashion-MNIST}
    \end{subfigure}
\end{figure*}

\begin{figure*}[t]
    \ContinuedFloat
    \centering
    \begin{subfigure}[b]{1\textwidth}
        \centering
        \begin{tabular}{llll}
            \toprule
            \textbf{Category} & \textbf{Query} & \textbf{True NN} & \textbf{Returned NN} \\
            \midrule
            \textbf{Animals} & dog    & dogs     & dogs \\
                             & lion   & wolf     & wolf \\
            \textbf{Colors}  & blue   & pink     & purple \\
                             & red    & pink     & purple \\
            \textbf{Clothing} & pants & jeans    & jeans \\
                             & shirt  & shirts   & worn \\
            \textbf{Tools}   & drill  & drilling & drilling \\
                             & hammer & throw    & flame \\
            \bottomrule
        \end{tabular}
        \caption{GloVe}
        \label{fig:retrieval_glove}
    \end{subfigure}
\end{figure*}



\begin{figure*}[!htbp]
    \ContinuedFloat
    \centering
    \begin{subfigure}[b]{0.7\textwidth}
        \centering
        \includegraphics[
            width=\linewidth,
            height=1\textheight,
            keepaspectratio
        ]{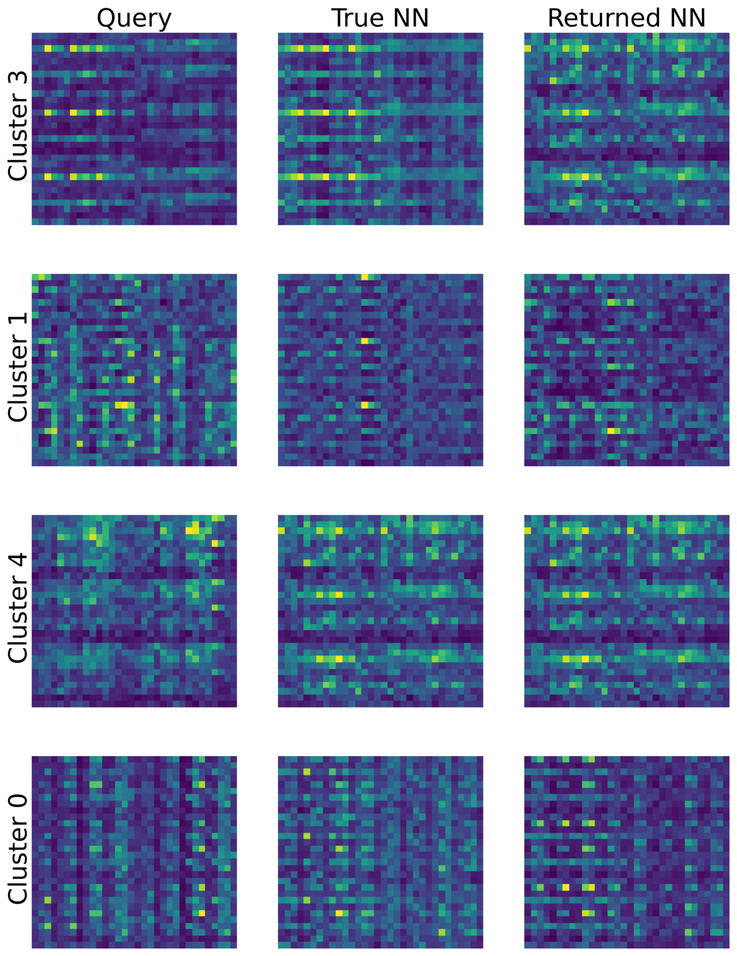}
        \caption{GIST}
    \end{subfigure}
    \caption{Retrieval of dataset items with Infinity Search ($q=5$).}
    \label{fig:retrieval_main}
\end{figure*}

\subsection{Two-Stage Infinity Search}\label{app:two_stage}
\begin{figure*}[!htbp]

    \begin{subfigure}[b]{1\textwidth}
        \centering
        \includegraphics[
            width=\linewidth,
            height=1\textheight,
            keepaspectratio
        ]{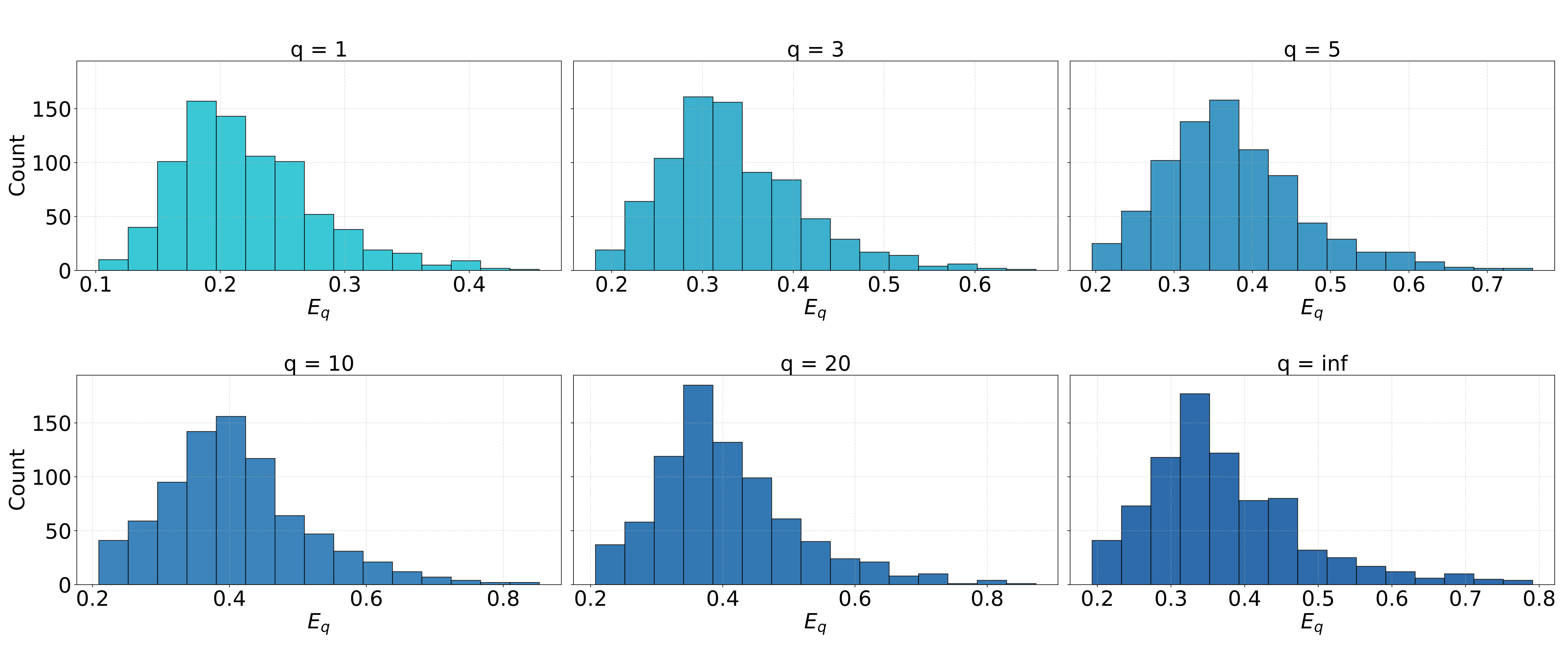}
        \caption{Train}
    \end{subfigure}%
    
    \vspace{0.5cm}

    \begin{subfigure}[b]{1\textwidth}
        \centering
        \includegraphics[
            width=\linewidth,
            height=1\textheight,
            keepaspectratio
        ]{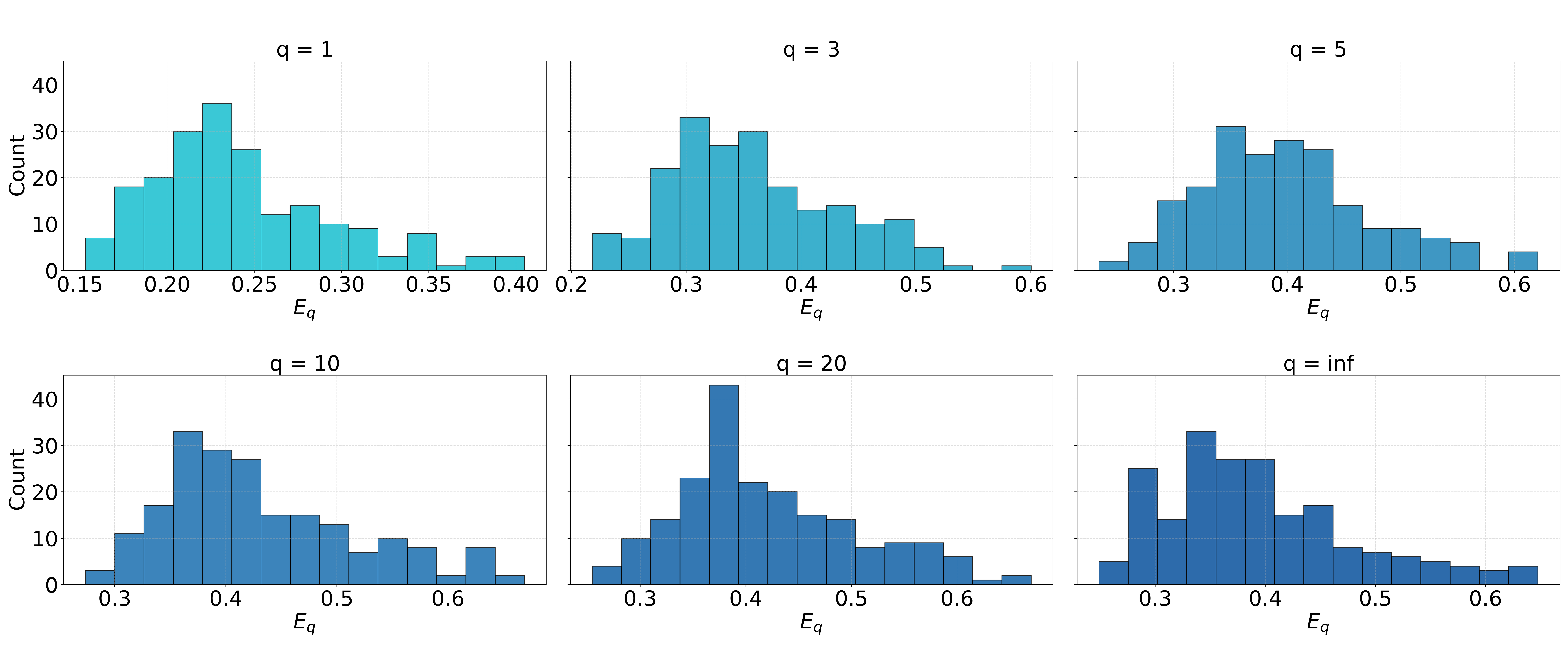}
        \caption{Test}
    \end{subfigure}%

    \caption{Histogram of the distance to Nearest Neighbor. In the X-axis $E_q$ depicts distance values obtained after projecting when a query point is added. The Y-axis shows number of counts for that distance bin.}
    \label{app:hist}
\end{figure*}

Although Proposition~\ref{eqn_nearest_neighbor_is_preserved} ensures preservation of the nearest neighbor structure, it also notes that the projected nearest neighbor may not remain unique. This ambiguity can lead to mismatches during retrieval. The effect appears in theoretical settings, as seen in Figures~\ref{app:haven1} and~\ref{app:haven2}, where rank order the rank order exhibits a sharp increase at $q = \infty$. A similar trend can be observed in the approximate setting, shown in Figures~\ref{app:inductive_haven1} and~\ref{app:inductive_haven2}. In Figure~\ref{app:hist}, the distribution of projected $q$-metrics shows a clustering effect as $q$ increases. Distances become more concentrated around mean, while the frequency of close values also increases. This suggests that distances between points become closer, making true nearest neighbors more difficult to discern.

To address this, one can extend the nearest neighbor set using the Canonical Projection and then prune it to avoid loss in accuracy. This motivates a two-stage modification of the Infinity Search algorithm:

 \begin{itemize}
     \item \textbf{Broad Search}: The Infinity Search algorithm is used to retrieve an initial candidate set of $K$ nearest neighbors. This will retrieve close neighbors in the $q$-metric space.

     \item \textbf{Specific Search}: Once the list of $K$ candidate neighbors is available, the original distance $D$ is used to retrieve the $k$ real nearest neighbors.

 \end{itemize}

This two-stage retrieval strategy is used in some ANN methods, including HNSW~\citep{hnsw}. As shown in the theoretical and approximate experiments of Sections~\ref{app:metric_representation} and~\ref{app:approximation}, the Canonical Projection preserves locality but not the exact order of nearest neighbors. This makes the two-stage approach suitable for improving accuracy.

Figures~\ref{app:best_inductive_haven1} and~\ref{app:best_inductive_haven2} confirm the improvement, showing higher recall and more accurate rank alignment compared to earlier approximations. The gains are particularly notable in rank order, with a 3 to 4 times reduction in error. As expected, this comes with a decrease in speedup, since the method processes a larger candidate set and computes original distances during Specific Search. Although the logarithmic comparison bound ($\log_2(n)$) no longer holds, the resulting speed remains competitive. The original Infinity Search can be recovered by setting $K = k$, while choosing $K > k$ offers additional flexibility to trade off speed and accuracy depending on the requirements of the searching problem.

\begin{figure*}[!htbp]

    \centering
    \includegraphics[width=1\textwidth]{legend_only.png}
    \vspace{-0.3cm}

    \begin{subfigure}[b]{1\textwidth}
        \centering
        \includegraphics[
            width=\linewidth
        ]{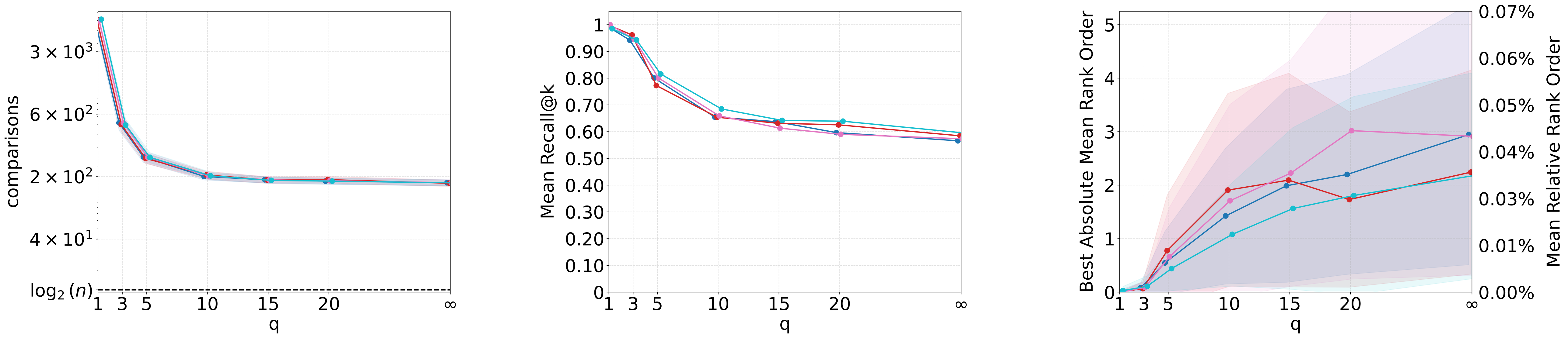}
        \vspace{0.3cm}
        \includegraphics[
            width=\linewidth
        ]{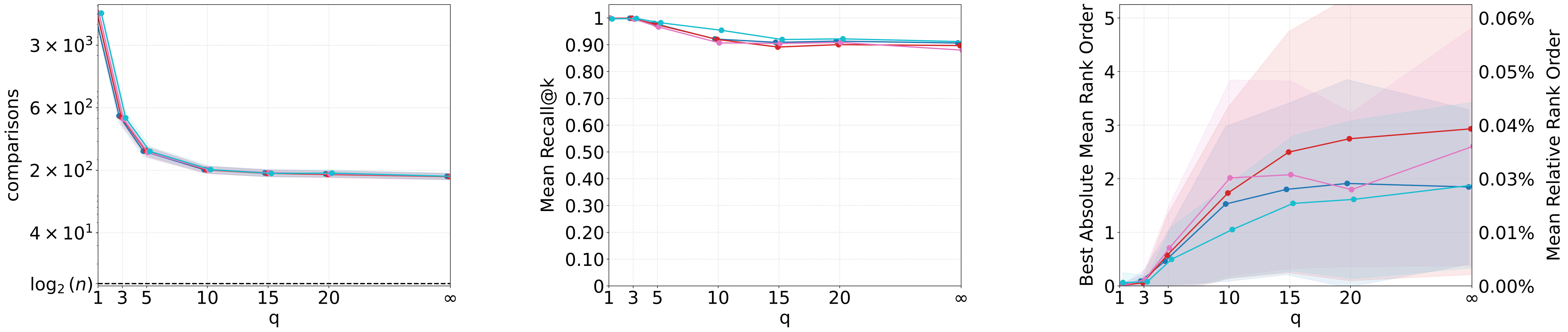}
        \vspace{0.3cm}
        \includegraphics[
            width=\linewidth
        ]{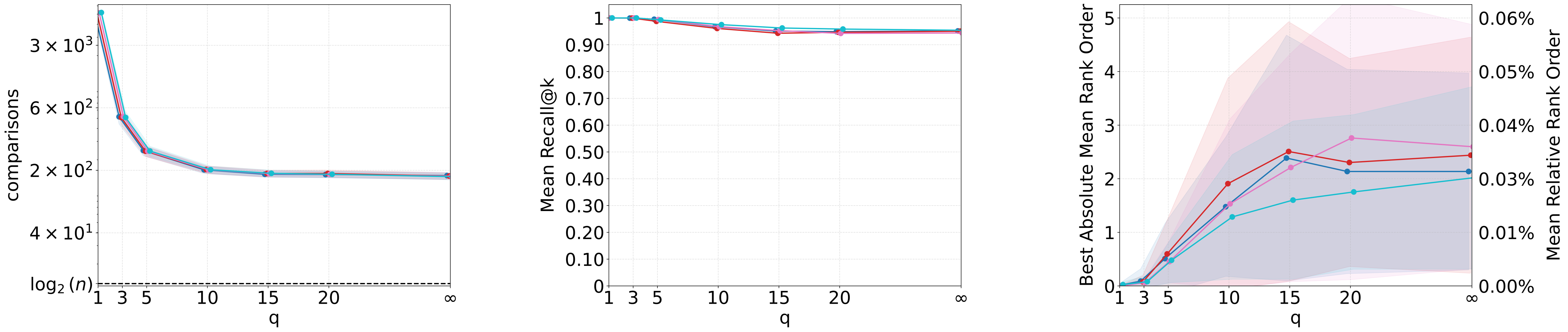}
        \caption{$n=10{,}000$ points of Fashion-MNIST}
        \label{app:best_inductive_haven1}
    \end{subfigure}%

    \vspace{0.5cm}

    \caption{Number of comparisons, Recall@k and Rank Order when searching with a two-stage retrieval Infinity Search. Solid lines denote the mean and shading the standard deviation computed across queries. The $k$-nearest neighbors are listed from top to bottom for $k=1,5,10$}

\end{figure*}

\begin{figure*}[!htbp]
    \ContinuedFloat
    \centering
    \includegraphics[width=1\textwidth]{legend_only.png}
    \vspace{-0.3cm}

    \begin{subfigure}[b]{1\textwidth}
        \centering
        \includegraphics[
            width=\linewidth,
            keepaspectratio
        ]{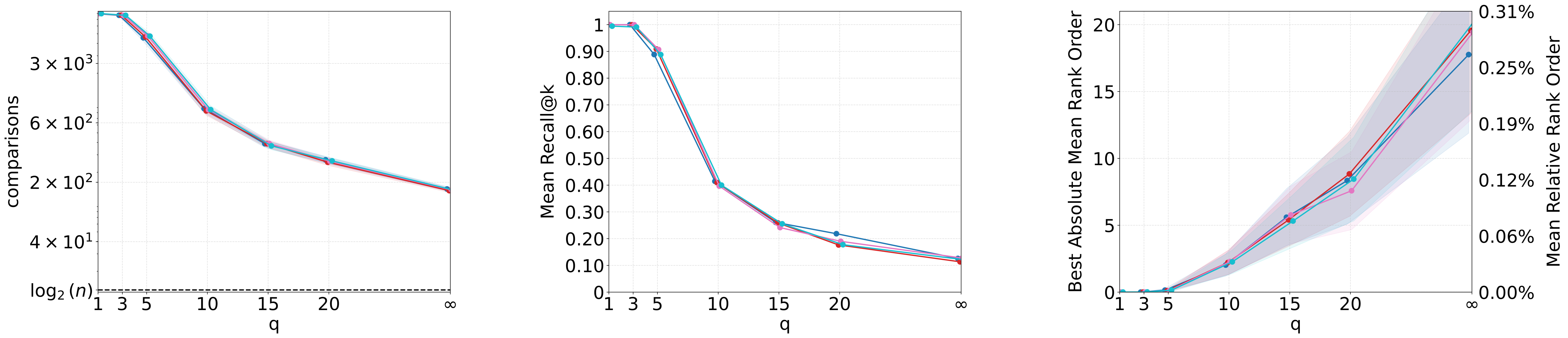}
        \vspace{0.3cm}
        \includegraphics[
            width=\linewidth,
            keepaspectratio
        ]{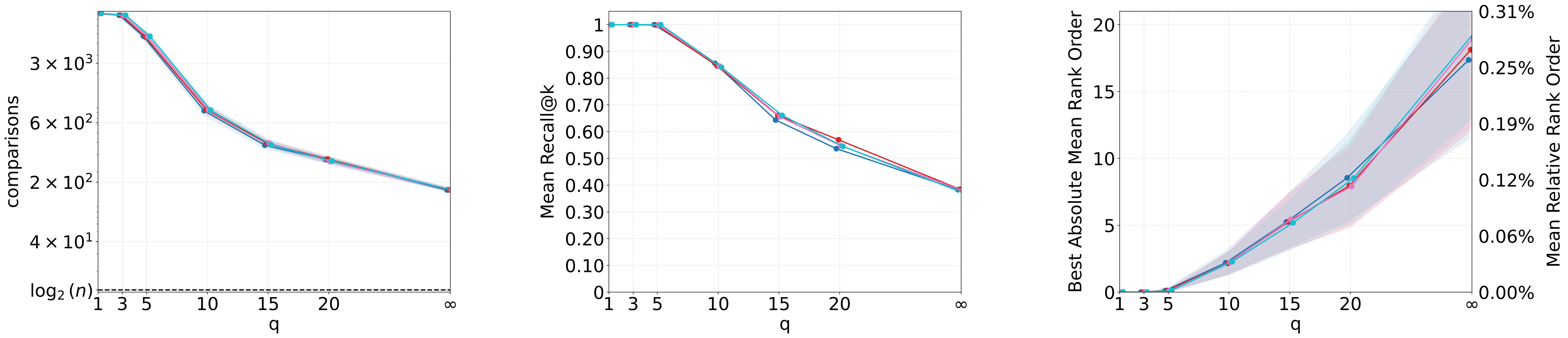}
        \vspace{0.3cm}
        \includegraphics[
            width=\linewidth,
            keepaspectratio
        ]{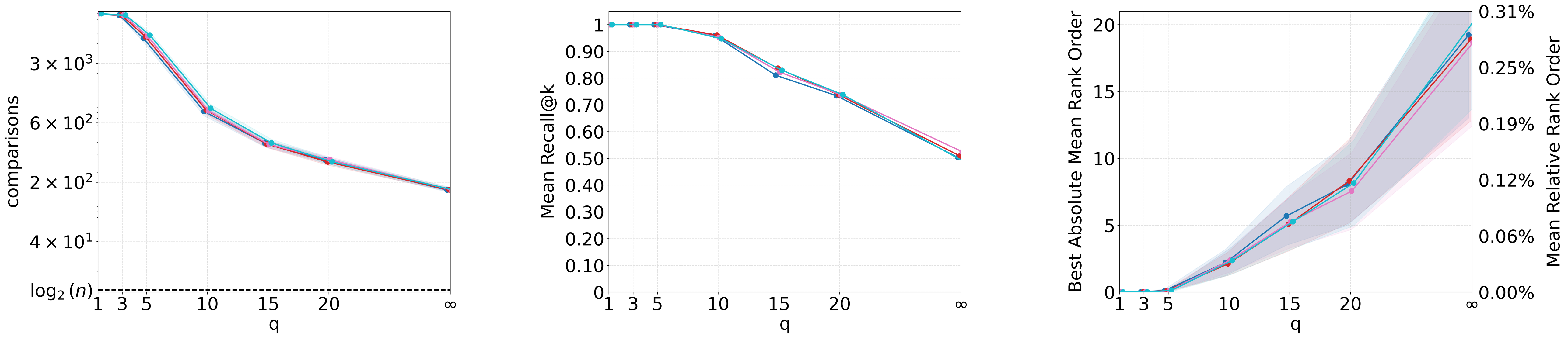}
        \caption{$n=10{,}000$ points of GloVe200}
        \label{app:best_inductive_haven2}
    \end{subfigure}%

    \vspace{0.5cm}

    \caption{Number of comparisons, Recall@k and Rank Order when searching with a two-stage retrieval Infinity Search. Solid lines denote the mean and shading the standard deviation computed across queries. The $k$-nearest neighbors are listed from top to bottom for $k=1,5,10$}

\end{figure*}

\subsection{VP-Forests}

The VP-Tree structure can be naturally extended to a forest of smaller VP-Trees, allowing the index to scale without substantially degrading search performance. We split each dataset into \(T\) independent subsets and train a corresponding embedding \(\Phi_1,\dots,\Phi_T\) for each of them before constructing their respective VP-Trees. At query time, the query is embedded using every \(\Phi_t\), and the search is performed across all trees in the forest, collecting the candidates returned by each tree. Because the trees are built using independent embeddings, their errors are distributed across the forest, and taking the union of their candidate sets can recover neighbors that may be missed by any individual tree. Moreover, each tree can be constructed independently, providing opportunities for parallelization.

Empirically, we find that using a small number of trees becomes beneficial as the number of points grows. In particular, we observe that the approximation error remains smaller than the median pruning threshold between vantage points and their children (Figure~\ref{fig:margin}). This separation between the learned distances and the pruning thresholds enables the VP-Trees to maintain effective pruning as the dataset scales.

\subsection{Extended Experimental reults}

\subsubsection{Infinity Search competes in ANN-Benchmarks}
\label{app:ann}

Infinity Search offers a configurable trade-off between query throughput and recall. To assess its competitiveness against modern ANN methods, we evaluated it within the ANN-Benchmarks framework \citep{DBLP:journals/corr/abs-1807-05614} with batch mode and allowing parallelization. The analysis was lmited to 10k points due to computational limitations . We ran experiments on five datasets provided by the library and compared against a wide set of algorithms chosen for their balance of speed, accuracy, and open-source availability.

\begin{figure*}[!htbp]
  \centering

  \includegraphics[width=\textwidth]{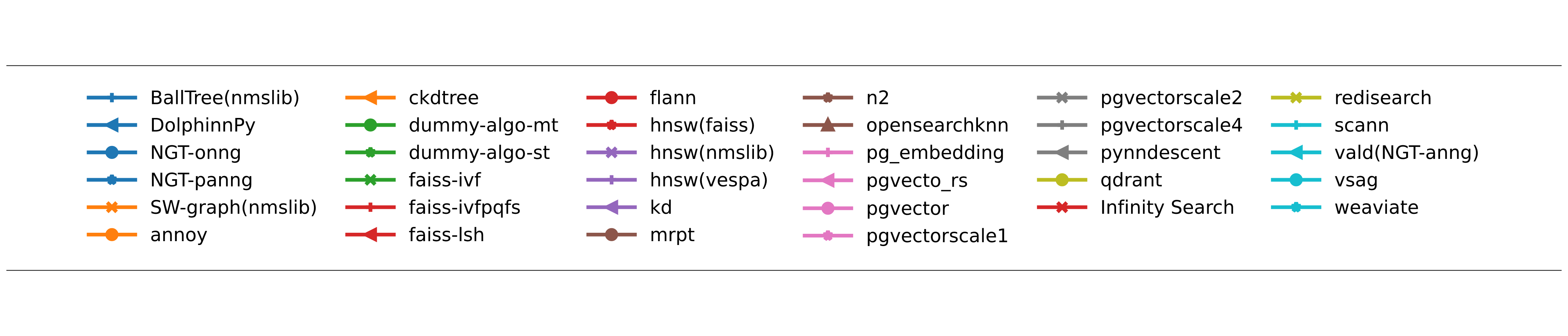}
  \vspace{-0.3cm}

  \begin{subfigure}[t]{0.8\textwidth}
    \centering
    \includegraphics[width=0.30\linewidth]{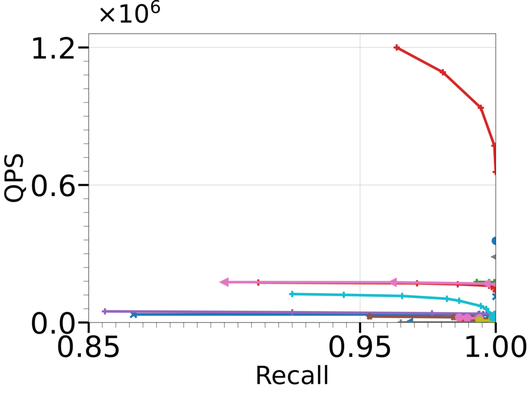}
    \includegraphics[width=0.30\linewidth]{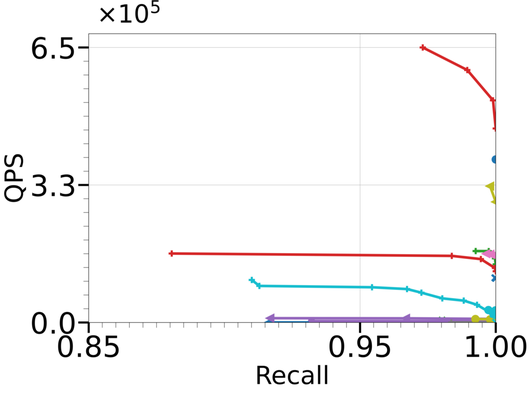}
    \includegraphics[width=0.30\linewidth]{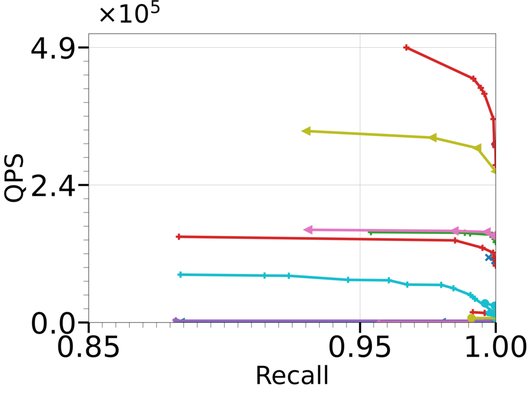}
    \caption{Fashion-MNIST-784 (Euclidean)}
    \label{app:best-fashion}
  \end{subfigure}

  \vspace{0.5cm}

  \begin{subfigure}[t]{0.8\textwidth}
    \centering
    \includegraphics[width=0.30\linewidth]{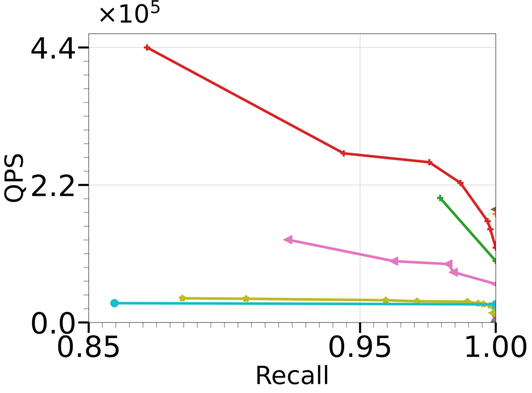}
    \includegraphics[width=0.30\linewidth]{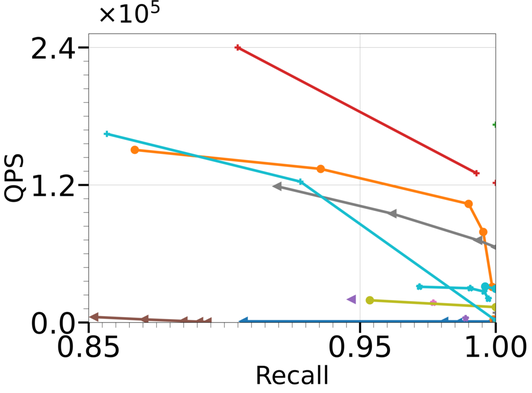}
    \includegraphics[width=0.30\linewidth]{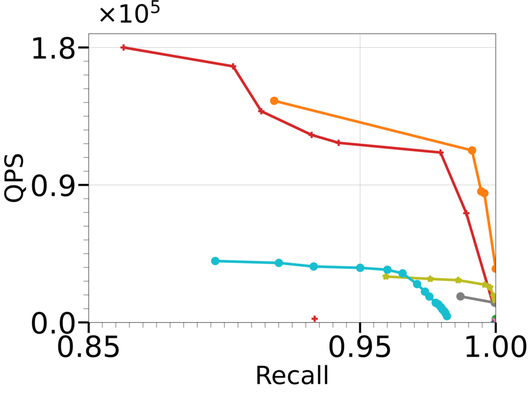}
    \caption{GloVe-200 (Cosine)}
    \label{app:best-glove}
  \end{subfigure}

  \vspace{0.5cm}

  \begin{subfigure}[t]{0.8\textwidth}
    \centering
    \includegraphics[width=0.30\linewidth]{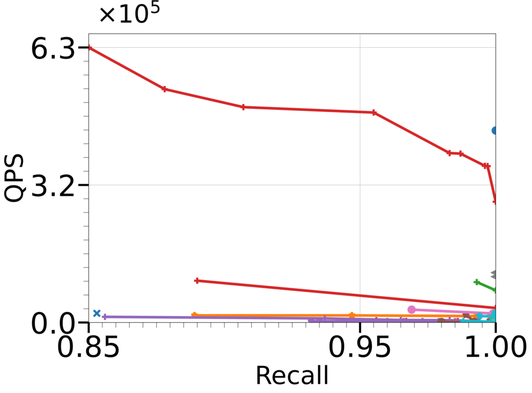}
    \includegraphics[width=0.30\linewidth]{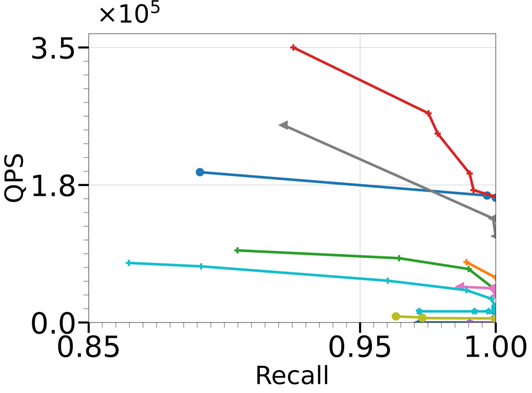}
    \includegraphics[width=0.30\linewidth]{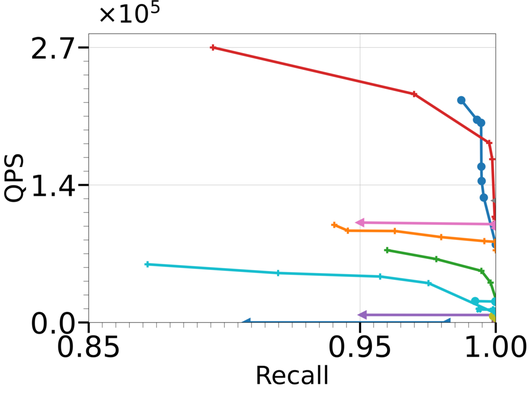}
    \caption{GIST-960 (Euclidean)}
    \label{app:best-gist}
  \end{subfigure}

  \vspace{0.5cm}

  \begin{subfigure}[t]{0.8\textwidth}
    \centering
    \includegraphics[width=0.30\linewidth]{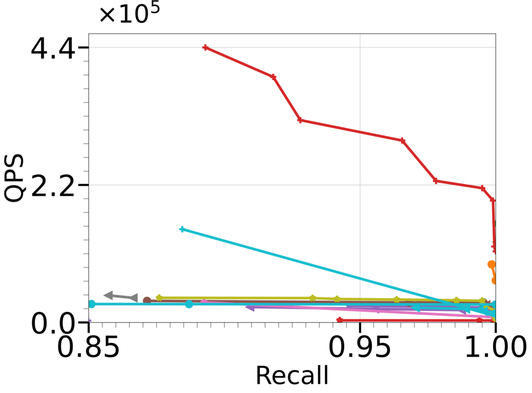}
    \includegraphics[width=0.30\linewidth]{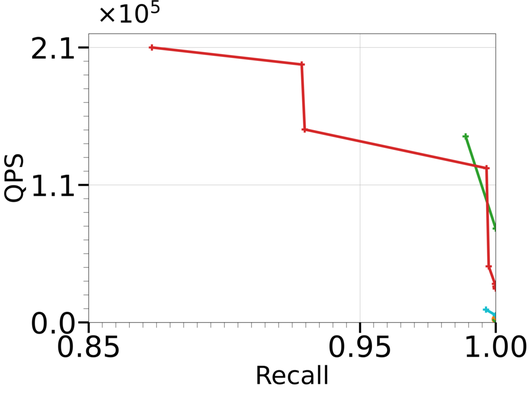}
    \includegraphics[width=0.30\linewidth]{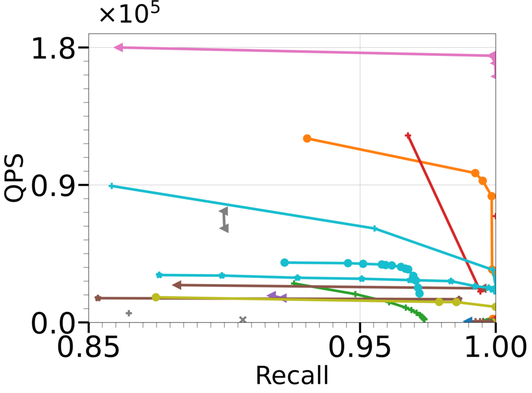}
    \caption{NYTimes-256 (Cosine)}
    \label{app:best-nytimes}
  \end{subfigure}

  \vspace{0.5cm}

  \begin{subfigure}[t]{0.8\textwidth}
    \centering
    \includegraphics[width=0.30\linewidth]{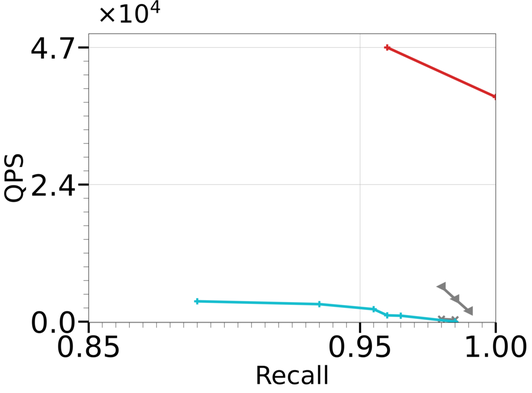}
    \includegraphics[width=0.30\linewidth]{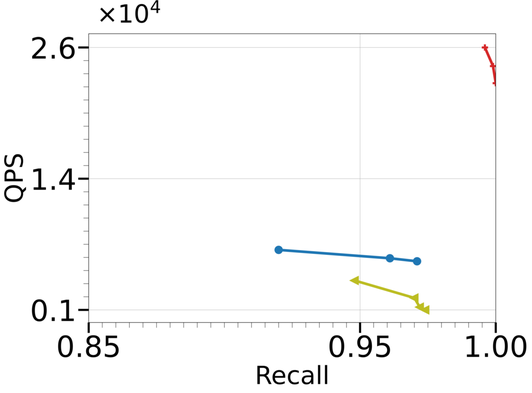}
    \includegraphics[width=0.30\linewidth]{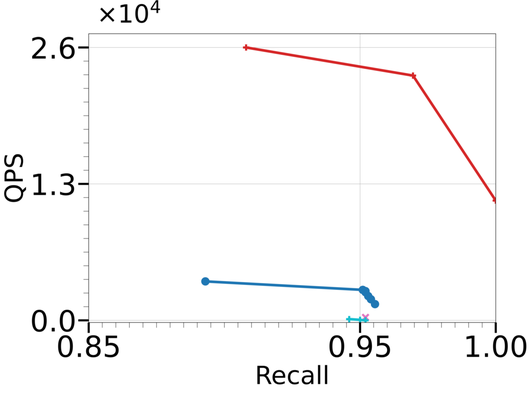}
    \caption{Kosarak-41,000 (Jaccard)}
    \label{app:best-kosarak}
  \end{subfigure}

  \caption{ANN-Benchmarks comparison for $n=10K$. Recall@k vs queries-per-second (QPS) across datasets. Columns within each row are $k=1,5,10$.}
  \label{app:ann-best-recall-qps}
\end{figure*}

In both theoretical analysis and empirical benchmarks, Infinity Search consistently accelerates nearest‐neighbor queries across all tested dissimilarities. On moderate‐dimensional datasets such as Fashion‐MNIST and GIST (Figure~\ref{app:ann-best-recall-qps}), it delivers a clear speedup by, in some cases, sacrificing perfect accuracy. Remarkably, on the high‐dimensional Kosarak dataset—with Jaccard dissimilarity—Infinity Search outperforms competing methods by an even wider margin. This supports the flexibility of the method when less popular dissimilarities are required.

Across all datasets, it offers a favorable speed–accuracy trade‐off for the $k=1$ nearest‐neighbor task. For larger neighborhood sizes ($k \in \{5,10\}$), the increased comparison overhead prevents it from always leading in Recall@k; nevertheless, its performance remains competitive. Note that, since $n = 10{,}000$ in these experiments, modest rank‐order errors at extreme speeds still correspond to few misplaced neighbors.

Overall, Infinity Search is a viable alternative when fast retrieval is required or non‐Euclidean or less structured similarity are used. 

\subsubsection{Infinity Search competes in larger scale benchmarks}
We provide an extension of the results in Section~\ref{section:experiments}. In particular, we show the Pareto curves for retrieving $k\in{5,10}$ nearest neighbors. As expected, the performance of Infinity Search embeddings degrades slightly when retrieving a larger number of neighbors; however, the overall performance remains comparable to that observed for $k=1$.

\begin{figure}[H]
\centering
\includegraphics[width=1\linewidth]{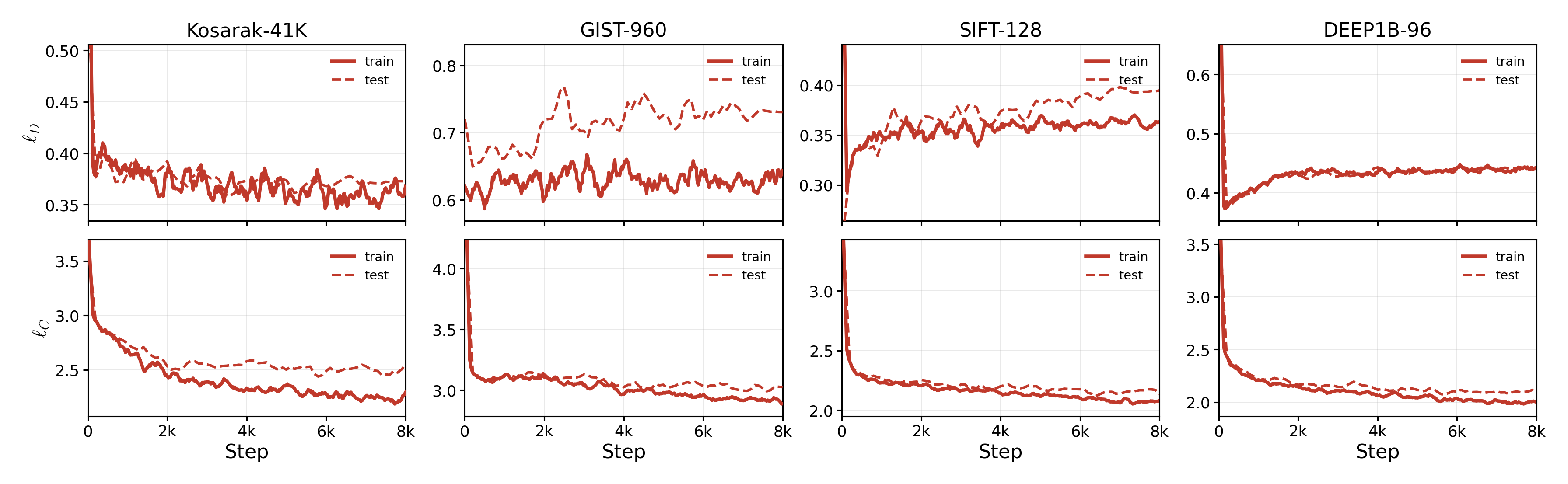}
\caption{\textbf{Training loss curves.} We observe a sustained decrease in both $\ell_D$ and $\ell_C$ training loss curves.}
\label{app:fig:loss}
\end{figure}

Additionally, we provide the training loss curves. Figure \ref{app:fig:loss} depicts a sustained decrease in both training and test contrastive and distance losses across datasets. 
\begin{figure}[H]
    \centering
    \includegraphics[width=1\linewidth]{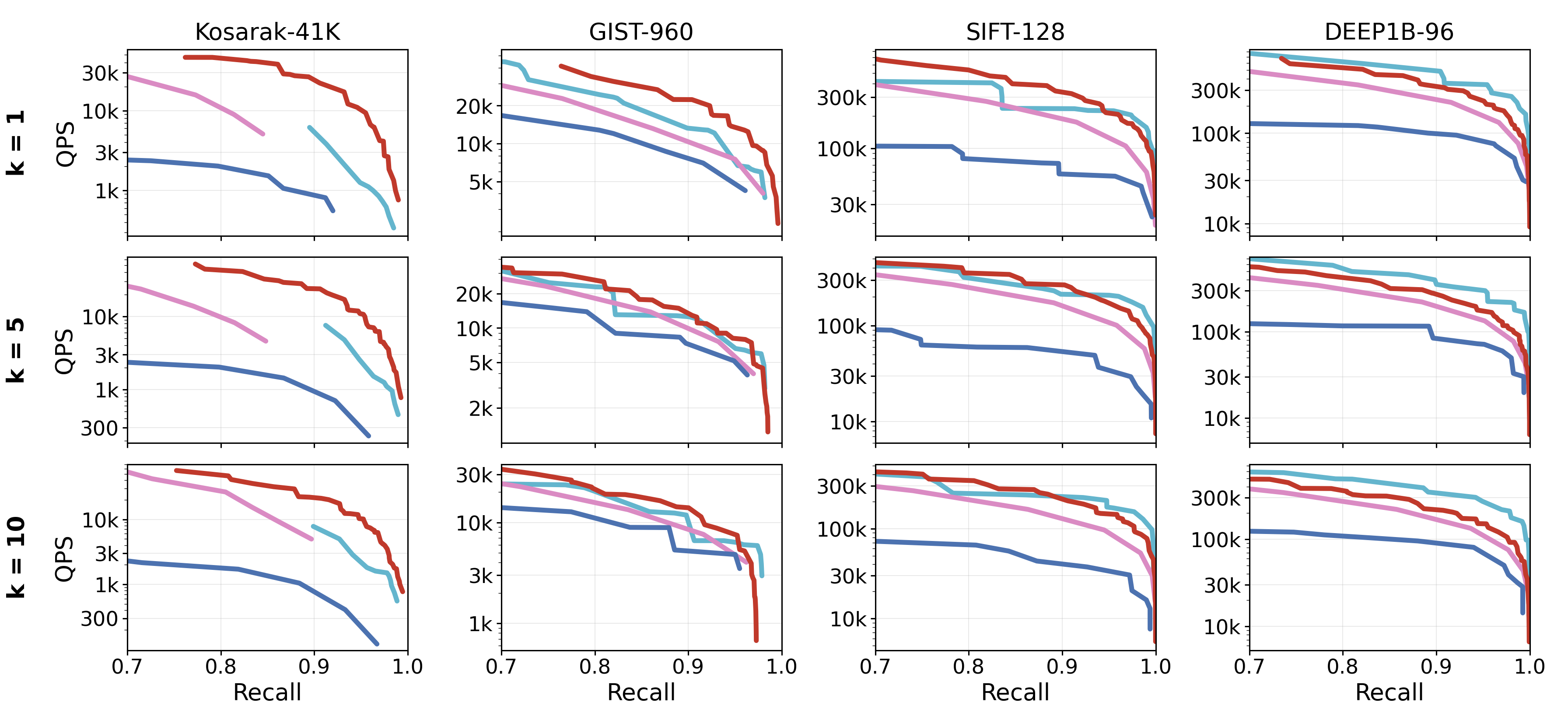}
\caption{\textbf{Infinity Search supports retrieving multiple neighbors.} Pareto curves for retrieving $k\in\{5,10\}$ nearest neighbors. Performance degrades only slightly compared with retrieving a single nearest neighbor.}
    \label{app:fig:extended}
\end{figure}

\end{document}